\documentclass[aps,prd,preprintnumbers,groupedaddress,nofootinbib,amssymb,eqsecnum,notitlepage]{revtex4-1}
\usepackage{here}
\usepackage[dvipdfmx]{graphicx}
\usepackage{amsmath,amsthm,amssymb}
\usepackage{bm}
\usepackage{color}

\usepackage{amsfonts}
\usepackage{dcolumn}
\usepackage{hyperref}
\allowdisplaybreaks[1]
\usepackage{stackengine}

\begin{document}

\newcommand{\newc}{\newcommand}
\newcommand{\rk}{\textcolor{red}}
\newc{\be}{\begin{equation}}
\newc{\ee}{\end{equation}}
\newc{\ba}{\begin{eqnarray}}
\newc{\ea}{\end{eqnarray}}
\newc{\D}{\partial}
\newc{\rH}{{\rm H}}
\newc{\rd}{{\rm d}}
\newc{\Mpl}{M_{\rm Pl}}
\newcommand{\rBH}{r_{s}}
\newcommand{\rc}{r_{c}}
\newcommand{\rh}{r_{h}}
\newcommand{\Xh}{X_{h}}
\newcommand{\hr}{\hat{r}}
\newcommand{\ma}[1]{\textcolor{magenta}{#1}}
\newcommand{\cy}[1]{\textcolor{cyan}{#1}}
\newcommand{\mm}[1]{\textcolor{red}{#1}}

\begin{flushright}
WUCG-22-11 \\
\end{flushright}

\title{Primordial black holes from Higgs inflation 
with a Gauss-Bonnet coupling}

\author{Ryodai Kawaguchi\footnote{{\tt ryodai0602@fuji.waseda.jp}} and Shinji Tsujikawa\footnote{{\tt tsujikawa@.waseda.jp}}}

\affiliation{Department of Physics, Waseda University, 3-4-1 Okubo, Shinjuku, Tokyo 169-8555, Japan} 

\begin{abstract}

Primordial black holes (PBHs) can be the source for all 
or a part of today's dark matter density. 
Inflation provides a mechanism for generating the seeds 
of PBHs in the presence of a temporal period where 
the velocity of an inflaton field $\phi$ rapidly decreases 
toward 0. We compute the primordial power spectra of 
curvature perturbations generated during Gauss-Bonnet (GB) 
corrected Higgs inflation in which the inflaton field
has not only a nonminimal coupling to gravity  
but also a GB coupling.  
For a scalar-GB coupling exhibiting a rapid change 
during inflation, we show that curvature perturbations 
are sufficiently enhanced by the appearance of an 
effective potential $V_{\rm eff}(\phi)$ 
containing the structures of plateau-type, 
bump-type, and their 
intermediate type. 
We find that there are parameter 
spaces in which PBHs can constitute 
all dark matter for these 
three types of $V_{\rm eff}(\phi)$. 
In particular, models with bump- and intermediate-types 
give rise to the primordial scalar and tensor power spectra 
consistent with the recent Planck data  
on scales relevant to the observations of 
cosmic microwave background.
This property is attributed to the fact that the number of 
e-foldings $\Delta N_c$ acquired around the bump region 
of $V_{\rm eff}(\phi)$ can be as small as a few, 
in contrast to the plateau-type where $\Delta N_c$ 
typically exceeds the order of 10.

\end{abstract}

\date{\today}


\maketitle

\section{Introduction}
\label{introsec}

If there were over-density regions in the early universe, 
primordial black holes (PBHs) may form as a result of the gravitational 
collapse during the radiation-dominated 
epoch \cite{zel1967hypothesis,Hawking:1971ei,Carr:1974nx}. 
Unlike astrophysical black holes, the PBHs can have a wide range of masses 
and can be the source for all or a part of dark matter 
(DM) \cite{Chapline:1975ojl,Meszaros:1975ef} (see Refs.~\cite{Khlopov:2008qy,Sasaki:2018dmp,Carr:2020xqk,Green:2020jor,Villanueva-Domingo:2021spv,
Carr:2021bzv,Escriva:2022duf,Karam:2022nym} for recent reviews). 
Although PBHs have not been observationally discovered yet, the detection of gravitational waves from binary black holes \cite{LIGOScientific:2016aoc,LIGOScientific:2018mvr,LIGOScientific:2020ibl,LIGOScientific:2021djp} suggested a possibility that they may arise from non-stellar origins \cite{Bird:2016dcv,Sasaki:2016jop,Clesse:2016vqa,Wang:2016ana}.
In addition, PBHs have been also considered as possible seeds of supermassive 
black holes in the center of galaxies \cite{Bean:2002kx}.
Various observations have given the upper limit to their abundance 
$f$ as a function of mass $M$.
In particular, the mass window in which all DM can be explained 
by PBHs exists in the range 
$10^{-16}M_{\odot} \lesssim M \lesssim 10^{-11}M_{\odot}$, 
where $M_{\odot}$ is a solar mass 
(see Ref.~\cite{Carr:2020gox} for a recent study).

Inflation can provide a possible framework for generating 
the seed of PBHs on scales smaller than the observed 
cosmic microwave background (CMB) temperature anisotropies
\cite{Ivanov:1994pa,Garcia-Bellido:1996mdl,Bullock:1996at,Yokoyama:1995ex,Yokoyama:1998pt,Kawasaki:1997ju,Kawasaki:2006zv,Kohri:2007qn,Saito:2008em,
Bugaev:2008bi,Alabidi:2009bk,Drees:2011hb,Drees:2011yz,Martin:2012pe,Kohri:2012yw,Kawasaki:2012wr,Clesse:2015wea,Kawasaki:2015ppx,Kawasaki:2016pql,Pi:2017gih,Garcia-Bellido:2017mdw,Kannike:2017bxn,Germani:2017bcs,Ando:2017veq,Ezquiaga:2017fvi,
Motohashi:2017kbs,Di:2017ndc,Ballesteros:2017fsr,Garcia-Bellido:2017aan,Hertzberg:2017dkh,Inomata:2018cht,Cai:2018tuh,Drees:2019xpp,Atal:2019cdz,Atal:2019erb,Mishra:2019pzq,Cheong:2019vzl,Fu:2019ttf,Dalianis:2019vit,Ashoorioon:2019xqc,Lin:2020goi,Yi:2020kmq,Palma:2020ejf,Braglia:2020eai,
Kefala:2020xsx,Ballesteros:2020qam,Aldabergenov:2020bpt,Aldabergenov:2020yok,Inomata:2021uqj,Inomata:2021tpx,Dalianis:2021iig,Cai:2021zsp,
Lin:2021vwc,Kawai:2021edk,Zhang:2021rqs,Ahmed:2021ucx,Cai:2022erk,Pi:2022zxs,Cheong:2022gfc,Kawai:2022emp}. 
If there is an intermediate stage in which the velocity 
$\dot{\phi}$ of an inflaton field rapidly decreases
toward 0 during inflation, it is possible to 
enhance curvature perturbations at particular scales.
In the presence of an inflection point in the inflaton potential $V(\phi)$ 
around which the derivative ${\rm d}V/{\rm d}\phi$ is close 
to 0, the field velocity decreases as $\dot{\phi} \propto a^{-3}$ 
in the ultra-slow-roll (USR) regime, where $a$ is a scale factor 
\cite{Garcia-Bellido:2017mdw,Kannike:2017bxn,Germani:2017bcs,Ezquiaga:2017fvi,Motohashi:2017kbs,Di:2017ndc,
Ballesteros:2017fsr,Hertzberg:2017dkh,Drees:2019xpp,Ballesteros:2020qam}. 
In many of these models, we require a tuning of model parameters 
to generate a plateau region of the potential. 
Moreover, if the number of e-foldings acquired in the  
USR regime exceeds the order 10, the scalar spectral index $n_s$ 
on CMB scales tends to be inconsistent with the value constrained by 
the Planck data \cite{Planck:2018jri}.

There are also models containing one or more bumps/dips or steps in the potential 
\cite{Atal:2019cdz,Atal:2019erb,Mishra:2019pzq,Kefala:2020xsx,Inomata:2021uqj,Inomata:2021tpx,Dalianis:2021iig,Cai:2021zsp,Cai:2022erk,Pi:2022zxs}. 
In this case, the scalar field rapidly loses its kinetic energy around them, resulting in 
a strong enhancement of curvature perturbations.
It is also known that oscillating features can appear 
in the scalar power spectrum especially for the step-type potential.
One advantage of these models is that the number of e-foldings 
during transition can be as small as the order 1. 
This allows a possibility for the compatibility of models 
with the observed values of 
$n_s$ and tensor-to-scalar ratio $r$. 
There are also multi-field inflationary models leading to the enhancement of 
curvature perturbations at particular scales \cite{Yokoyama:1995ex,Garcia-Bellido:1996mdl,Kawasaki:1997ju,
Kawasaki:2012wr,Kohri:2012yw,Ando:2017veq,Aldabergenov:2020bpt,Aldabergenov:2020yok,Palma:2020ejf,Braglia:2020eai,Cheong:2022gfc,Kawai:2022emp}. 
In this case, we need to address whether the presence of entropy perturbations
does not contradict with CMB constraints on the isocurvature mode.
The possibility for producing the seed of PBHs during preheating 
after inflation was also discussed in 
Refs.~\cite{Green:2000he,Bassett:2000ha,Suyama:2004mz,Martin:2019nuw}.

One of the advantages of PBHs as DM is that the origin of DM can be explained 
within the framework of Standard Model (SM) of particle physics. 
A minimal model of inflation without introducing additional scalar degrees 
of freedom to those appearing in SM is known as Higgs inflation, 
in which the Higgs field $\phi$ is nonminimally coupled to 
gravity \cite{Bezrukov:2007ep,Bezrukov:2008ej,Bezrukov:2010jz} 
(see Refs.~\cite{Futamase:1987ua,Fakir:1990eg} for early related works).
Indeed, this model is perfectly consistent with observational bounds 
on $n_s$ and $r$ constrained by the CMB 
data \cite{Komatsu:1999mt,Tsujikawa:2004my,Linde:2011nh,Planck:2013jfk,Tsujikawa:2013ila}.
If we allow the runnings of Higgs self-coupling $\lambda(\phi)$ and nonminimal 
coupling $\xi (\phi)$, then it is possible to have an inflection point 
in the Higgs potential. This gives rise to a plateau region in which the enhancement 
of curvature perturbations occurs to generate the seed of 
PBHs \cite{Ezquiaga:2017fvi,Drees:2019xpp,Yi:2020kmq,Lin:2021vwc}.
In this scenario the typical number of e-foldings acquired during the USR regime 
is of order 10, which results in the values of CMB observables 
deviating from those in standard Higgs inflation. 
The model can be consistent with the current CMB observations,  
but it is typically outside the $1\sigma$ contour constrained by 
the Planck data \cite{Ezquiaga:2017fvi}.

Recently, Kawai and Kim \cite{Kawai:2021edk} proposed a single-field inflationary scenario in which the inflaton field is coupled to a Gauss-Bonnet (GB) curvature invariant $R_{\rm GB}^2$ of the form 
$\mu(\phi)R_{\rm GB}^2$. 
A scalar-field dependent GB coupling $\mu(\phi)$ 
can give rise to an inflection point $\phi=\phi_c$ in an effective potential of the inflaton. On the other hand, we have to caution that a large contribution 
from the scalar-GB coupling to the inflaton energy density 
modifies the primordial scalar power spectrum on 
CMB scales \cite{Hwang:2005hb,Guo:2006ct,Satoh:2008ck,Guo:2009uk,Guo:2010jr,Kawai:2021bye}.
Moreover, the scalar-GB coupling leads to the propagation speeds of 
scalar and tensor perturbations different from 
the speed of light \cite{KYY,Kase:2018aps}. 
Since the Laplacian instabilities associated with negative 
squared propagation speeds may arise, we need to make sure whether 
the stability conditions are not violated during inflation.

In Ref.~\cite{Kawai:2021edk}, the scalar-GB coupling 
$\mu(\phi)=\mu_0\tanh [\mu_1(\phi-\phi_c)]$ was proposed to generate 
the seed of PBHs around the inflection point $\phi=\phi_c$. 
Since $\mu(\phi)$ approaches constants in the two asymptotic regimes 
$\phi \ll \phi_c$ and $\phi \gg \phi_c$, the scalar-GB coupling is 
important only in the vicinity of $\phi=\phi_c$. 
A temporal USR region can arise from the balance between 
the GB term and the scalar potential. 
The enhancement of curvature perturbations in such a transient epoch 
was studied for natural inflation \cite{Kawai:2021edk} and 
for $\alpha$-attractor \cite{Zhang:2021rqs}. 
In these papers, the authors mostly focused on the USR regime 
realized by a plateau-type effective potential. 
In this case, the CMB observables are subject to modifications 
by the presence of a plateau-region with the number of 
e-foldings of order 10. Hence it is nontrivial to produce the large 
amplitude of primordial scalar perturbations responsible for the 
seed of PBHs, while satisfying CMB constraints on $n_s$ and $r$. 

In this paper, we will address this issue in Higgs inflation with a scalar-GB coupling mentioned above. We will not incorporate the runnings of Higgs and nonminimal couplings 
to focus on effects of the scalar-GB coupling on the background and perturbations.
We show that, besides the plateau-type effective potential, it is possible to realize a bump- or step-type effective potential. In this latter case, the field velocity $\dot{\phi}$ 
around $\phi=\phi_c$ decreases faster in comparison to the USR regime with 
a smaller number of e-foldings of order 1.  
The primordial scalar power spectrum can also have a sharp feature 
with a peak amplitude enhanced by a factor of $10^7$. 
In such cases, PBHs can be the source for all DM 
in the mass range $10^{-16}M_{\odot} \lesssim M \lesssim 
10^{-13}M_{\odot}$. Moreover, the bump-type effective potential 
can give rise to the values of $n_s$ and $r$ inside the $1\sigma$ 
observational contour constrained by the Planck CMB data.  
There are also intermediate-type potentials between plateau- and 
bump-types consistent with the CMB constraints, while generating the 
seed of PBHs. 
Thus, our inflationary scenario provides a versatile possibility 
for realizing various shapes of the effective scalar potential. 
We note that each shape of potentials was discussed separately 
in different contexts in the literature.

This paper is organized as follows.
In Sec.~\ref{model}, we obtain the background equations of motion in 
Higgs inflation with a scalar-GB coupling $\mu(\phi)R_{\rm GB}^2$ and 
revisit the scalar and tensor power spectra generated in slow-roll 
Higgs inflation with $\mu(\phi)=0$.
In Sec.~\ref{PBHformation}, we derive an effective potential 
$V_{\rm eff}(\phi)$ of the inflaton field and classify it into three classes:
(1) plateau-type, 
(2) bump-type, and 
(3) intermediate-type.
In Sec.~\ref{seedsec}, we compute the primordial scalar power spectra for three sets of model parameters with which there 
are neither ghost nor Laplacian instabilities. 
We show that the bump-type is favored over the plateau-type 
for the consistency with CMB observables.
In Sec.~\ref{pbhabundance}, we calculate the PBH abundance 
relative to 
the relic DM density and show that our model produces a sufficient 
amount of PBHs that can be the source for all DM.
Sec.~\ref{consec} is devoted to conclusions.
Throughout the paper, we use the natural units ($c=\hbar=1$).

\section{Inflationary model with a Gauss-Bonnet term}
\label{model}

We begin with theories given by the action 
\be
{\cal S}=\int {\rm d}^4 x \sqrt{-g} \left[\left( \frac{\Mpl^{2}}{2}
+\frac{1}{2}\xi\phi^{2}\right)R
-\frac{1}{2}g^{\mu \nu}\nabla_{\mu}\phi
\nabla_{\nu}\phi-V(\phi)+\mu(\phi)R_{\rm GB}^{2}\right] ,
\label{action1}
\ee
where $g$ is a determinant of the metric tensor $g_{\mu \nu}$, 
$\Mpl$ is the reduced Planck mass, $\xi$ is a nonminimal coupling 
constant, $\phi$ is a scalar field with the covariant derivative 
operator $\nabla_{\mu}$, and $R$ is the Ricci scalar. 
The scalar field has a potential of the form 
\be
V(\phi)=\frac{\lambda}{4}\phi^4\,,
\label{potential}
\ee
where $\lambda$ is a positive coupling constant.
The dynamics of nonminimally coupled inflation with the potential 
(\ref{potential}) was originally addressed 
in Refs.~\cite{Futamase:1987ua,Fakir:1990eg} 
(see also Refs.~\cite{Salopek:1988qh,Makino:1991sg,Kaiser:1994vs,Komatsu:1999mt,Tsujikawa:2004my}).
It can also accommodate 
the Higgs potential 
$V(\phi)=\lambda (\phi^2-v^2)^2/4$ in the large field 
regime $\phi^2 \gg v^2$, where 
$v={\cal O}(10^2)$~GeV \cite{Bezrukov:2007ep,Bezrukov:2008ej,Bezrukov:2010jz}. 
Provided that the nonminimal coupling is in the range 
\be
\xi \gg 1\,,
\ee
the self-coupling of order $\lambda=0.01 \sim 0.1$ can be consistent with the 
amplitude of observed CMB temperature anisotropies\footnote{If we consider 
quantum corrections arising from the renormalization group 
running of the standard model, the Higgs self-coupling $\lambda$ can be 
much smaller than 0.01 or even negative at inflationary energy 
scales \cite{DeSimone:2008ei,Hamada:2014iga,Hamada:2014wna,Bezrukov:2014ipa,Bezrukov:2017dyv}. 
In this paper, we do not consider the runnings of 
coupling constants $\lambda$ or $\xi$.}. 

The scalar field is coupled to a GB curvature 
invariant defined by 
\be
R_{\rm GB}^{2}\equiv{R^{2}-4R_{\mu\nu}R^{\mu\nu}
+R_{\mu\nu\rho\sigma}R^{\mu\nu\rho\sigma}}\,,
\label{GBterm}
\ee
with a $\phi$-dependent coupling function $\mu(\phi)$, 
where $R_{\mu\nu}$ and $R_{\mu\nu\rho\sigma}$ are the Ricci and Riemann tensors, respectively. 
The action (\ref{action1}) belongs to 
a subclass of Horndeski theories 
with second-order field equations of 
motion \cite{Horndeski,KYY,Def11,Charmousis:2011bf} 
(see Appendix.~\ref{Horndeskitheory}). 
In this case, there is only one propagating scalar 
degree of freedom besides two tensor polarizations. 
As we will study in Sec.~\ref{PBHformation}, 
it is possible to enhance scalar perturbations at 
particular scales for a specific choice of $\mu(\phi)$.
The action (\ref{action1}) corresponds to Higgs inflation 
corrected by the Higgs-GB coupling. 
This allows a possibility for generating the seed of 
PBHs as the source for all DM within the framework of 
SM of particle physics.

We note that the $\xi \to 0$ limit in the action (\ref{action1}) 
with the potential of natural inflation corresponds to the 
model studied by Kawai and Kim \cite{Kawai:2021edk}. 
In our model, the basic mechanism for the generation of 
seeds of PBHs is similar to that advocated in Ref.~\cite{Kawai:2021edk}.
However, natural inflation is in tension with  
the observation of CMB temperature anisotropies \cite{Planck:2018jri}. 
Instead, we would like to construct an explicit inflationary model 
consistent with CMB observations, while enhancing curvature 
perturbations on scales relevant to PBHs. 
As we will show later, this is indeed possible for Higgs inflation 
with $\xi \gg 1$ in the presence of the scalar-GB coupling.

For the background, we consider a spatially flat 
Friedmann-Lema\^{i}tre-Robertson-Walker 
(FLRW) line element given by 
\be
{\rm d}s^2=-{\rm d}t^2+a^2(t) \delta_{ij}{\rm d}x^i {\rm d}x^j\,,
\ee
where $a(t)$ is a time-dependent scale factor.
On this background, the Friedmann and scalar-field 
equations of motion are
\ba
3 \left( \Mpl^{2}+\xi\phi^2 \right)H^{2}
=\frac{1}{2}\dot{\phi}^{2}+V(\phi)-6\left( \xi\phi+4H^2\mu_{,\phi}
\right)H\dot{\phi}\,,
\label{backgroundFriedmanneq}\\
\ddot{\phi}+3H\dot{\phi}+V_{,\phi}
-6 \xi \phi \left( 2H^2+\dot{H} \right)
-24 \mu_{,\phi}H^2 \left( H^2+\dot{H} \right)=0\,,
\label{backgroundFieldeq}
\ea
where a dot represents a derivative with respect to 
$t$, $H=\dot{a}/a$ is the Hubble expansion rate, 
and we use the notations 
$\mu_{,\phi}={\rm d}\mu/{\rm d}\phi$ and 
$V_{,\phi}={\rm d}V/{\rm d}\phi$. 
Taking the time derivative of Eq.~(\ref{backgroundFriedmanneq}) and 
using Eq.~(\ref{backgroundFieldeq}), we obtain the closed-form 
differential equations
\ba
& &
\dot{H}=\frac{1}{2{\cal D}}
[ 8 \mu_{,\phi}H^2 (V_{,\phi}-18 \xi \phi H^2 -24 \mu_{,\phi}H^4)
+2\xi \phi (V_{,\phi}-12\xi \phi H^2)-\dot{\phi}^2 
(2\xi+1+8 \mu_{,\phi \phi}H^2)\nonumber \\
& &
\qquad \qquad
+8 H \dot{\phi} (\xi \phi+4\mu_{,\phi}H^2)]\,,
\label{tH}\\
& &
\ddot{\phi}+3H \dot{\phi}+V_{{\rm eff},\phi}=0\,,
\label{ddotphi}
\ea
where 
\ba
V_{{\rm eff},\phi} 
&\equiv&\frac{1}{{\cal D}}
[ (\xi \phi^2+\Mpl^2)(V_{,\phi}-12 \xi \phi H^2-24 \mu_{,\phi}H^4)
+3 \dot{\phi}^2 (2\xi+1+8\mu_{,\phi \phi}H^2)(\xi \phi+4\mu_{,\phi}H^2)
\nonumber \\
& &
\quad
+8 H \dot{\phi} (\mu_{,\phi}V_{,\phi}-3\xi^2 \phi^2)
-288 H^3 \dot{\phi} \mu_{,\phi} (\xi \phi+2H^2 \mu_{,\phi})]\,,
\label{Veffphi} \\
{\cal D} &\equiv& (6\xi+1)\xi \phi^2+\Mpl^2
+8 H \mu_{,\phi} (\dot{\phi}+6\xi H \phi+12H^3 \mu_{,\phi})\,.
\ea
Notice that $V_{\rm eff}$ is an effective potential of the scalar 
field. Numerically, we solve Eqs.~(\ref{tH}) and (\ref{ddotphi}) 
for $H$ and $\phi$ with the initial conditions of $H$, $\phi$, and $\dot{\phi}$ consistent with 
the Hamiltonian constraint (\ref{backgroundFriedmanneq}).

\subsection{Linear perturbations and stability conditions}
\label{linearsec}

To study the evolution of cosmological perturbations during inflation, we 
consider a perturbed line element containing scalar perturbations 
$\alpha$, $\psi$, $\zeta$ and tensor perturbations $h_{ij}$ as 
\be
\rd s^2=-(1+2\alpha)\rd t^2+2 \partial_i \psi \rd t \rd x^i
+a^2(t) \left[ (1+2\zeta) \delta_{ij}+h_{ij} \right] 
\rd x^i \rd x^j\,.
\ee
In full Horndeski theories including the action (\ref{action1}) as a 
special case, the linear perturbation equations of motion were already 
derived in the literature \cite{KYY}. 
On using the Hamiltonian and momentum constraints to 
eliminate $\alpha$ and $\psi$ and integrating the 
action (\ref{action1}) by parts, the  
second-order action of scalar perturbations 
is given by \cite{KYY,DeFelice:2011zh,Kawai:2021bye}
\be
{\cal S}_s^{(2)}=\int \rd t \rd^3 x\,a^3 Q_s \left[\dot{\zeta}^2
-\frac{c_s^2}{a^2}(\nabla \zeta)^2\right]\,,
\label{action2}
\ee
where
\be
Q_s=16\frac{\Sigma}{\Theta^2}Q_t^2+12Q_t\,,\qquad 
c_s^2=\frac{1}{Q_s}\left[ \frac{16}{a}\frac{\rd}{\rd t}
\left(\frac{a}{\Theta}Q_t^2\right)-4c_t^2 Q_t\right]\,,
\label{cs}
\ee
with 
\ba
& &
Q_t=\frac{1}{4}\left(8H\mu_{,\phi}\dot{\phi}+\Mpl^2+\xi\phi^2\right)\,,
\qquad 
c_t^2=\frac{1}{4Q_t}
\left(\Mpl^2+\xi\phi^2+8\mu_{,\phi\phi}\dot{\phi}^2
+8\mu_{,\phi}\ddot{\phi}\right)\,,\label{cT}\\
& &
\Sigma=\frac{1}{2}\dot{\phi}^2-3H^2(\Mpl^2+\xi\phi^2)-6\xi H\phi\dot{\phi}-48H^3 \mu_{,\phi}\dot{\phi}
\label{sigma}\,,\qquad 
\Theta=H(\Mpl^2+\xi\phi^2)+\xi\phi\dot{\phi}+12H^2 \mu_\phi\dot{\phi}\,.
\ea
In the tensor sector, the reduced action is of the form 
\be
{\cal S}_t^{(2)}=\frac{1}{2}\int {\rm d} t {\rm d}^3 x\hspace{0.1cm} a^3 Q_t \left[\dot{h}_{ij}^2-\frac{c_t^2}{a^2}(\nabla h_{ij})^2\right] ,
\label{actionT}
\ee
where $Q_t$ and $c_t^2$ are defined in Eq.~(\ref{cT}). 
To avoid the ghost and Laplacian instabilities of scalar and tensor perturbations, 
we require the following conditions
\be
Q_s>0\,,\qquad c_s^2>0\,,\qquad 
Q_t>0\,,\qquad c_t^2>0\,.
\label{avoidinstability}
\ee
On using the background Eq.~(\ref{backgroundFriedmanneq}), 
we can express $Q_t$ and $Q_s$ in the forms
\be
Q_t=\frac{1}{12 H^2}\left[ \frac{1}{2}\dot{\phi}^2+V(\phi)
-6\xi H\phi\dot{\phi} \right]\,,\qquad
Q_s=\frac{4\dot{\phi}^2}{\Theta^2}Q_t
\left[ 2Q_t+3(\xi\phi+4H^2\mu_{,\phi})^2 \right]\,.
\label{QS2}
\ee
Provided that $\phi$ decreases during inflation in the 
region $\phi>0$, we have 
$-6\xi H \phi \dot{\phi}>0$ for $\xi>0$. 
In such cases, both $Q_t$ and $Q_s$ are positive and hence 
the ghost instabilities are absent. 
In the absence of the scalar-GB coupling, both $c_t^2$ and $c_s^2$ 
are equivalent to 1. However, the deviations of $c_t^2$ and $c_s^2$ 
from 1 arise in theories with $\mu(\phi) \neq 0$, 
so we need to numerically compute $c_t^2$ and $c_s^2$  for a given coupling 
$\mu(\phi)$ to ensure the absence of Laplacian instabilities.

\subsection{Higgs slow-roll inflation with $\mu(\phi)=0$}
\label{standardhiggs}

We briefly revisit the background dynamics and perturbation spectra 
generated during Higgs slow-roll inflation for $\mu(\phi)=0$. 
{}From Eqs.~(\ref{backgroundFriedmanneq}) and (\ref{ddotphi}), 
we have
\ba
& &
H^2= \frac{1}{3(\Mpl^2+\xi \phi^2)} 
\left( \frac12 \dot{\phi}^2
+\frac{1}{4}\lambda \phi^4-6 \xi H \phi \dot{\phi} 
\right)\,,
\label{H0} \\
& &
\ddot{\phi}+3H \dot{\phi}+\frac{(6 \xi+1)\xi \phi \dot{\phi}^2}
{(6\xi+1)\xi \phi^2+\Mpl^2}
+\frac{\lambda \phi^3 \Mpl^2}
{(6\xi+1)\xi\phi^2+\Mpl^2}=0\,.
\label{phi0}
\ea
Let us consider the large coupling regime with $\xi \gg 1$ and $\xi \phi^2 \gg \Mpl^2$.
During slow-roll inflation, the dominant term in Eq.~(\ref{H0}) 
is the potential 
$V(\phi)=\lambda \phi^4/4$, 
while the dominant contributions to Eq.~(\ref{phi0}) are
second and fourth terms. 
Then, Eqs.~(\ref{H0}) and (\ref{phi0}) approximately 
reduce to 
\be
H^2 \simeq \frac{\lambda \phi^2}{12\xi}\,,\qquad
H \dot{\phi} \simeq -\frac{\lambda \phi \Mpl^2}{18 \xi^2}\,.
\label{sloweq}
\ee
The field value $\phi_f$ at the end of inflation is determined by 
the condition $\epsilon_H \equiv -\dot{H}/H^2=1$.
Using the two equations in (\ref{sloweq}), we have 
$\epsilon_H \simeq 2\Mpl^2/(3\xi \phi^2)$ and hence
$\phi_f=\sqrt{2/(3\xi)}\Mpl$. 
The number of e-foldings counted 
backward from the end of 
inflation can be estimated as 
\be
N=\int_{\phi}^{\phi_f} \frac{H}{\dot{\phi}} \rd \phi 
\simeq \frac{3\xi \phi^2}{4\Mpl^2}-\frac{1}{2}\,,
\label{Nnum}
\ee
where we exploited Eq.~(\ref{sloweq}) in the second approximate equality.
For $N \gg 1$, we obtain the simple relation $\phi^2 \simeq 4\Mpl^2 N/(3\xi)$.

For perturbations deep inside the Hubble radius (with the wavenumber $k \gg aH$), they are initially in the Bunch-Davies vacuum state.
On the inflationary background, $\zeta$ and $h_{ij}$ 
approach constants after the sound horizon crossing.
The power spectra of scalar and tensor perturbations generated 
during the quasi de Sitter period are given, 
respectively, by \cite{KYY}
\be
{\cal P}_\zeta=\frac{H^2}{8\pi^2 Q_s c_s^3}\biggr|_{c_s k=a H}\,,
\qquad 
{\cal P}_h=\frac{H^2}{2\pi^2 Q_t c_t^3}\biggr|_{c_t k=a H}\,.
\label{power0}
\ee
In theories with $\mu(\phi)=0$, we have $c_s^2=1=c_t^2$ and hence 
both ${\cal P}_\zeta$ and ${\cal P}_h$ should be evaluated at $k=aH$.

Since $Q_s \simeq 3 \xi \dot{\phi}^2/H^2$ and $Q_t \simeq \xi \phi^2/4$
in the regimes $\xi \gg 1$ and $\xi \phi^2 \gg \Mpl^2$, 
the power spectra (\ref{power0}) reduce to
\ba
& &
{\cal P}_\zeta \simeq \frac{H^4}{24\pi^2 \xi \dot{\phi}^2}
\simeq \frac{\lambda \phi^4}{128 \pi^2 \Mpl^4}
\simeq \frac{\lambda N^2}{72 \pi^2 \xi^2}
\label{power1}\,,\\
& &
{\cal P}_h \simeq \frac{2H^2}{\pi^2 \xi \phi^2}
\simeq \frac{\lambda}{6\pi^2 \xi^2}\,,
\label{power2}
\ea
where we used the approximate background Eq.~(\ref{sloweq}). 
We note that the subscript $k=aH$ is omitted 
in Eqs.~(\ref{power1}) and (\ref{power2}).
Then, we obtain the scalar spectral index $n_s$ 
and the tensor-to-scalar ratio $r$, as
\ba
n_s-1 &=& \frac{\rd \ln {\cal P}_\zeta}{\rd \ln k}\biggr|_{k=aH}
\simeq \frac{4\dot{\phi}}{H \phi} \simeq -\frac{8 \Mpl^2}{3\xi \phi^2}
\simeq -\frac{2}{N}\label{ns}\,,\\
r &=& \frac{{\cal P}_h}{{\cal P}_\zeta} \simeq \frac{12}{N^2}\label{r}\,.
\ea
Taking $N=60$ for scales relevant to the observed CMB 
temperature anisotropies, we obtain $n_s=0.9667$ and 
$r=3.3 \times 10^{-3}$. 
These values are consistent 
with the bounds 
$n_s=0.9661\pm 0.0040$ (68\,\%\,CL) and 
$r<0.066$ (95\,\%\,CL)  constrained by the 
Planck 2018 data \cite{Planck:2018jri}.
The Planck normalization ${\cal P}_{\zeta}=2.1 \times 10^{-9}$ 
with $N=60$ gives the constraint $\lambda/\xi^2=4.1 \times 10^{-10}$.
If $\lambda=0.1$, then $\xi=1.6 \times 10^4$. 

The above results are valid for slow-roll inflation with $\mu(\phi)=0$. 
In the presence of the scalar-GB coupling, the background dynamics and 
perturbation spectra are subject to modifications.
In subsequent sections, we will address this issue 
along with the problem of generating the source for PBHs.

\section{Effective potentials with plateau and bump}
\label{PBHformation}

Let us proceed to the case in which the scalar-GB coupling 
$\mu (\phi)R_{\rm GB}^2$ is present. 
If $\mu(\phi)$ is a smooth function whose time variation 
is small during inflation, the inflaton field $\phi$ can slowly 
evolve along the potential. 
In such a case, the primordial power spectra of scalar and 
tensor perturbations are given by Eq.~(\ref{power0}). 
Since $Q_s$ is proportional to $\dot{\phi}^2$, a smaller 
inflaton velocity generally leads to a larger amplitude of 
${\cal P}_{\zeta}$. In the context of slow-roll inflation, however, this enhancement of ${\cal P}_{\zeta}$ is limited by 
a small time variation of $\dot{\phi}$.

If the scalar-GB coupling generates a period in which 
the field velocity $\dot{\phi}$ temporally approaches 0, 
it is possible to realize the large enhancement of  
${\cal P}_{\zeta}$ for scales smaller than those of 
the observed CMB temperature anisotropies.
One possible choice of $\mu(\phi)$ is a dilatonic coupling 
of the form 
$\mu(\phi)=\mu_0 e^{-\lambda \phi}$ \cite{Gross:1986mw,Metsaev:1987zx,Gasperini:1996fu,Kawai:1998ab,Cartier:2001is,Calcagni:2005im,Guo:2006ct}. 
However, this type of continuously varying functions 
affects not only the scalar perturbation on 
particular scales but also that on other scales 
including CMB. Moreover, it can happen that 
the dominance of the scalar-GB coupling over the potential and nonminimal couplings leads to the violation of 
stability conditions (\ref{avoidinstability}).

Instead, we consider the step-like coupling given by 
\cite{Kawai:2021edk,Zhang:2021rqs,Khan:2022odn}
\be
\mu(\phi)=\mu_{0}\tanh{[\mu_1(\phi-\phi_{c})]},
\label{couplingfunction}
\ee
where $\mu_{0}$, $\mu_{1}$, and $\phi_{c}$ are constants. 
Around the field value $\phi=\phi_c$, this coupling rapidly changes from the asymptotic constant $\mu(\phi)=-\mu_0$ 
(for $\phi \ll \phi_c$) 
to the other asymptotic constant $\mu(\phi)=+\mu_0$ 
(for $\phi \gg \phi_c$). 
Since the GB curvature invariant is a topological term, the scalar-GB coupling does not affect the cosmological dynamics in the two asymptotic regimes with constant $\mu(\phi)$.
Provided that $\phi_c$ is in the range 
$\phi_f<\phi_c<\phi_{\rm CMB}$, where $\phi_{\rm CMB}$ 
is the field value about 60 e-foldings before the end of inflation (with the field value $\phi_f$), 
it should be possible to enhance the scalar power 
spectrum for scales smaller than those of observed 
CMB temperature anisotropies. 
In Sec.~\ref{seedsec}, we will study whether the sufficient generation of seeds for PBHs is possible, while satisfying observational constraints on 
$n_s$ and $r$ on CMB scales. 

Provided that the field kinetic term is sufficiently small during inflation, we can ignore the $\dot{\phi}$-dependent terms in Eq.~(\ref{Veffphi}). 
Then, the $\phi$-derivative of the effective 
potential $V_{\rm eff}$ 
is approximately given by 
\be
V_{{\rm eff},\phi} \simeq 
\frac{(\xi \phi^2+\Mpl^2)
(V_{,\phi}-12\xi \phi H^2 -24 \mu_{,\phi} H^4)}
{(6\xi+1)\xi \phi^2+\Mpl^2+48 H^2 \mu_{,\phi}
(\xi \phi+2H^2 \mu_{,\phi})}\,.
\label{Veffp}
\ee
{}From Eq.~(\ref{backgroundFriedmanneq}), 
the Hubble parameter is approximately given by 
\be
H^2 \simeq \frac{V(\phi)}{3(\Mpl^2+\xi \phi^2)}\,.
\label{Ha}
\ee
Substituting Eq.~(\ref{Ha}) into Eq.~(\ref{Veffp}) and 
integrating it with respect to $\phi$, 
the effective potential $V_{\rm eff}$ can be numerically 
known as a function of $\phi$.

In the following, we classify the effective potential into 
three classes: 
(1) plateau-type, 
(2) bump-type, and 
(3) intermediate-type. 
The set 1, 2, 3 model parameters shown in Table~\ref{table1}
are the typical examples of plateau-, bump-, and 
intermediate-types, respectively.
The nonminimal coupling constant is fixed to be $\xi=5000$ 
in all cases. 

Since PBHs are treated as the main component of DM in this paper, we consider the PBH mass range $10^{-16} M_{\odot} \lesssim M 
\lesssim 10^{-11}M_{\odot}$. 
This gives a constraint on the value of $\phi_c$. 
The two constants $\mu_0$ and $\mu_1$ determine the types of 
$V_{\rm eff}(\phi)$ mentioned 
above. Instead of the parameter $\mu_0$, we will use the combination 
\be
\tilde{\mu}_0 \equiv \frac{1}{4} 
\mu_0 \mu_1 \lambda\,,
\ee
which appears later in Eq.~(\ref{balance3}).
The values of $\tilde{\mu}_0$ are chosen to be 
close or not far away from the right hand side (RHS) of Eq.~(\ref{balance3}), see the last two 
columns in Table~\ref{table1}. 
The Higgs self-coupling $\lambda$ is determined by the observed amplitude of primordial curvature perturbations on CMB scales. As we see in Table~\ref{table1}, $\lambda$ is of order 0.01 for three sets of model parameters.

\vskip-\baselineskip
\begin{table}[ht]
\newcommand\xrowht[2][0]{\addstackgap[.5\dimexpr#2\relax]{\vphantom{#1}}}
\newcolumntype{C}[1]{>{\hfil}m{#1}<{\hfil}}
\centering
\caption{Three sets of model parameters. For dimensionfull parameters, the units are 
given inside the squared 
parenthesis.}
\begin{tabular}{C{15mm}C{18mm}C{21mm}C{21mm}C{21mm}C{26mm}C{40mm}} \hline\hline \xrowht{12pt}
      & $\xi$ & $\lambda$ & $\phi_c\hspace{0.2cm}[\Mpl]$ & $\mu_1\hspace{0.2cm}[\Mpl^{-1}]$ & $\tilde{\mu}_0 \hspace{0.2cm} [10^8 \Mpl^{-1}]$ & RHS of Eq.~(\ref{balance3})$\hspace{0.2cm} [10^8 \Mpl^{-1}]$ \\ \hline \xrowht{12pt}
     Set 1 & 5000 & 0.0244211 &0.0380   & 1000 & 1.564709 & 1.556127 \\
    \xrowht{12pt}
     Set 2 & 5000 & 0.0110810 & 0.0760 & 5000 & 0.249909 & 0.176768 \\
     \xrowht{12pt}
     Set 3 & 5000 & 0.0152511 & 0.0600 & 1600 & 0.376125 & 0.366512 \\
     \hline\hline
\end{tabular}\label{table1}
\end{table}
%


\subsection{Plateau type}
\label{BackgroundCase1}

Thanks to the existence of the scalar-GB coupling, there is a stationary 
fixed point $\phi=\phi_*$ at which $V_{{\rm eff},\phi}$ 
vanishes \cite{Kawai:2021edk}.  
{}From Eq.~(\ref{Veffp}) with Eq.~(\ref{Ha}), 
there is the following relation 
\be
V_{,\phi}-\frac{4\xi \phi V}{\Mpl^2+\xi \phi^2}
-\frac{8 \mu_{,\phi}V^2}{3(\Mpl^2+\xi \phi^2)^2}
\biggr|_{\phi=\phi_*} 
=0\,.
\label{balance}
\ee
This is known as the USR regime in which the scalar-field 
equation (\ref{ddotphi}) reduces 
to \cite{Garcia-Bellido:2017mdw,Kannike:2017bxn,Germani:2017bcs,Motohashi:2017kbs} 
\be
\ddot{\phi}+3H\dot{\phi} 
\simeq 0
\qquad
\text{(around $\phi=\phi_*$)}\,.
\label{USR}
\ee
The solution to this equation is given by 
\be
\dot{\phi} \propto a^{-3} \propto e^{-3n}
\qquad
\text{(around $\phi=\phi_*$)}\,,
\label{dotphiUSR}
\ee
where $n=\ln a$ is the number of e-foldings counted forward.
Hence $\dot{\phi}$ rapidly decreases toward 0 
in the USR regime.

Setting $\phi_*=\phi_c$ for the coupling (\ref{couplingfunction}), 
the moment at which 
$V_{{\rm eff},\phi}$ vanishes
coincides with the instant 
at transition of $\mu(\phi)$. 
For the potential $V(\phi)=\lambda \phi^4/4$, the condition (\ref{balance}) 
translates to 
\be
\tilde{\mu}_0
=\frac{3\Mpl^{2}(\Mpl^{2}+\xi\phi_c^2)}{2\phi_c^5}\,,
\label{balance3}
\ee
which gives a constraint between $\mu_0, \mu_1, \phi_c$ 
and $\lambda, \xi$. 
During the USR regime, the variation of 
the scalar field is of order $1/\mu_1$. 
On using Eqs.~(\ref{sloweq}) and (\ref{dotphiUSR}), we can estimate 
the order of $1/\mu_1$ as 
\be
\frac{1}{\mu_1}\simeq\int\frac{|\dot{\phi}|}{H}{\rm d}n\simeq\int^{\Delta N_c}_{0}\frac{2\Mpl^2}{3\xi\phi_c}e^{-3n}{\rm d}n=\frac{2\Mpl^2}{9\xi\phi_c}\left(1-e^{-3\Delta N_c}\right)\,,
\label{mu1constraint}
\ee
where $\Delta N_c$ is the number of
e-foldings acquired during the USR phase.

\begin{figure}[ht]
\begin{center}
\includegraphics[height=2.0in,width=2.3in]{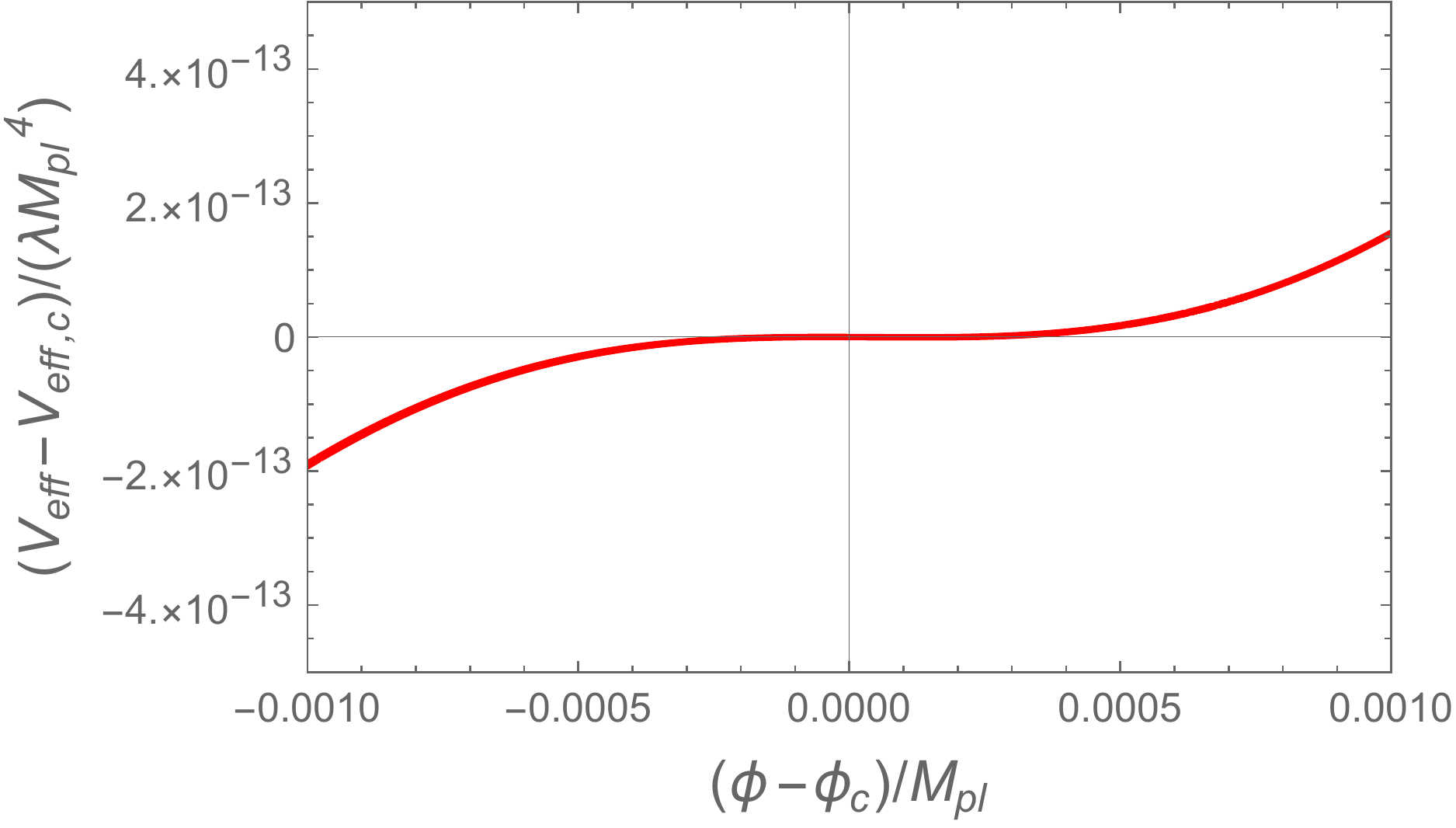}
\includegraphics[height=2.0in,width=2.3in]{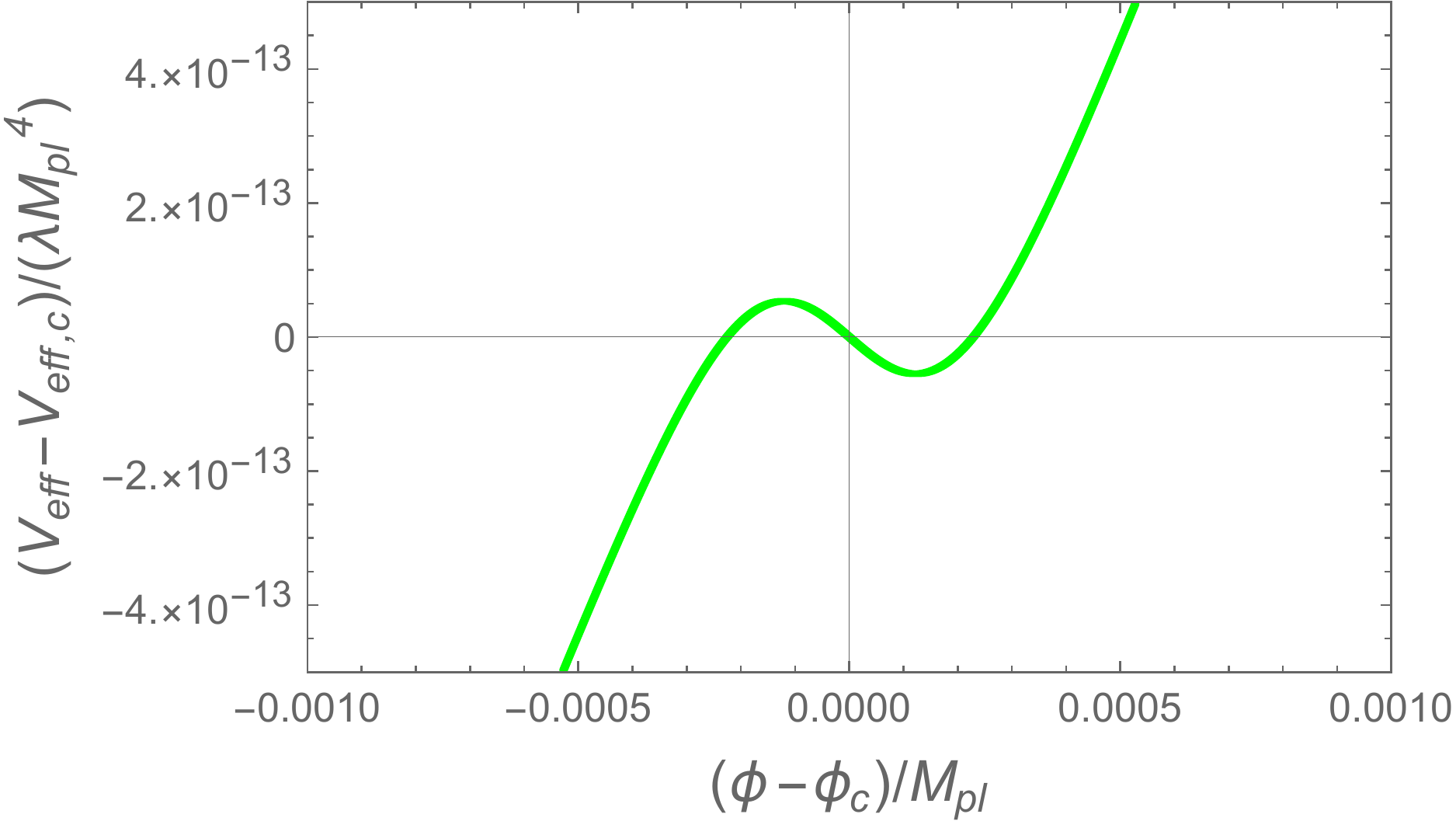}
\includegraphics[height=2.0in,width=2.3in]{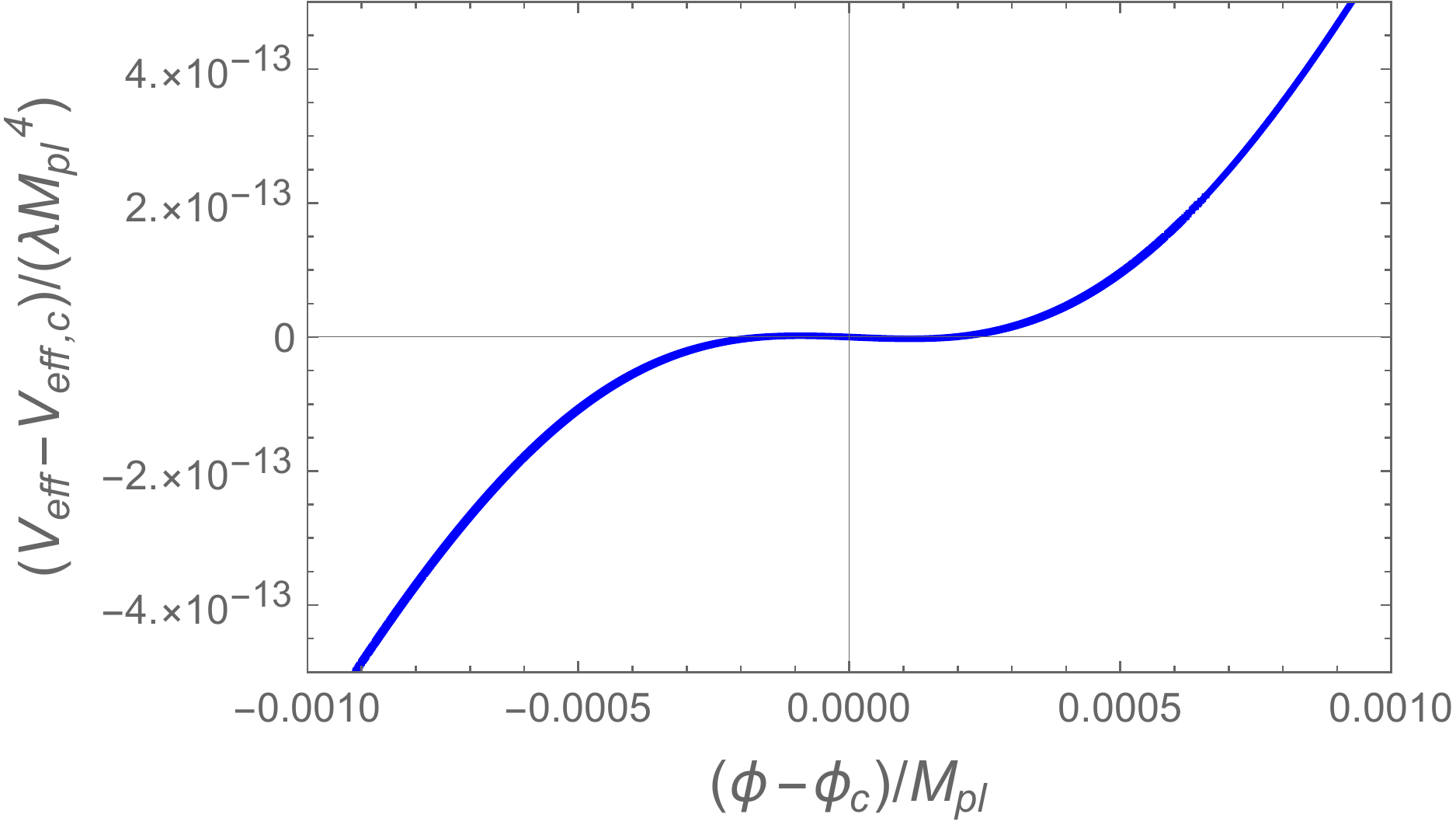}
\end{center}\vspace{-0.5cm}
\caption{\label{figpo}
Shapes of the effective potential $V_{\rm eff}(\phi)$
around $\phi=\phi_c$, where 
$V_{{\rm eff},c}=V_{\rm eff}(\phi_c)$.
Each plot shows 
(1) plateau-type (left),
(2) bump-type (middle), and 
(3) intermediate-type (right).
}
\end{figure}

\begin{figure}[ht]
\begin{center}
\includegraphics[height=2.0in,width=3.0in]{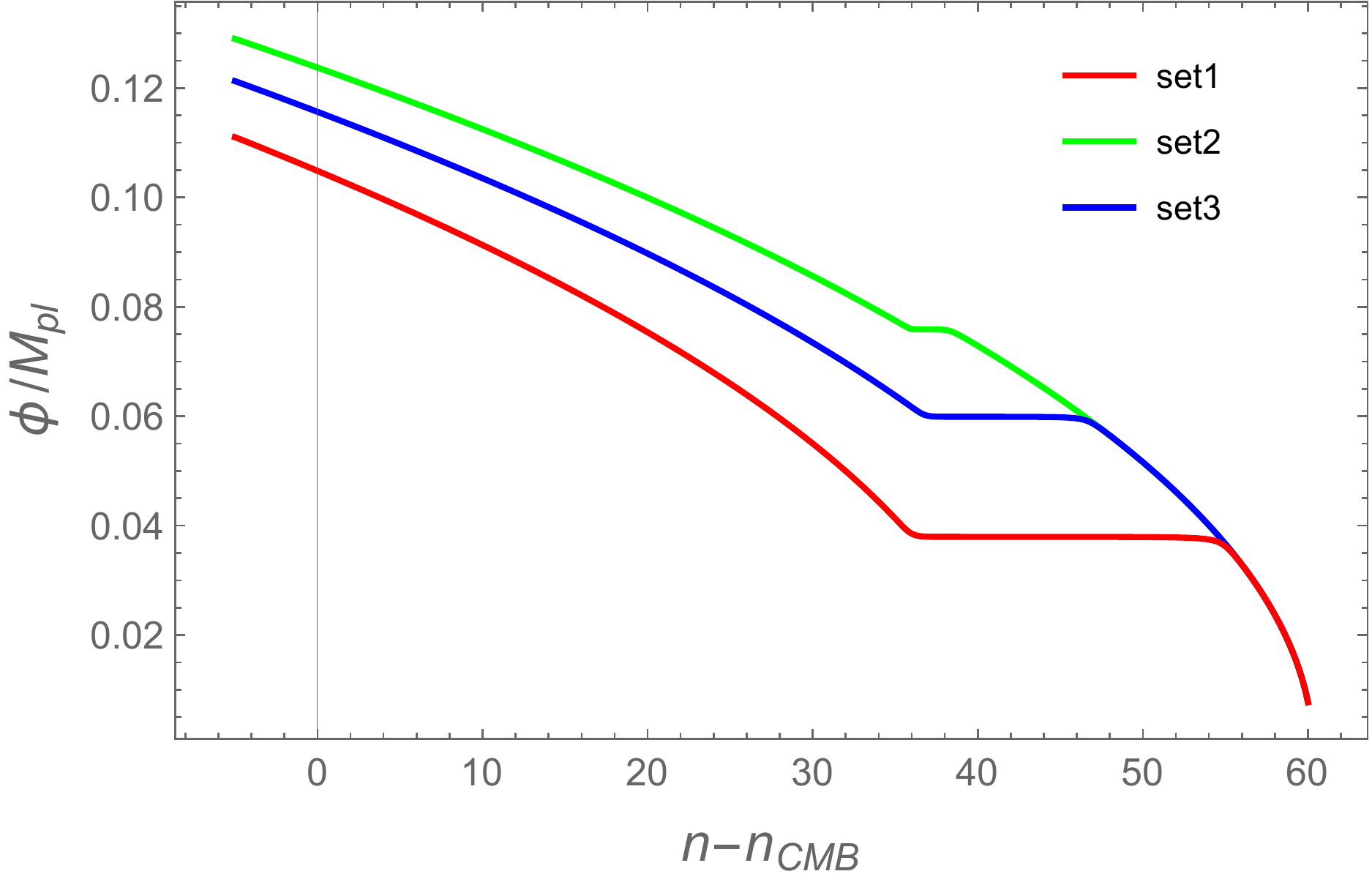}
\hspace{0.7cm}
\includegraphics[height=2.0in,width=3.0in]{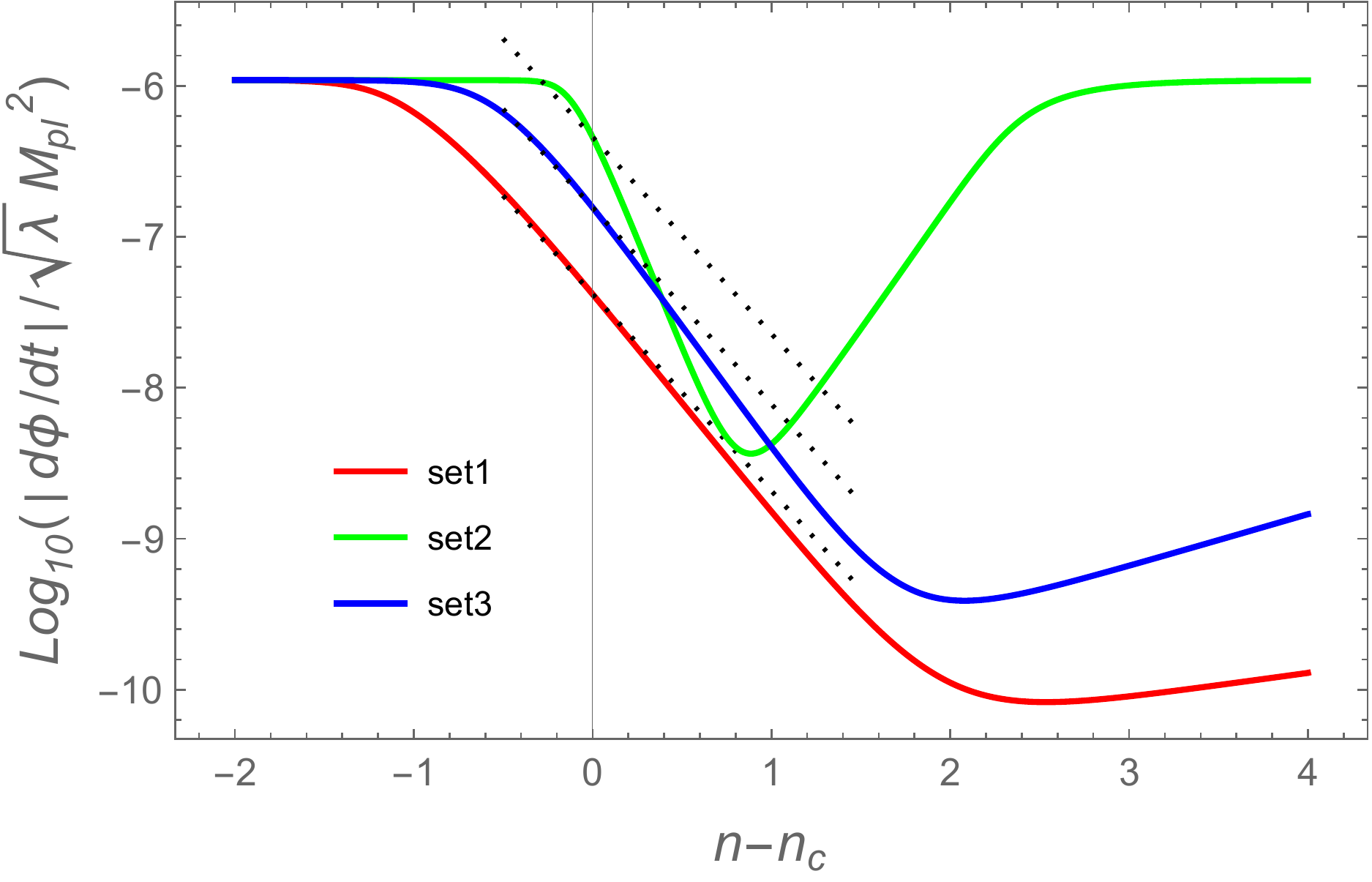}
\end{center}\vspace{-0.5cm}
\caption{\label{figevo}
Evolution of $\phi$ during inflation (left) and 
$|\dot{\phi}|$ around $\phi=\phi_c$ (right) 
for the three sets of parameters given in Table~\ref{table1}. 
Here, $n$ is the number of e-foldings counted forward 
toward the end of inflation, where $n_{\rm CMB}$ and 
$n_c$ are the e-foldings corresponding to 
the field values $\phi_{\rm CMB}$ 
and $\phi_c$ respectively. 
In the right panel, the dotted lines represent the evolution 
of $\dot{\phi}$ in the USR 
regime, i.e., $\dot{\phi} \propto e^{-3n}$.
}
\end{figure}

In the left panel of Fig.~\ref{figpo}, we plot $V_{\rm eff}(\phi)$  
in the vicinity of $\phi=\phi_c$ for the Set 1 model parameters 
given in Table~\ref{table1}. 
In this case, $\tilde{\mu}_0$ is 
chosen to be close to the RHS 
of Eq.~(\ref{balance3}). 
There is a plateau of $V_{\rm eff}(\phi)$ in the region 
$|\phi-\phi_c| \lesssim 
5 \times 10^{-4}\Mpl$, where 
$\phi_c=0.0380 \Mpl$. 
In the left panel of Fig.~\ref{figevo}, we show the evolution of $\phi$
versus the number of e-foldings for the parameter Set 1 as a 
red line.
The field value $\phi_f$ at the end of inflation is numerically derived by the condition $\epsilon_H=1$, which gives 
$\phi_f=0.0077 \Mpl$.

The initial field value for realizing the total number of e-foldings $n-n_{\rm CMB}=60$ by the end of inflation 
corresponds to 
$\phi_{\rm CMB} \simeq 0.105 \Mpl$ (at which $n=n_{\rm CMB}$).
In this case, the first slow-roll stage of inflation is followed by 
the USR period starting at $n-n_{\rm CMB} \simeq 36$. 
After the inflaton approaches the plateau of $V_{\rm eff}(\phi)$ 
around $\phi=\phi_c$, the field derivative rapidly 
decreases as $|\dot{\phi}| \propto e^{-3n}$, see 
the right panel of Fig.~\ref{figevo}.
For the model parameters of Set 1, the number of e-foldings
acquired during the USR epoch 
is 18.6.
Finally, the scalar field exits from the USR regime, 
after which $|\dot{\phi}|$ starts to increase.

As the plateau region of $V_{\rm eff}(\phi)$ gets wider, 
the number of e-foldings $\Delta N_c$ acquired during the USR phase tends to be larger.
For $\Delta N_c$ exceeding the order 10, the CMB 
observables like $n_s$ and $r$ are subject to modifications 
in comparison to those derived 
for $\mu(\phi)=0$. 
We will discuss this issue 
in Sec.~\ref{seedsec}.

\subsection{Bump type}
\label{BackgroundCase2}

For the plateau-type effective potential, the scalar-GB coupling balances the contributions 
arising from the potential and nonminimal couplings in the 
scalar-field equation of motion. 
On the other hand, it should be possible that the scalar-GB term  
temporarily becomes larger than the contributions from other terms. This causes an instantaneous slowdown of the 
inflaton velocity 
in a manner different from the USR case discussed 
in Sec.~\ref{BackgroundCase1}.  
We call this class as a bump-type model, in which 
the $\phi$ derivative of $V_{\rm eff}(\phi)$
has the following feature 
\be
V_{{\rm eff},\phi}(\phi)=
\begin{cases}
\hspace{0.2cm} >0 \hspace{0.7cm} {\rm for}~\phi\gg\phi_{c}\,, \\ 
\hspace{0.2cm} <0 \hspace{0.7cm} {\rm around}~ \phi=\phi_c\,,\\
\hspace{0.2cm} >0 \hspace{0.7cm} {\rm for}~\phi\ll\phi_{c}.
\end{cases}
\ee
The parameter Set 2 in Table \ref{table1} gives rise to an 
effective potential of the bump-type, which is 
illustrated in the middle panel of Fig.~\ref{figpo}. 
In this case, $\tilde{\mu}_0$ exhibits some deviation 
from the value on the RHS of Eq.~(\ref{balance3}).
The effective potential has a local maximum as well as a local minimum in the vicinity of $\phi=\phi_c$.
We note that similar toy models  have been studied in Refs.~\cite{Atal:2019cdz,Atal:2019erb,Inomata:2021tpx,Cai:2021zsp,Cai:2022erk} in different contexts.

If $\tilde{\mu}_0$ is larger than the RHS of Eq.~(\ref{balance3}), 
the scalar-GB coupling dominates over the contributions from 
the potential and nonminimal couplings.
However, the period of the dominance must be sufficiently 
short to end inflation properly. 
This requires that the parameter $\mu_1$ is quite large.
In the limit $\mu_1\rightarrow\infty$, $\mu_{,\phi}$ and $\mu_{,\phi\phi}$ behave as
\ba
& &
\lim_{\mu_1\to\infty}\mu_{,\phi}=2\mu_0\delta(\phi-\phi_c)\,,
\label{limit1}\\
& &
\lim_{\mu_1\to\infty}\mu_{,\phi\phi}=\mu_0\mu_1
\left[ \delta(\phi-(\phi_c-\epsilon))-\delta(\phi-(\phi_c+\epsilon))\right]\,,
\label{limit2}
\ea
where $\epsilon=\text{arctanh}(1/\sqrt{3})/\mu_1$. 
This type of step-like behavior for large $\mu_1$ may induce Laplacian 
instabilities of cosmological perturbations, so we will address this issue 
in Sec.~\ref{seedsec} by computing the values of 
$c_s^2$ and $c_t^2$.

For the Set 2 model parameters corresponding to the bump-type 
effective potential, we plot the evolution of $\phi$ and $|\dot{\phi}|$ 
as a green line in Fig.~\ref{figevo}. 
Unlike the plateau model, the field velocity decreases 
faster than 
$|\dot{\phi}| \propto e^{-3n}$ due to the existence of the region 
$V_{{\rm eff},\phi}<0$ around $\phi=\phi_c$. 
When the field reaches a local maximum of $V_{\rm eff}(\phi)$, 
however, this period of the rapid decrease of $\dot{\phi}$ 
soon comes to end. After this short epoch, the scalar field 
quickly returns back to the slow-roll evolution. 
As we see in the left panel of Fig.~\ref{figevo}, the number of 
e-foldings $\Delta N_c$ acquired during the transient phase around $\phi=\phi_c$ is only a few, which is much smaller than 
$\Delta N_c$ in the USR case.

\subsection{Intermediate type}
\label{BackgroundCase3}

Besides the two types of $V_{\rm eff}(\phi)$ discussed above, 
there is also an intermediate case between the plateau- and 
bump-types. In this case the effective potential is not exactly 
flat in the vicinity of $\phi=\phi_c$, but it has a small peak and 
trough with a slight negative value of $V_{{\rm eff},\phi}$ around $\phi=\phi_c$. 
The Set 3 parameters in Table \ref{table1} give rise to such a 
shape of $V_{\rm eff}(\phi)$, see the right panel of Fig.~\ref{figpo}.

As we plot as a blue line in Fig.~\ref{figevo}, the field derivative initially decreases in proportion to $|\dot{\phi}| \propto e^{-3n}$, 
which is followed by a temporal period in which the 
decreasing rate of $|\dot{\phi}|$ becomes faster than 
$|\dot{\phi}| \propto e^{-3n}$.  
In this latter regime, the scalar field loses its velocity by climbing up the potential hill with $V_{{\rm eff},\phi}<0$. 
After the field reaches the local maximum of $V_{\rm eff}(\phi)$, 
$|\dot{\phi}|$ starts to grow toward the slow-roll regime. 
Since the effective potential has neither an exactly flat region nor a sharp bump, 
the number of e-foldings $\Delta N_c$ acquired around $\phi=\phi_c$ is between those of plateau- 
and bump-types.
In Set 3 model parameters, we have $\Delta N_c \simeq 9.7$. 

\section{Generation of the seed for PBHs}
\label{seedsec}

In this section, we study how the power spectrum of 
curvature perturbations is enhanced by the presence of the scalar-GB coupling (\ref{couplingfunction}).
As we discussed in Sec.~\ref{PBHformation}, 
the inflaton effective potential $V_{\rm eff}(\phi)$ 
can be classified into three classes: 
(1) plateau-type, (2) bump-type, and (3) 
intermediate-type.
Examples of the model parameters corresponding to 
each type of $V_{\rm eff}(\phi)$ are given in Table~\ref{table1} 
as Sets 1, 2, 3, respectively. 
In Sec.~\ref{stasec}, we first discuss whether each model satisfies the stability conditions (\ref{avoidinstability}).  
In Sec.~\ref{Prisca}, we compute the primordial scalar power 
spectra by paying particular attention to the enhancement 
of curvature perturbations around $\phi=\phi_c$. 
In Sec.~\ref{CMBsec}, we confront our models with the observed 
scalar spectral index and tensor-to-scalar ratio 
on CMB scales.

\subsection{Stability conditions}
\label{stasec}

Let us first study whether the stability conditions (\ref{avoidinstability}) 
can be satisfied during the whole stage of inflation. 
As we alluded in Sec.~\ref{linearsec}, 
the no-ghost parameter $Q_t$ 
in the tensor sector is positive for $\xi H \phi \dot{\phi}<0$.
For the three sets of model parameters in Table~\ref{table1}, 
the field derivative $\dot{\phi}$ is 
negative without reaching 0, 
see the right panel of Fig.~\ref{figevo}. 
Then, we have $Q_t>0$ even during the transient epoch 
around $\phi=\phi_c$. 
{}From Eq.~(\ref{QS2}), the other no-ghost 
parameter $Q_s$ is also positive for $Q_t>0$. 
Thus, the ghost instabilities are absent for both 
tensor and scalar perturbations, whose property is also 
confirmed numerically.

\begin{figure}[ht]
\begin{center}
\includegraphics[height=2.2in,width=3.3in]{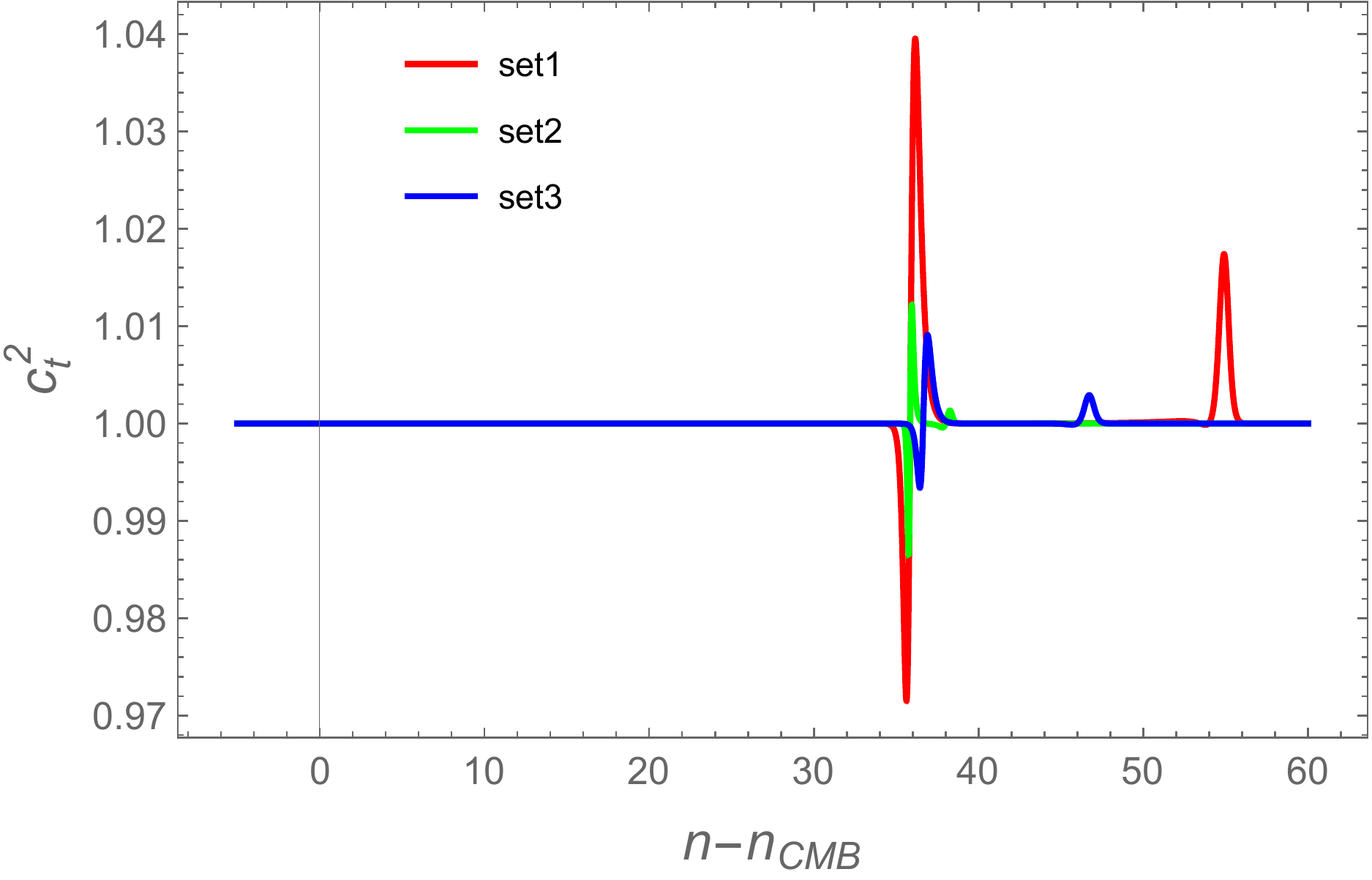}
\hspace{0.7cm}
\includegraphics[height=2.2in,width=3.3in]{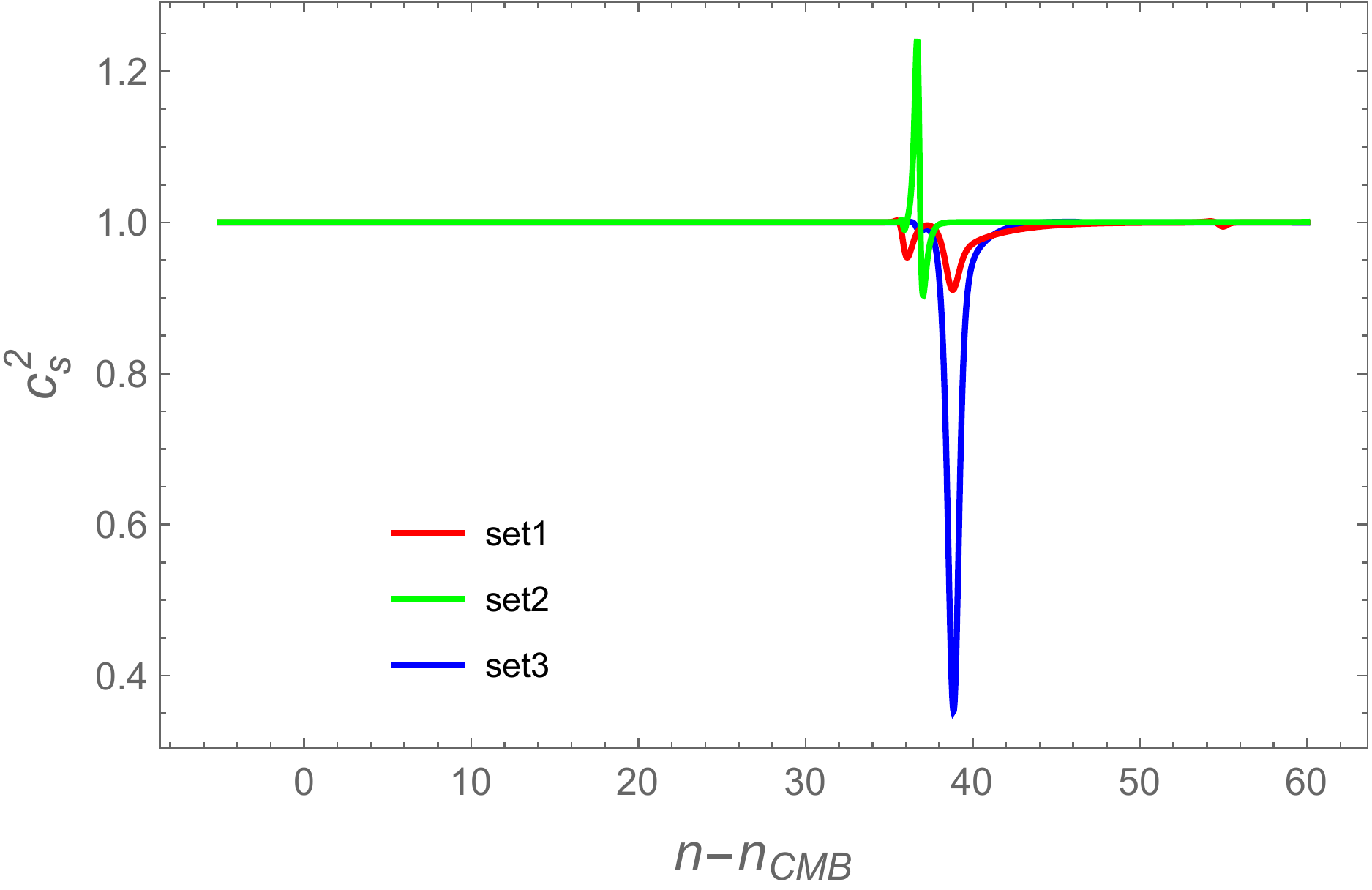}
\end{center}\vspace{-0.5cm}
\caption{\label{cts}
Evolutions of $c_t^2$ (left) and $c_s^2$ (right) for 
the three sets of model parameters given in Table \ref{table1}.
}
\end{figure}

The scalar-GB coupling generally gives rise to the tensor and 
scalar propagation speeds different from 1. 
In the left panel of Fig.~\ref{cts}, we show the evolution of $c_t^2$ versus the number of e-foldings $n-n_{\rm CMB}$ for the three sets of model parameters in Table~\ref{table1}. 
Just after the entry to the transient regime around 
$\phi=\phi_c$, $c_t^2$ exhibits the deviation from 1. 
During the USR stage realized for the Set 1 model 
parameters, the field velocity is significantly 
suppressed and hence $c_t^2$ approaches 1. 
Just after the inflaton field exits from 
the USR regime, there is the temporal deviation of $c_t^2$ 
from 1. In the subsequent slow-roll regime,  
$c_t^2$ again goes back to 1.
For the bump-type potential, which corresponds to the 
Set 2 model parameters, the time variation of $c_t^2$ 
occurs instantly in the vicinity of $\phi=\phi_c$. 
In all the cases shown in the left panel of Fig.~\ref{cts}, 
the deviation of $c_t^2$ from 1 is insignificant 
($|c_t^2-1| \lesssim 0.04$) and hence the stability condition $c_t^2>0$ is always satisfied.

In the right panel of Fig.~\ref{cts}, we plot $c_s^2$ 
versus $n-n_{\rm CMB}$ for the same sets of model parameters as those used in the left. 
For scalar perturbations, the no-ghost quantity $Q_s$ is proportional to 
$\dot{\phi}^2$. Since $Q_s$ appears in the denominator of $c_s^2$ 
in Eq.~(\ref{cs}), 
a smaller $\dot{\phi}^2$ does not imply a value of $c_s^2$ closer to 1. 
Indeed, during the transient regime around $\phi=\phi_c$, the 
scalar propagation speeds deviate from 1 
in all the three cases shown in Fig.~\ref{cts}. 
For Set 1, the minimum value of $c_s^2$ reached 
during the USR regime is about 0.91. 
The bump-type potential corresponds to the Set 2 model parameters, 
in which case $c_s^2$ temporally increases to the superluminal region 
and then quickly evolves to the minimum around $c_s^2=0.90$. 
For Set 3, $c_s^2$ decreases to the minimum around 0.35 and 
returns back to the value close to 1 in the slow-roll regime.
Since the positivity of $c_s^2$ holds in all these cases, there are no 
Laplacian instabilities for scalar perturbations. 

Depending on the model parameters, there are cases 
in which the scalar sound speed enters the region $c_s^2<0$. 
Since such models should be excluded by the Laplacian 
instability, we will focus on the case $c_s^2>0$ 
in subsequent sections.

\subsection{Primordial scalar power spectrum}
\label{Prisca}

To study the evolution of curvature perturbations during inflation, we introduce the ``sound-horizon'' time defined by 
\be
\tau_s=\int {\rm d}t \frac{c_s}{a}\,,
\ee
where $c_s>0$.
Then, the second-order action (\ref{action2}) of scalar perturbations reduces to 
\be
{\cal S}_s^{(2)}=\int \rd \tau_s \rd^3 x\,a^2 Q_s c_s 
\left[ \zeta'^2-(\nabla \zeta)^2\right]\,,
\label{action3}
\ee
where a prime represents the derivative with respect to $\tau_s$.
We decompose the curvature perturbation into 
the Fourier modes 
$\zeta_k$ as  
\be
\zeta(\tau_s, {\bm x})=\frac{1}{(2\pi)^3} \int \rd^3 k \left[ 
\zeta_k (\tau_s, {\bm k}) a ({\bm k})+
\zeta_k^* (\tau_s, -{\bm k}) a^\dagger (-{\bm k}) 
\right] e^{i {\bm k} \cdot {\bm x}}\,,
\ee
where ${\bm k}$ is a comoving wavenumber, and 
$a ({\bm k})$ and $a^\dagger ({\bm k})$ are 
annihilation and creation operators, respectively. 
Introducing the rescaled field 
\be
u_k=Z_s \zeta_k\,,\qquad {\rm with}
\qquad 
Z_s=a \sqrt{2Q_s c_s}\,,
\label{ZS}
\ee
we obtain the following differential equation 
\be
u_k''+\left( k^2-\frac{Z_s''}{Z_s}\right)u_k=0\,.
\label{curvatureperturbEOM}
\ee
For the perturbations deep inside the sound-horizon 
($k^2 \gg a^2 H^2/c_s^2$), $Z_s''/Z_s$ is 
suppressed relative to $k^2$. 
For such modes, we choose a positive-frequency solution 
in a Bunch-Davies vacuum state, i.e., 
\be
u_k=\frac{1}{\sqrt{2 k}}e^{-ik\tau_s}\,,
\label{BunchDavis}
\ee
as an initial condition.
In practice, we set the initial time for each wavenumber 
that corresponds to 6 e-foldings before the 
sound-horizon crossing, i.e., at $k =e^6 aH/c_s$. 
Numerically we integrate Eq.~(\ref{curvatureperturbEOM}) by 
the end of inflation (characterized by the 
time $\tau_s=\tau_{sf}$) 
and compute the scalar power spectrum given by 
\be
{\cal P}_{\zeta}(k)=
\frac{k^3}{2\pi^2}
\left| \zeta_k (\tau_{sf},{\bm k}) 
\right|^2\,.
\ee

We evaluate ${\cal P}_{\zeta}(k)$ for three sets of model parameters given in Table \ref{table1}. 
At the 60 e-foldings before the end of inflation, 
the observed amplitude 
${\cal P}_{\zeta}=2.1 \times 10^{-9}$ on CMB scales is 
used to give a constraint among the model parameters.
The two constants $\mu_0$ and $\mu_1$ in Table \ref{table1} are chosen to realize the sufficient enhancement 
of ${\cal P}_{\zeta}$ around $\phi=\phi_c$.

\begin{figure}[ht]
\begin{center}
\includegraphics[height=3.3in,width=4.8in]{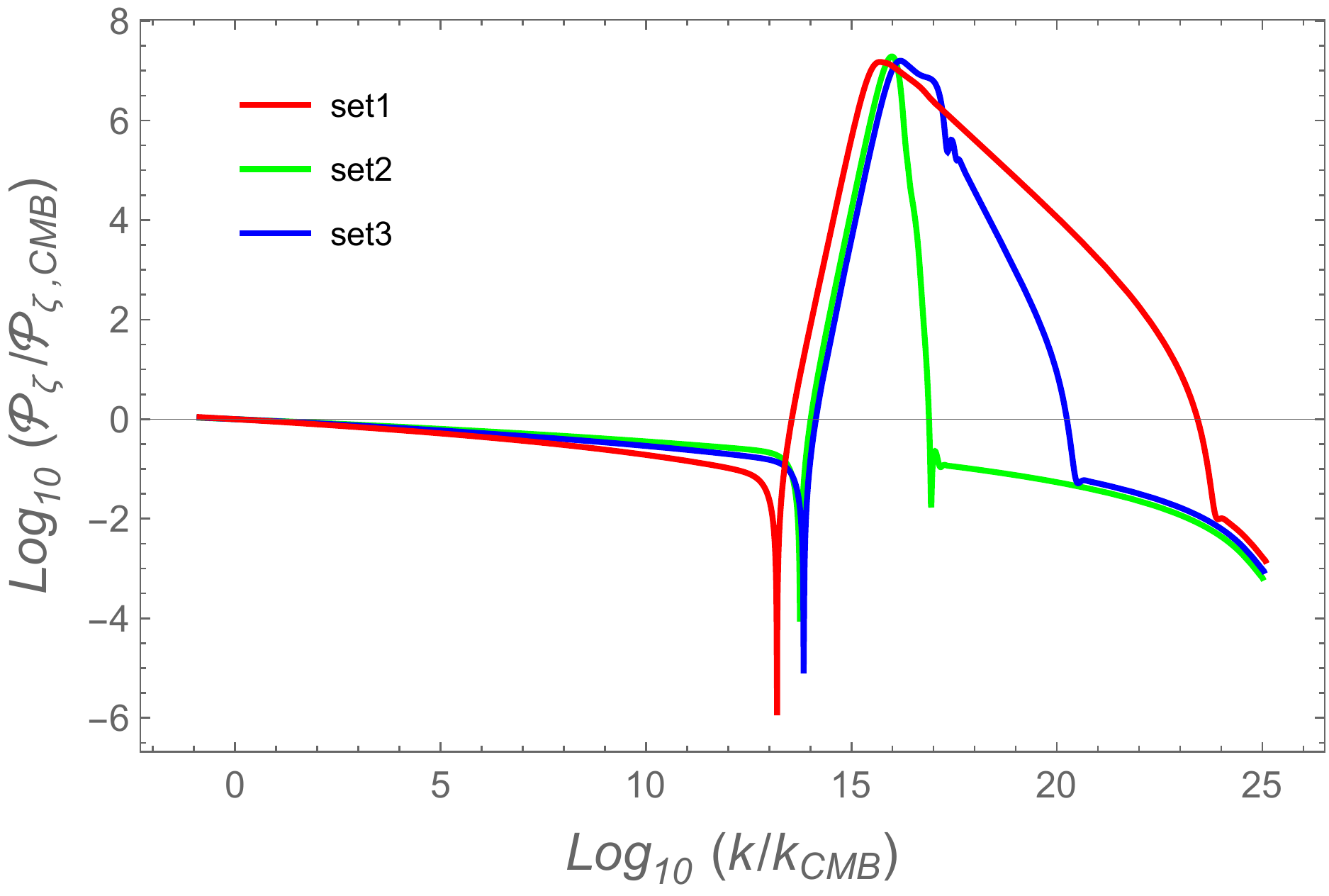}
\end{center}
\caption{\label{Pzfig}
The scalar power spectrum ${\cal P}_{\zeta}$ 
(normalized by its value on 
the CMB scale ${\cal P}_{\zeta,{\rm CMB}}$) 
versus the comoving wavenumber $k$ (normalized by 
the wavenumber $k_{\rm CMB}$ on the CMB scale). 
Each plot corresponds to the 
three sets of model parameters presented in 
Table \ref{table1}.}
\end{figure}

In Fig.~\ref{Pzfig}, we show the primordial scalar power spectrum ${\cal P}_{\zeta}(k)$ for three sets of model parameters.  
In comparison to the value of ${\cal P}_{\zeta}$ on the CMB scale 
(characterized by the comoving wavenumber of order 
$k_{\rm CMB}=10^{-3}~$Mpc), there is an enhancement of 
${\cal P}_{\zeta}$ by a factor of $10^7$. 
While the heights of peaks are similar between the three sets 
of parameters, the widths of ${\cal P}_{\zeta}$ are different from 
each other. The latter property is mostly attributed to the fact that 
the number of e-foldings 
$\Delta N_c$ acquired around 
$\phi=\phi_c$ are different between the plateau-, bump-, 
and intermediate-types.

Let us try to understand how the enhancement of ${\cal P}_{\zeta}$ around $\phi=\phi_c$ occurs in detail. 
For the perturbations outside the sound horizon 
($k^2 \ll a^2 H^2/c_s^2$), Eq.~(\ref{curvatureperturbEOM}) 
is approximately given by 
\be
u_k''-\frac{Z_s''}{Z_s}u_k \simeq 0\,.
\label{curvatureperturbEOM2}
\ee
The solution to this equation can be generally expressed 
in the form 
\be
\zeta_k(\tau_s)=\frac{u_k(\tau_s)}{Z_s(\tau_s)}=
A_k +B_k 
\int^{\tau_s} \frac{\rd \tilde{\tau}_s}
{Z_s^2(\tilde{\tau}_s)}\,,
\label{uksolution}
\ee
where $A_k$ and $B_k$ are constants. 
On using the approximations $\Theta \simeq H \xi \phi^2$ 
and $Q_t \simeq V/(12H^2)$ in the regimes 
$\xi \phi^2 \gg \Mpl^2$ and $\xi \gg 1$, the 
no-ghost parameter $Q_s$ in Eq.~(\ref{QS2}) approximately 
reduces to 
\be
Q_s \simeq \frac{4(3\xi^2+\lambda \mu_{,\phi}\phi)^2}
{\lambda \xi^2 \phi^2}\dot{\phi}^2\,,
\ee
which is positive. 
Since we are considering the scalar-field evolution with 
$\dot{\phi}<0$, the quantity $Z_s$ has the dependence
\be
Z_s \simeq -2 \sqrt{\frac{2c_s}{\lambda}} 
\frac{3\xi^2+\lambda \mu_{,\phi}\phi}{\xi \phi}
a \dot{\phi}\,.
\ee
During the slow-roll regime in which the contribution from the 
scalar-GB coupling is negligible, the field derivative $\dot{\phi}$ changes slowly with $c_s \simeq 1$. 
In the limit that $\dot{\phi}$ is constant, we have $Z_s \propto a$ and hence the last term of Eq.~(\ref{uksolution}) decays in proportion to $a^{-3}$ on the quasi de-Sitter 
background (where $a \simeq -(H \tau_s)^{-1}$).
Then, after the sound horizon crossing, 
$\zeta_k$ approaches a constant value $A_k$.

On the other hand, the field derivative exhibits a temporal rapid decrease 
during the transition around $\phi=\phi_c$. 
Since the scalar-field solution in the USR regime 
arising from the plateau-type effective 
potential is given by $\dot{\phi} \propto a^{-3}$, 
neglecting the time dependence of $c_s$ leads to the approximate relation 
$Z_s \propto a^{-2}$. Then, the last term in Eq.~(\ref{uksolution}) 
increases in proportion to $a^3$.
This means that the decaying mode in the slow-roll regime
is replaced by the rapidly growing mode in the USR regime. 
Since $\dot{\phi}$ also decreases rapidly for 
bump- and intermediate-type potentials, the curvature 
perturbation can be strongly enhanced around $\phi=\phi_c$ 
as well.

\begin{figure}[ht]
\begin{center}
\includegraphics[height=2.3in,width=3.3in]{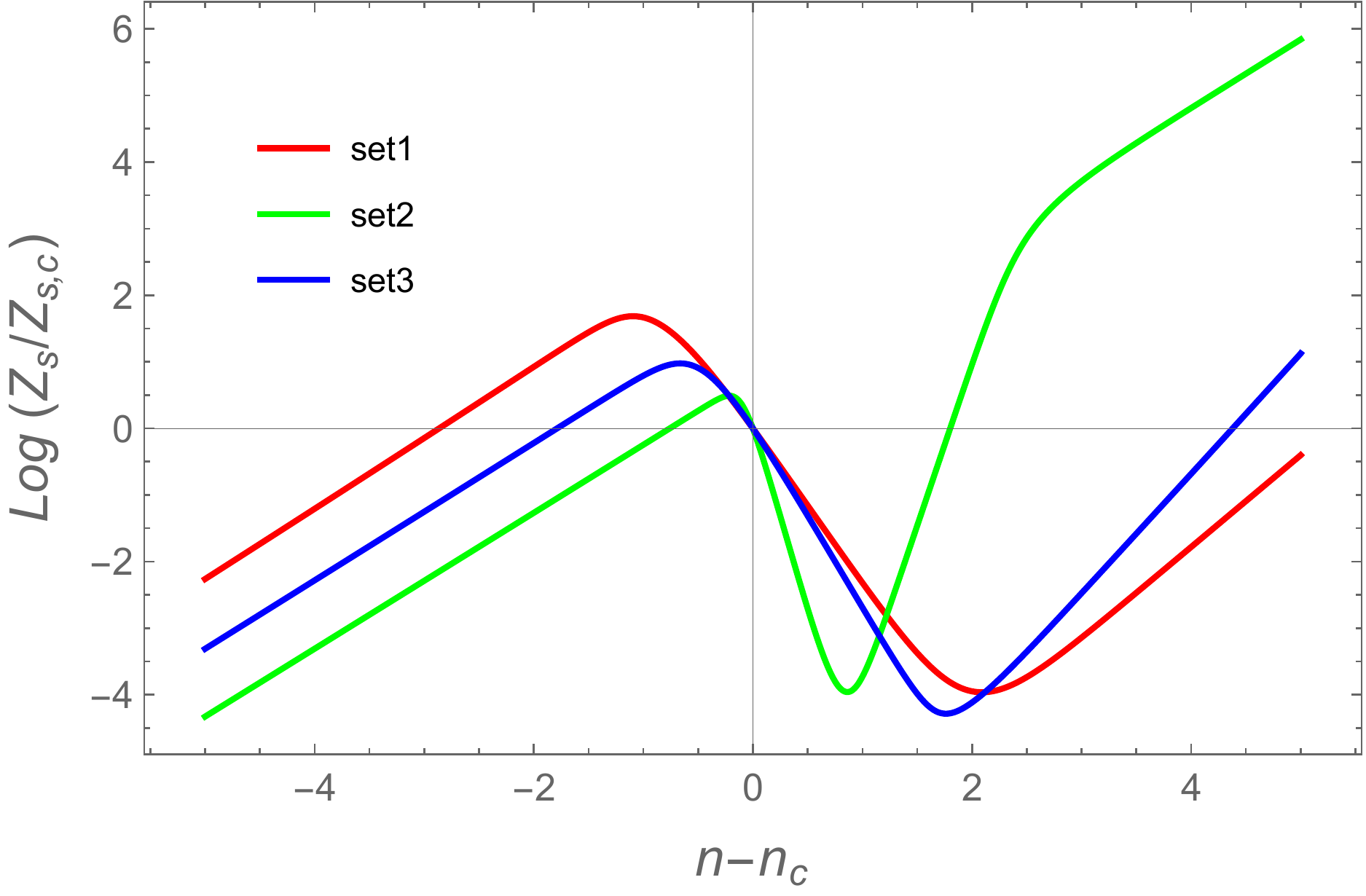}
\hspace{0.3cm}
\includegraphics[height=2.3in,width=3.3in]{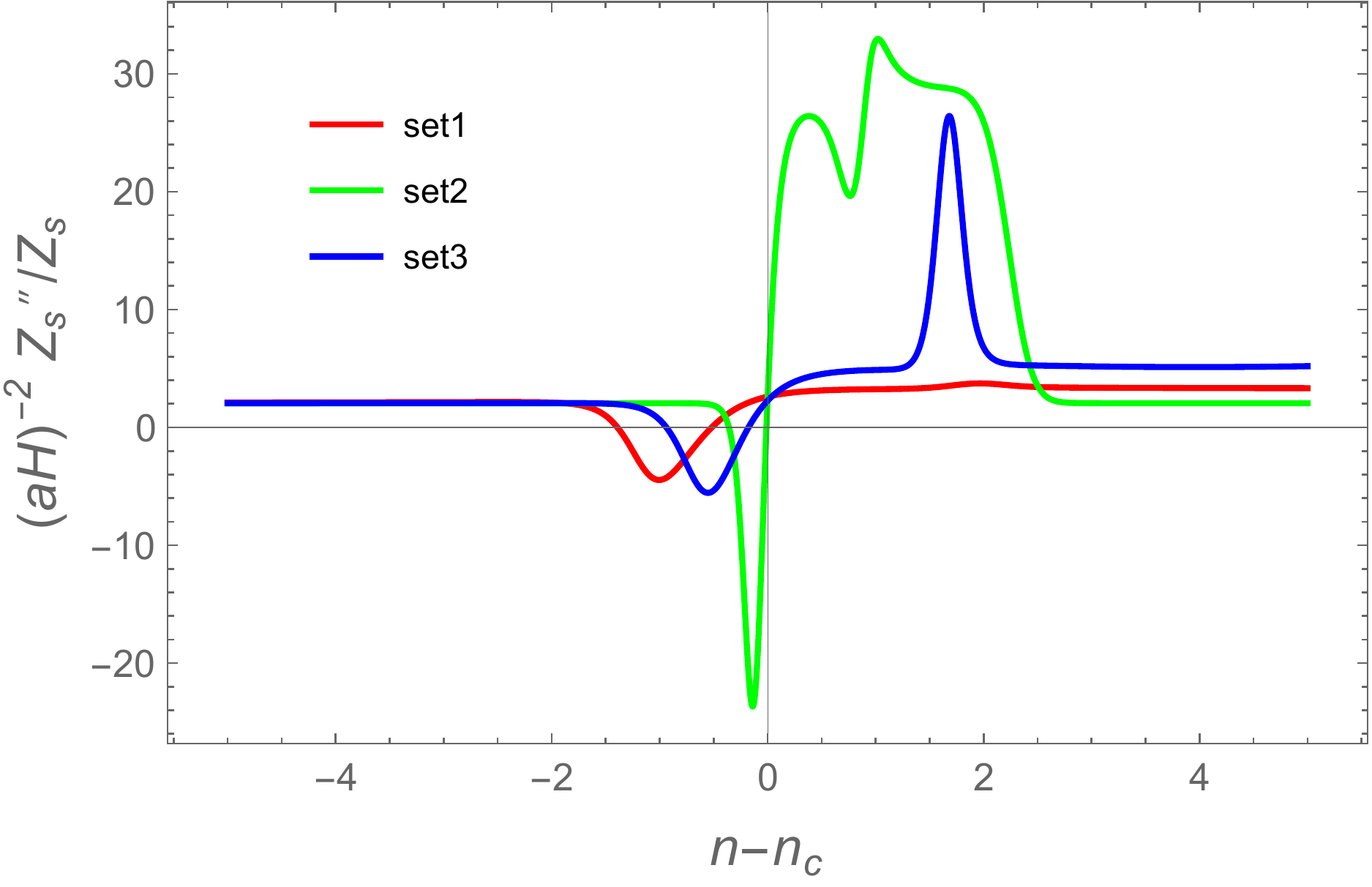}
\end{center}\vspace{-0.5cm}
\caption{\label{zfig}
Evolution of $Z_s$ (left) and $(aH)^{-2}Z_s''/Z_s$ 
(right) around $n=n_c$ for three sets of model 
parameters given in Table~\ref{table1}.
Note that $Z_s$ is normalized by the value $Z_{s,c}$ 
at $n=n_c$. }
\end{figure}

In the left panel of Fig.~\ref{zfig}, we plot $Z_s$ versus $n-n_c$ for 
three sets of model parameters in Table \ref{table1}. 
During the initial slow-roll regime, the evolution of $Z_s$ is 
approximately given by $Z_s \propto a$ in all these three cases.
For Set 1, which corresponds to the plateau-type potential, 
we find that $Z_s$ decreases as $Z_s \propto a^{-2}$ 
in the vicinity of $\phi=\phi_c$ as expected. 
As we see in the right panel of Fig.~\ref{cts}, 
the deviation of $c_s^2$ from 1 is insignificant 
for Set 1 model parameters and hence the evolution 
of $Z_s$ is hardly affected by the variation of $c_s^2$.

The bump-type potential realized by Set 2 model parameters 
leads to a larger decreasing rate of $Z_s$ around $\phi=\phi_c$ in comparison to the plateau-type 
because of the stronger suppression of 
$\dot{\phi}$ (see the right panel of Fig.~\ref{figevo}). 
For Set 2, the scalar sound speed temporally 
enters the region $c_s^2>1$ with 
a shorter transient period around 
$\phi=\phi_c$ in comparison to Set 1. 
For Set 3, which corresponds to the intermediate-type 
potential, the decreasing rate of $Z_s$ is 
between those of the plateau- and bump-types. 
In all these cases, $Z_s$'s begin to increase 
after reaching their minima, whose behavior is 
correlated with the evolution of $\dot{\phi}$ 
plotted in Fig.~\ref{figevo}.

In the superhorizon regime, the enhancement 
of $\zeta_k$ depends on the integral of 
$1/Z_s^2 (\tau_s)$ with respect to 
$\tau_s=\int {\rm d}t\,c_s/a$. 
On the other hand, we recall that we ignored 
the $k^2$ term in Eq.~(\ref{curvatureperturbEOM}) 
relative to $Z_s''/Z_s$ in the 
above argument. 
To understand which modes of $\zeta_k$ are subject to 
the amplification, 
we explicitly compute $Z_s''/Z_s$ as 
\be
\frac{Z_s''}{Z_s}=
\frac{(aH)^2}{c_s^2}\left[2-\epsilon_H+\frac{3\epsilon_Q}{2}+\frac{\epsilon_c}{2}-\frac{1}{4}(\epsilon_c+\epsilon_Q)(2\epsilon_H-\epsilon_Q+\epsilon_c)+\frac{1}{2}(\epsilon_c\eta_c+\epsilon_Q\eta_Q)\right]\,,
\label{ZS2}
\ee
where 
\be
\epsilon_c\equiv\frac{\dot{c_s}}{Hc_s}\,,\qquad \epsilon_Q\equiv\frac{\dot{Q_s}}{HQ_s}\,,\qquad 
\eta_c\equiv\frac{\dot{\epsilon_c}}{H\epsilon_c}\,,\qquad \eta_Q\equiv\frac{\dot{\epsilon_Q}}{H\epsilon_Q}\,.
\label{epsiloneta}
\ee
In the regime of slow-roll inflation, we have 
the approximate relation $Z_s''/Z_s \simeq 2(aH)^2$ 
and hence $\zeta_k$ soon approaches a constant 
after the Hubble radius crossing ($k \lesssim aH$). 
During the transient regime around $\phi=\phi_c$, 
the quantities defined in Eq.~(\ref{epsiloneta}) can be 
larger than order 1. 
In the right panel of Fig.~\ref{zfig}, we plot 
${\cal Z} \equiv (aH)^{-2} Z_s''/Z_s$ 
versus $n-n_c$ for three sets of model parameters.
For Set 1, ${\cal Z}$ starts to evolve from the value 
close to 2, temporally shows some decrease, and again increases to 
the value around 4. This means that, for the wavenumbers 
in the range $k \lesssim 2aH$, there is the enhancement of 
$\zeta_k$ when the scalar field evolves along the plateau region 
of $V_{\rm eff}(\phi)$.
For the perturbations which crossed the Hubble radius 
in the preceding slow-roll period, the last integral 
in Eq.~(\ref{uksolution}) has already decayed sufficiently around
the time at which $\phi$ approaches $\phi_c$. 
Hence the enhanced modes of $\zeta_k$ are those crossed 
the Hubble radius during the transient epoch 
around $\phi=\phi_c$.

For Set 2, we observe in Fig.~\ref{zfig} that ${\cal Z}$ quickly 
reaches the value around 25 in the range $0 \lesssim n-n_c \lesssim 2$. 
This means that curvature perturbations up to 
the wavenumber $k \lesssim 5 aH$, which include 
subhorizon modes, can be amplified. 
In comparison to Set 1, the shorter transient period around $\phi=\phi_c$ still limits the range of $k$ for enhanced modes, see Fig.~\ref{Pzfig}. 
Nevertheless, the height of peak of ${\cal P}_{\zeta}(k)$ in Set 2 is similar to that in Set 1 thanks to the rapid increase of ${\cal Z}$.
For Set 3, the enhanced scalar power spectrum spans 
in the ranges of $k$ between those of Sets 1 and 2. 
In this case, the rapid decrease of $c_s^2$ seen in 
the right panel of Fig.~\ref{cts} generates a 
sharp peak in ${\cal Z}$. This gives rise to 
a shape of ${\cal P}_{\zeta}(k)$ whose peak structure 
is different from those in other two cases.
As we will see in Sec.~\ref{pbhabundance}, 
the PBH abundance generated 
by these three sets of primordial power spectra can be 
sufficiently large to 
serve as almost all DM.

\subsection{CMB constraints}
\label{CMBsec}

The existence of the transient regime with strongly suppressed values of 
$\dot{\phi}$ modifies the spectra of scalar and tensor perturbations 
on scales relevant to the observed CMB 
temperature anisotropies. 
Besides 
Eq.~(\ref{curvatureperturbEOM}), 
we also numerically solve the equation of tensor perturbations 
following from the second-order action (\ref{actionT}).
We then compute the CMB observables like $n_s$, $r$, and 
${\cal P}_{\zeta}$ around the $N_{\rm CMB}=60$ e-foldings 
backward from the end of inflation.

As we see in Fig.~\ref{figevo}, the inflaton field stays nearly constant 
in the region around $\phi=\phi_c$. In this transient regime, 
the acquired number of e-foldings $\Delta N_c$ is different 
depending on the model parameters.
For larger $\Delta N_c$, the field value $\phi_{\rm CMB}$ 
around the CMB scale tends to be smaller. 
Since $\phi$ stays nearly constant during the transient regime, 
we can simply replace the relation 
$\phi^2 \simeq 4\Mpl^2 N/(3\xi)$ (which was 
derived for $\mu(\phi)=0$ in the limit $N \gg 1$) with 
$\phi \to \phi_{\rm CMB}$ and 
$N \to N_{\rm CMB}-\Delta N_c$. 
Then, it follows that 
\be
\phi_{\rm CMB} \simeq 
2 \sqrt{\frac{N_{\rm CMB}-\Delta N_c}{3\xi}}\Mpl\,.
\label{phiCMB}
\ee
In the USR case corresponding to the Set 1 model parameters 
in Fig.~\ref{figevo}, we have $\Delta N_c=18.6$ with 
$N_{\rm CMB}=60$ and 
hence $\phi_{\rm CMB} \simeq 0.105\Mpl$ from 
Eq.~(\ref{phiCMB}). 
This shows good agreement with 
the numerical value of $\phi_{\rm CMB}$. 

{}From Eqs.~(\ref{power1}) and (\ref{power2}), the scalar power spectrum on CMB scales is known by 
the replacement 
$N \to N_{\rm CMB}-\Delta N_c$, while the tensor power spectrum is hardly subject to modifications 
by the scalar-GB coupling. 
Applying the change $N \to N_{\rm CMB}-\Delta N_c$ 
to Eqs.~(\ref{power1}), (\ref{ns}), and (\ref{r}), 
we obtain 
\ba
{\cal P}_\zeta\bigr|_{\text{CMB}} &\simeq&
\frac{\lambda(N_{\text{CMB}}-\Delta N_c)^2}{72 \pi^2 \xi^2}
\label{power3}\,,\\
n_s-1 &\simeq& -\frac{2}{N_{\text{CMB}}-\Delta N_c}\label{ns2}\,,\\
r &\simeq& \frac{12}{(N_{\text{CMB}}-\Delta N_c)^2}\label{r2}\,.
\ea
For larger $\Delta N_c$, these observables exhibit more significant 
deviations from those in standard Higgs inflation. 
The values of $\lambda$ in Table \ref{table1} are chosen to 
match the amplitude ${\cal P}_{\zeta}|_{\text{CMB}}
=2.1\times10^{-9}$ constrained by 
the Planck CMB data.

\begin{table}[h]
\newcommand\xrowht[2][0]{\addstackgap[.5\dimexpr#2\relax]{\vphantom{#1}}}
\newcolumntype{C}[1]{>{\hfil}m{#1}<{\hfil}}
\centering
\caption{Numerical values of the scalar spectral index 
$n_s$ and the tensor-to-scalar ratio $r$ 
for three sets of model parameters 
given in Table \ref{table1} and 
Higgs inflation with $\mu(\phi)=0$.
The number of e-foldings on the CMB scale is 
fixed to be $N_{\rm CMB}=60$.
}
\begin{tabular}{C{28mm}C{28mm}C{28
mm}} \hline\hline \xrowht{12pt}
      & $n_s$ & $r$  \\ \hline \xrowht{12pt}
     Set 1 & 0.951415 &  0.00757526   \\
    \xrowht{12pt}
     Set 2 & 0.965142 &  0.00347235  \\
      \xrowht{12pt}
    Set 3 & 0.960055 & 0.00476115 \\
     \xrowht{12pt}
    Higgs inflation & 0.966527 & 0.00323724 \\
   \hline\hline
\end{tabular}\label{table2}
\end{table}
\begin{figure}[ht]
\begin{center}
\includegraphics[height=3.8in,width=4.8in]{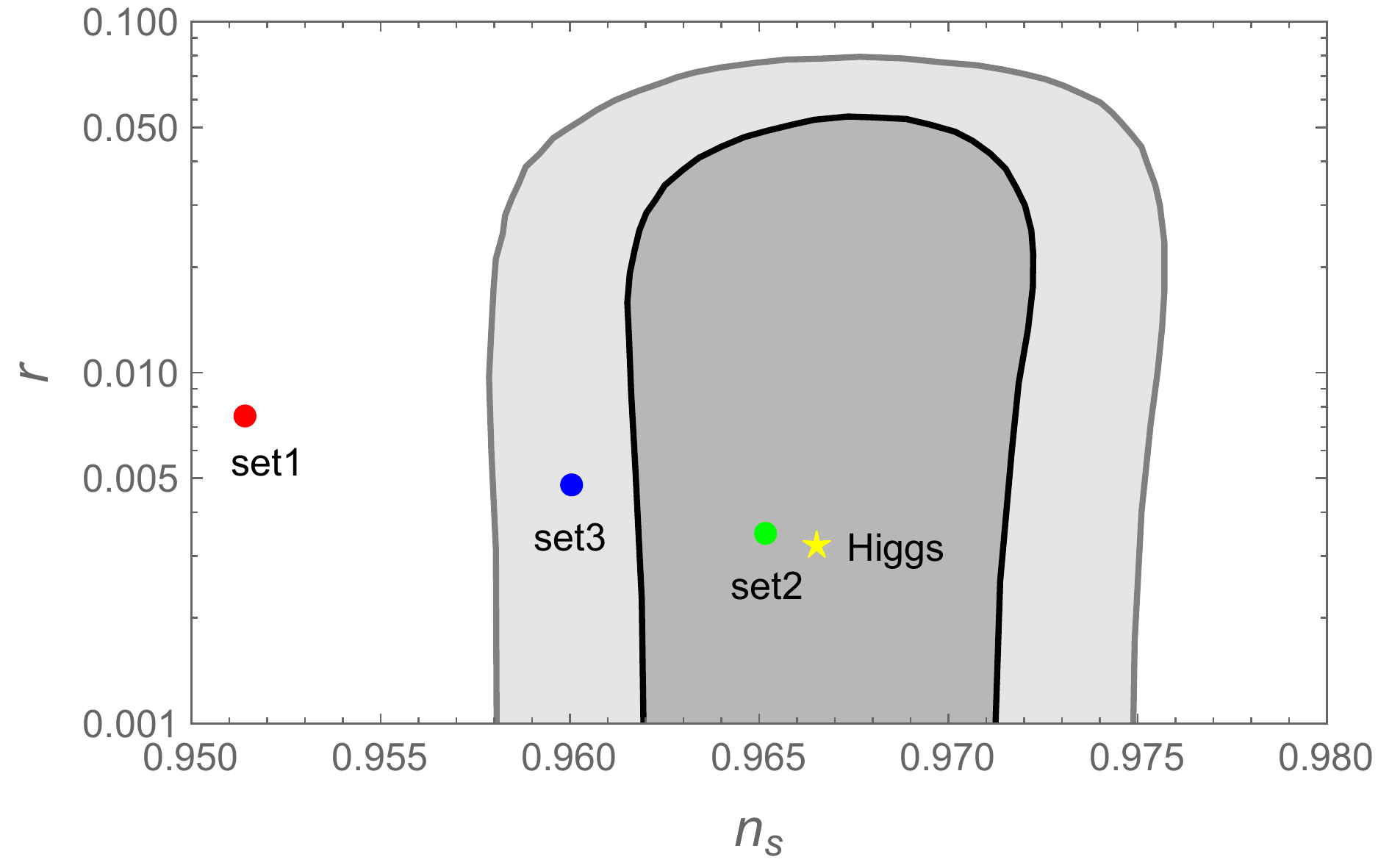}
\end{center}\vspace{-0.5cm}
\caption{\label{fignsr}
Dark grey and light grey areas represent the $1\sigma$ and 
$2\sigma$ observational regions, respectively, 
constrained by the joint data analysis 
of Planck 2018 + BK14 + BAO at 
$k=0.002$~Mpc$^{-1}$. 
The red, green, and blue points correspond to the 
theoretical predictions of $n_s$ and $r$ for the model 
parameters of Sets 1, 2, 3, respectively, while the 
yellow star represents those of Higgs inflation. 
The number of e-foldings on the CMB scale is 
fixed to be $N_{\rm CMB}=60$.
}
\end{figure}

For the Set 1 model parameters in Table \ref{table1}, substituting 
$\Delta N_c=20$ and $N_{\rm CMB}=60$ into 
the analytic estimations of Eqs.~(\ref{ns2}) and (\ref{r2}) 
gives $n_s=0.950$ and $r=0.00750$. 
These are close to the numerically derived values presented in the first column of Table \ref{table2}. 
In Fig.~\ref{fignsr}, we show the $1\sigma$ and $2\sigma$
observational bounds constrained from Planck 2018 data
combined with the data of B-mode polarizations available from 
the BICEP2/Keck field (BK14) and baryon acoustic oscillations 
(BAO) \cite{Planck:2018jri}.
The theoretical values of $n_s$ and $r$ for Set 1 are outside 
the $2\sigma$ observational contour, so the plateau-type effective potential 
with $\Delta N_c=18.6$ is disfavored from the data.
Unless we choose an unusually large value of $N_{\rm CMB}$ exceeding 68, 
the model with $\Delta N_c \simeq 20$ does not enter the inside of 
the $2\sigma$ contour.

For Set 2, the number of e-foldings acquired around the bump region 
of $V_{\rm eff}(\phi)$ is as small as $\Delta N_c=2.3$, so 
the CMB observables are similar to those 
in standard Higgs inflation, 
see the second and fourth columns 
in Table \ref{table2}.
As we observe in Fig.~\ref{fignsr}, the model with Set 2 is inside the $1\sigma$ observational contour. 
For Set 3, we numerically obtain the values of $n_s$ and $r$ given 
in the third column of Table \ref{table2}, 
with $\Delta N_c=9.7$. 
In this case, the model is between $1\sigma$ and $2\sigma$ 
observational contours in the ($n_s, r$) plane. 
For $N_{\rm CMB}=60$, the number of e-foldings acquired 
around $\phi=\phi_c$ should be in the ranges
\ba
& &
0 \le \Delta N_c < 7.6 
\qquad \quad~(1\sigma)\,,
\label{const1}\\
& &
7.6 \le\Delta N_c < 12.5\qquad (2\sigma)\,,
\label{const2}
\ea
for the consistency with observations 
of $n_s$ and $r$ at $1\sigma$ and $2\sigma$ 
confidence levels, respectively. 
Thus, the bump-type model with $\Delta N_c=$~a few 
is favored over the plateau-type model with 
$\Delta N_c>12.5$ from the viewpoint of CMB constraints.

\begin{figure}[ht]
\begin{center}
\includegraphics[height=3.7in,width=3.7in]{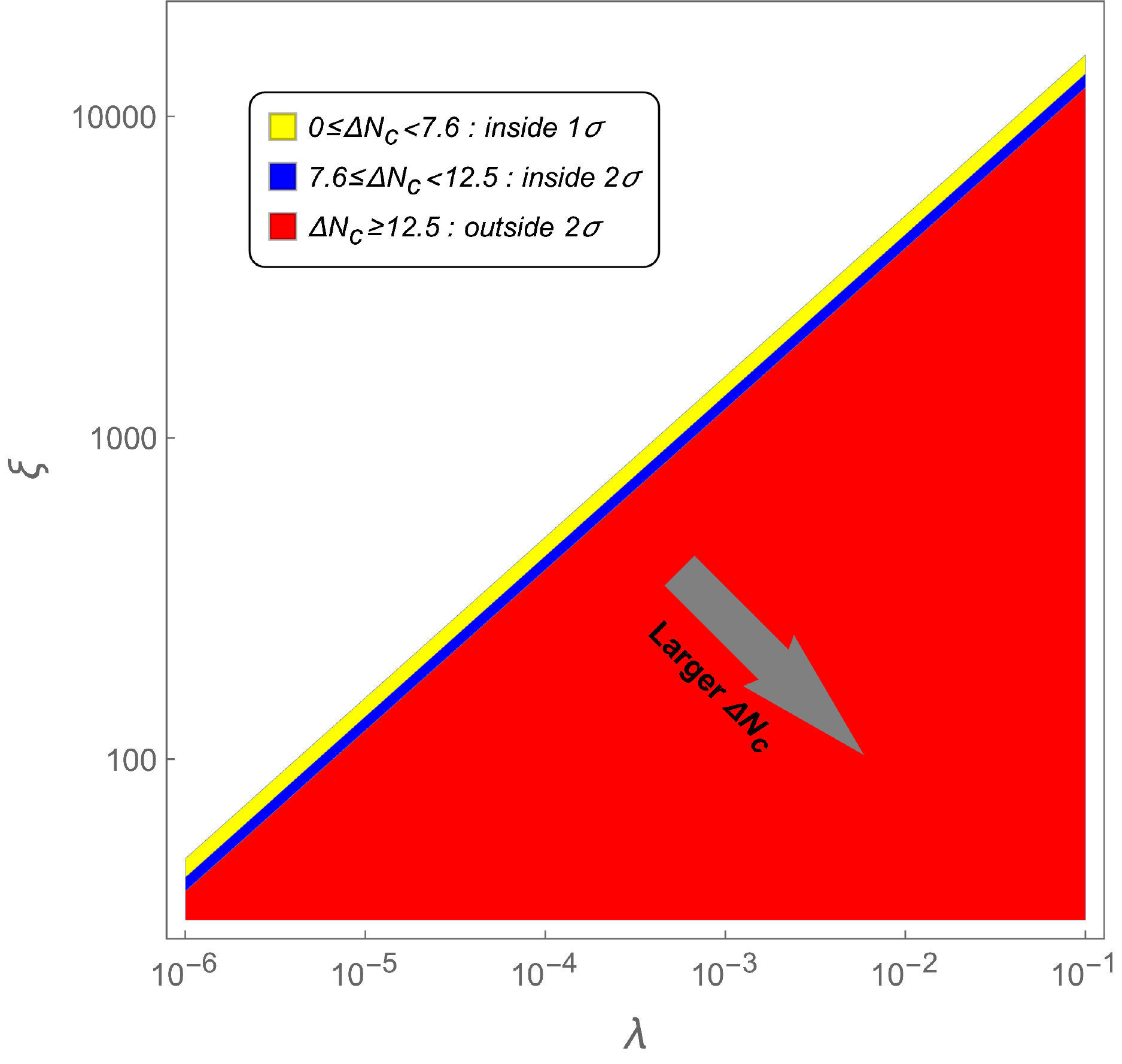}
\end{center}
\caption{$1\sigma$ (yellow) and $2\sigma$ (blue) 
confidence level parameter spaces in the 
$(\lambda, \xi)$ 
plane constrained from the 
observed CMB temperature 
anisotropies with ${\cal{P}}_{\zeta}=2.1\times 10^{-9}$ and $N_{\rm CMB}=60$. The $1\sigma$ and 
$2\sigma$ regions correspond 
to $\Delta N_c$ in the ranges (\ref{const1}) and (\ref{const2}), 
respectively.
The red region is outside the $2\sigma$ 
observational contour.
\label{xilambda}}
\end{figure}

On using the Planck normalized value 
${\cal P}_{\zeta |{\rm CMB}}=2.1 \times 10^{-9}$ 
in Eq.~(\ref{power3}) together with Eq.~(\ref{ns2}), 
we obtain the following relations
\be
\frac{\lambda}{\xi^2} \simeq 
\frac{1.49 \times 10^{-6}}
{(N_{\rm CMB}-\Delta N_c)^2} \simeq
3.73 \times 10^{-7} (n_s-1)^2\,.
\label{lamxi}
\ee
Fixing $N_{\rm CMB}$ to be 60, the scalar spectral 
index is in the range $n_s=1-2/(60-\Delta N_c) 
\leq 0.9667$ for $\Delta N_c \geq 0$.
This means that, from Eq.~(\ref{lamxi}), 
the ratio $\lambda/\xi^2$ is in the range 
$\lambda/\xi^2 \geq 4.1 \times 10^{-10}$. 
As $\Delta N_c$ increases, 
$\lambda/\xi^2$ gets larger.
The $1\sigma$ and $2\sigma$ confidence 
regions of $\Delta N_c$, which are given by 
Eqs.~(\ref{const1}) and (\ref{const2}) respectively,   
translate to
\ba
& &
4.1 \times 10^{-10} \le \frac{\lambda}{\xi^2} 
< 5.4 \times 10^{-10}\qquad (1\sigma)\,,\\
& &
5.4 \times 10^{-10} \le \frac{\lambda}{\xi^2} 
< 6.6 \times 10^{-10}\qquad (2\sigma)\,,
\ea
In Fig.~\ref{xilambda}, these parameter spaces are plotted as 
yellow ($1\sigma$) 
and blue ($2\sigma$) regions 
in the $(\lambda, \xi)$ 
plane. The red region is outside the $2\sigma$ observational contour. 
The allowed parameter spaces
shown in Fig.~\ref{xilambda} will be useful to put further constraints on 
the values of $\lambda$ and 
$\xi$ from future collider
experiments.

\section{PBH abundance}
\label{pbhabundance}

After inflation the perturbations reenter the Hubble radius, whose epoch depends on the comoving wavenumber $k$. 
In Sec.~\ref{seedsec}, we showed that the presence of the scalar-GB coupling can lead to the sufficient amplification 
of curvature perturbations during inflation for particular wavelengths  smaller than the CMB scale ($k^{-1} \simeq 10^3$~Mpc). 
Such overdense regions can collapse to form PBHs after 
the horizon reentry.  

The horizon mass associated with 
the Hubble distance $H^{-1}$ is given by $M_H=4\pi \Mpl^2 H^{-1}$.
The mass of PBHs at its formation time $t_{\text{form}}$ can 
be expressed as \cite{Green:2004wb}
\be
M=\gamma M_H=4\pi \gamma \Mpl^2 
H(t_{\rm form})^{-1}\,, 
\label{Mgam}
\ee
where $\gamma$ is the ratio of how much of the inner region of the Hubble radius collapses into PBHs. 
We will consider the case in which the 
formation of PBHs occurs during the 
radiation-dominated epoch.
The Hubble parameter at $t=t_{\rm form}$ 
is related to today's Hubble constant 
$H_0$ as \cite{Kawai:2021edk}
\be
\frac{H(t_{\rm form})}{H_0}
=\sqrt{\Omega_{r0}} 
\left[ \frac{a_0}{a(t_{\rm form})} 
\right]^2 \left[ \frac{g_{*0}}
{g_{*}(t_{\text{form}})} 
\right]^{1/6}\,,
\label{Hre}
\ee
where the subscript ``0'' represents 
today's values and $g_*$ is the 
relativistic degrees of freedom. 
The wavenumber at horizon reentry 
corresponds to $k=a(t_{\text{form}})
H(t_{\text{form}})$, whose relation 
can be used to eliminate $a(t_{\text{form}})$ in
Eq.~(\ref{Hre}). Solving Eq.~(\ref{Hre}) 
for $H(t_{\rm form})$ and substituting it into Eq.~(\ref{Mgam}), it follows that
\be
M(k)=10^{-13}M_{\odot}\left(\frac{\gamma}{0.2}\right)\left[ \frac{g_{*}(t_{\text{form}})}{106.75} 
\right]^{-1/6} \left(\frac{k}{4.9\times 10^{12}~\text{Mpc}^{-1}}\right)^{-2}\,,
\label{PBHmass}
\ee
where we used the values $g_{*0}=3.36$, 
$\Omega_{r0}=9 \times 10^{-5}$, and 
$H_0=10^{-42}$~GeV.

The abundance of PBHs can be estimated by using the  Press-Schechter theory.
Assuming a Gaussian distribution for the coarse-grained density fluctuation $\delta_g$, the probability that 
$\delta_g$ is higher than a certain threshold value $\delta_c$ at $t=t_{\text{form}}$ is given by \cite{Green:2004wb,Young:2014ana,Harada:2013epa,Germani:2018jgr}
\be
\beta(M(k))=\int^\infty_{\delta_{c}} \frac{1}{\sqrt{2\pi\sigma^2(k)}}
\exp\left[ -\frac{\delta_g^2}{2\sigma^2(k)} \right]
{\rm d}\delta_g\,,
\label{beta}
\ee
where
\be
\delta_g (\bm{x},R)\equiv \int 
W(|\bm{x}-\bm{y}|,R)\delta(\bm{y})
{\rm d}^3 y\,.
\label{deltacg}
\ee
The window function $W(|\bm{x}-\bm{y}|,R)$ determines how the density contrast $\delta=(\rho-\bar{\rho})/\bar{\rho}$ around $\bm{x}$ is coarse-grained, 
where $\rho$ is 
the density and $\bar{\rho}$ is its background part. 
We choose the Gaussian window function 
of the form 
\be
W(|\bm{x}-\bm{y}|,R)=\frac{1}{(2\pi)^{3/2
}R^3}\exp\left(-\frac{|\bm{x}-\bm{y}|^2}{2R^2}\right)\,,
\label{windowfunction}
\ee
where the distance $R$ is taken to be $k^{-1}$. Then, the variance of $\delta_{g}$ is given by 
\be
\sigma^2(k)=\frac{16}{81}\int 
\exp \left[ -\left(\frac{p}{k}\right)^2 \right] \left(\frac{p}{k}\right)^4 {\cal{P}}_\zeta (p)\,{\rm d}\ln p\,.
\label{sigma2}
\ee
Since the PBH density decreases as 
$\rho_{\rm PBH} \propto a^{-3}$ after 
its formation, today's PBH density 
can be estimated as 
\be
\rho_{{\rm PBH},0}=\rho_{\rm PBH} 
(t_{\rm form}) \left[ \frac{a_0}
{a(t_{\rm form})} \right]^{-3}
=\gamma \beta 
\rho_r (t_{\rm form})
\left[ \frac{a_0}
{a(t_{\rm form})} \right]^{-3}\,.
\label{rhoPBH}
\ee
Since $\rho_r (t_{\rm form})$ is related 
to $H(t_{\rm form})$ as 
$\rho_r (t_{\rm form})
=3\Mpl^2 H(t_{\rm form})^2$, today's 
density parameter of PBHs 
corresponding to Eq.~(\ref{rhoPBH}) is
\be
\Omega_{{\rm PBH},0}=
\frac{\rho_{{\rm PBH},0}}
{3\Mpl^2 H_0^2}=
\gamma \beta \frac{H(t_{\rm form})^2}
{H_0^2} \left[ \frac{a(t_{\rm form})}
{a_0} \right]^3\,,
\ee
with total PBH density parameter 
$\int \Omega_{{\rm PBH},0}\, {\rm d}\ln M$.
Then, we obtain the ratio of the PBH 
abundance in a mass range 
$[M, M+{\rm d}\ln M]$ to the entire density of cold DM (today's density parameter 
$\Omega_{{\rm CDM},0}$) as
\be
f(M)=\frac{\Omega_{{\rm PBH},0}}{\Omega_{{\rm CDM},0}} 
\simeq \left[ \frac{\beta(M)}{1.04\times10^{-14}} \right] \left(\frac{\gamma}{0.2}\right)^{3/2}
\left[ \frac{g_{*}(t_{\text{form}})}{106.75}\right]^{-1/4}
\left(\frac{\Omega_{\text{CDM},0}\,h^2}{0.12}\right)^{-1}\left(\frac{M}{10^{-13}M_{\odot}}\right)^{-1/2}\,,
\label{massfunction2}
\ee
where $H_0=100\,h$~km~sec$^{-1}$~Mpc$^{-1}$.

\begin{figure}[h]
\begin{center}
\includegraphics[height=3.6in,width=5in]{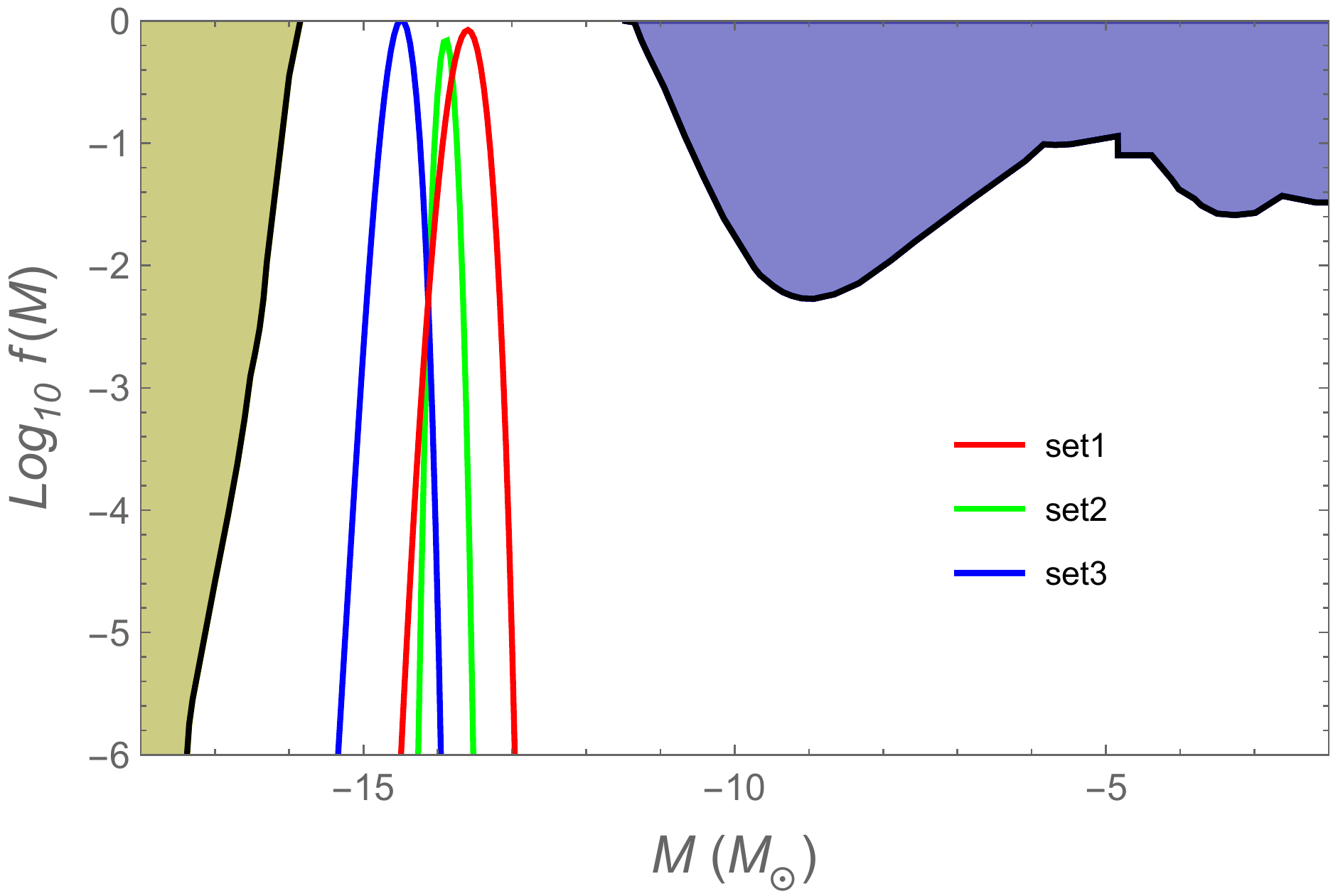}
\end{center}\vspace{-0.5cm}
\caption{
\label{figpbh}
The ratio of PBH abundances $f(M)$ relative to all the cold DM density 
as a function of the PBH mass $M$ 
(in the unit of solar mass $M_{\odot}$) for the 
three sets of model parameters presented 
in Table \ref{table1}. 
The colored areas correspond to excluded regions from 
the evaporation of black holes (dark yellow) 
and observational constraints from 
the microlensing (dark blue).
These observational constraints are taken from Refs.~\cite{bradley_j_kavanagh_2019_3538999,Green:2020jor}. 
The list of references used to derive these constraints can be 
found in  {\tt https://github.com/bradkav/PBHbounds/blob/master/bounds/README.md.}
}
\end{figure}

In the following, we use the values $\gamma=0.2$ \cite{Carr:1975qj}, $g_{*}(t_{\rm form})
=106.75$, $\Omega_{\text{CDM},0}h^2=0.12$, and $\delta_c=0.4$ 
for the computations of $M$ and $f(M)$. 
With a given wavenumber $k$ at horizon reentry, 
the PBH mass $M(k)$ is known from 
Eq.~(\ref{PBHmass}). 
The variance (\ref{sigma2}) is affected by 
the primordial spectrum 
${\cal P}_{\zeta}(k)$ 
enhanced at some particular 
scales during inflation. 
This modifies the 
PBH mass function $f(M)$ through the 
change of $\beta(M(k))$ 
in Eq.~(\ref{beta}).

In Fig.~\ref{figpbh}, we plot $f(M)$ versus $M$ 
for the three sets of model parameters presented
in Table \ref{table1}.
In these three cases, the 
mass functions $f(M)$ span in the range 
$10^{-16}M_{\odot} \lesssim M \lesssim 10^{-13}M_{\odot}$.
As we observe in Fig.~\ref{Pzfig}, 
the wavenumber $k$ corresponding to the peak positions 
of ${\cal P}_{\zeta}(k)$ is smallest 
for Set 1, while largest for Set 3. 
{}From Eq.~(\ref{PBHmass}), the PBH mass $M$ 
decreases for larger $k$.
Then, the mass $M$ corresponding to 
the peaks of $f(M)$ is smallest for 
Set 3, while largest for Set 1.
The maximum values of $f(M)$ are found to be 
$0.85$, $0.71$, and $1$ for Sets 1, 2, 3, respectively, 
so PBHs are the source for practically 
all cold DM in these three cases.

In Fig.~\ref{figpbh}, we also show the regions excluded 
by the black hole evaporation and by microlensing observations.
The models with Sets 1, 2, 3 are also consistent with 
such constraints. 
While we have considered the PBH mass range
$10^{-16}M_{\odot} \lesssim M \lesssim 10^{-13}M_{\odot}$, 
it is also possible to produce PBHs with 
the mass  
$M>10^{-13}M_{\odot}$ by choosing different sets of 
model parameters.
The heights of peaks of $f(M)$ can be also lower than the 
order 0.1 to be consistent with the microlensing data.
Thus, our model allows  versatile possibilities for generating PBHs in broad mass ranges.

Finally, we explicitly derive a relation between $M$ and the critical field value $\phi_c$. This is useful for estimating the mass of PBH and determining the model parameters.
After the transient regime the inflationary  dynamics rapidly approaches the 
slow-roll solution, so we can exploit Eq.~(\ref{Nnum}) as a good approximation.
Then, we obtain 
\be
\frac{3\xi\phi_c^2}{4\Mpl^2}\simeq N_{\rm CMB}
-\ln\frac{a(t_{\rm PBH})}{a(t_{\rm CMB})}-\Delta N_c\,,
\label{phic3}
\ee
where $t_{\rm CMB}$ and $t_{\rm PBH}$ represent the moments at which the CMB and PBH scales leave the Hubble horizon, respectively.
{}From Eq.~(\ref{PBHmass}), we can estimate 
$a(t_{\rm PBH})/a(t_{\rm CMB})$ as
\be
\frac{a(t_{\rm PBH})}{a(t_{\rm CMB})}\simeq 
\frac{k}{k_{\rm CMB}}
\simeq
\left[ \frac{10^{-13}M_{\odot}}{M(k)} \right]^{1/2} 
\left(\frac{\gamma}{0.2}\right)^{1/2}
\left[ \frac{g_{*}(t_{\text{form}})}{106.75}
\right]^{-1/12} 
\left(
\frac{4.9\times 10^{12}~\text{Mpc}^{-1}}{k_{\rm CMB}}
\right)
\simeq
\left[ \frac{6\times 10^{17}M_{\odot}}{M(k)}
\right]^{1/2}\,,
\label{apbh}
\ee
where we used the values  
$\gamma=0.2$, $g_{*}(t_{\text{form}})=106.75$, and 
$k_{\rm CMB}=0.002~\text{Mpc}^{-1}$ 
in the last equality. 
Substituting Eq.~(\ref{apbh}) into Eq.~(\ref{phic3}), 
it follows that 
\be
\phi_c
\simeq \Mpl \sqrt{\frac{4}{3\xi}\left\{ 
N_{\rm CMB}-\Delta N_{c}+\frac{1}{2}
\ln\left[ \frac{M(k)}{6\times 10^{17}M_{\odot}}
\right] \right\} }\,.
\label{phic2}
\ee
In the USR case corresponding to Set 1 model parameters, 
for example, we have $\Delta N_c=18.6$ and $M\simeq 4\times 10^{-14}M_{\odot}$, and hence 
$\phi_c \simeq 0.0383\Mpl$ from Eq.~(\ref{phic2}).
This is in good agreement with the exact value 
of $\phi_c$, i.e., $\phi_c=0.0380\Mpl$. 
For the PBH mass range $10^{-16}M_{\odot} \lesssim M \lesssim 10^{-11}M_{\odot}$ in which PBHs can be the source for all DM, 
the corresponding region of $\phi_c$ is given by
\be
\Mpl\sqrt{\frac{4}{3\xi}(21-\Delta N_c)}\lesssim \phi_c\lesssim \Mpl\sqrt{\frac{4}{3\xi}(27-\Delta N_c)}\,.
\ee
We recall that, for the plateau-type potential, $\tilde{\mu}_0$ and $\mu_1$ are related to $\phi_c$ according to 
Eqs.~(\ref{balance3}) and (\ref{mu1constraint}), respectively.
By using these relations with Eq.~(\ref{phic2}), 
the orders of parameters $\tilde{\mu}_0$ and $\mu_1$ are known 
for given values of $M(k)$ and $\Delta N_c$. 
We caution that the relations (\ref{balance3}) and (\ref{mu1constraint}) lose their accuracy for the 
bump-type potential, but they are still useful to estimate 
the orders of $\tilde{\mu}_0$ and $\mu_1$. 
To know the precise values of $\tilde{\mu}_0$ and $\mu_1$ 
with which curvature perturbations are sufficiently 
enhanced for scales relevant to PBHs, 
the numerical analysis is required as we 
performed in Sec.~\ref{seedsec}.
Basically, $\tilde{\mu}_0$ and $\mu_1$ determine the height 
of enhancement of $\zeta$ and the length of transient
period, respectively.

\section{Conclusions}
\label{consec}

We studied a mechanism for producing the seed of PBHs 
in Higgs inflation in the presence of a scalar-GB coupling. 
This provides a minimal scenario of inflation 
within a framework of standard model of particle physics, while allowing for the 
compatibility with observed CMB temperature anisotropies.
In comparison to original Higgs inflation, however, the scalar-GB coupling can modify theoretical predictions of the scalar spectral index 
and tensor-to-scalar ratio on CMB scales. 
We explored the possibility for enhancing 
curvature perturbations at some particular scales 
to generate the seed for PBHs, while the model 
is still compatible with the CMB observations.

The enhancement of curvature perturbations during inflation 
is possible when the inflaton velocity $\dot{\phi}$ 
rapidly decreases toward 0 during some transient epoch. 
If there is a period in which the scalar-GB coupling 
$\mu(\phi)$ quickly changes, it is possible to generate 
particular shapes in the inflaton 
effective potential $V_{\rm eff}(\phi)$.  
For this purpose, we considered a coupling function of the 
form $\mu(\phi)=\mu_0 \tanh[\mu_1(\phi-\phi_c)]$, 
where $\phi_c$ is the field value at transition.
Since $\mu(\phi)$ approaches constants in two 
asymptotic regimes $\phi \ll \phi_c$ and 
$\phi \gg \phi_c$, the scalar-GB coupling affects 
the dynamics of background and perturbations 
only in the vicinity of $\phi=\phi_c$.  
Depending on the model parameters, the effective potential 
can be classified into three classes: 
(1) plateau-type, (2) bump-type, (3) intermediate-type.

A typical example of the plateau-type is the Set 1 model parameters given in Table \ref{table1}, in which case the scalar-GB coupling almost balances the potential and nonminimal coupling terms around $\phi=\phi_c$.
For this type the effective potential has a nearly flat region, as seen in the left panel of Fig.~\ref{figpo}. 
During the USR regime, the field derivative decreases as $\dot{\phi} \propto a^{-3}$.
As we observe in the left 
panel of Fig.~\ref{figevo} for Set 1, the number of e-foldings $\Delta N_c$ acquired around $\phi=\phi_c$ is as large as 20. 
The Set 2 model parameters in Table \ref{table1} give rise to 
a bump-type potential illustrated in the middle panel of Fig.~\ref{figpo}.
In this case, the existence of an explicit peak in $V_{\rm eff}(\phi)$ leads 
to the approach of $\dot{\phi}$ toward 0 even faster than the USR case. 
For Set 2, the number of e-foldings $\Delta N_c$ acquired around $\phi=\phi_c$ is 
as small as a few.
The Set 3 model parameters in Table \ref{table1} correspond to the 
intermediate-type potential plotted in the right panel of Fig.~\ref{figpo}. 
For this type, both the decreasing rate of $\dot{\phi}$ and $\Delta N_c$ are between those of plateau- and bump-types.

The evolution of scalar perturbations during inflation crucially 
depends on the quantity $Z_s$ defined in Eq.~(\ref{ZS}). 
After the sound horizon crossing, the solution to the 
Fourier-transformed curvature perturbation $\zeta_k$ is expressed in 
the form (\ref{uksolution}). In all the three types of $V_{\rm eff}(\phi)$ 
mentioned above, $Z_s$ rapidly decreases in the vicinity of $\phi=\phi_c$. 
Since the last integral in Eq.~(\ref{uksolution}) becomes a rapidly 
growing mode during the transient regime, 
there is the strong enhancement of $\zeta_k$ 
for the modes exiting the sound horizon around $\phi=\phi_c$.
For the effective potential closer to the bump-type,  
the quantity $Z_s''/Z_s$, which appears in 
the equation of motion for $u_k=Z_s \zeta_k$, can exceed 
the order of $10(aH)^2$ during the rapid transition.
Then, the bump-type potential also leads to the amplification 
of some subhorizon modes. 
Still, the shapes of primordial scalar power spectra 
${\cal P}_{\zeta}(k)$ mostly depend on the number of 
e-foldings $\Delta N_c$, in such a way that the peak 
tends to be sharper for smaller $\Delta N_c$. 

The primordial power spectra ${\cal P}_{\zeta}(k)$ plotted in Fig.~\ref{Pzfig} 
correspond to those of Set 1, 2, 3 model parameters. 
In all these cases, the peak values of ${\cal P}_{\zeta}(k)$ are about $10^7$ 
times as large as the amplitude of ${\cal P}_{\zeta}(k)$ on CMB scales. 
With these model parameters, all the stability conditions given in Eq.~(\ref{avoidinstability}) 
are consistently satisfied and hence there are neither ghost nor Laplacian instabilities. 
We note that the 
inflaton-GB coupling gives rise to the scalar propagation speed 
different from 1, whose property also affects the shapes of ${\cal P}_{\zeta}(k)$.

The existence of the transient epoch around $\phi=\phi_c$ modifies the scalar power 
spectrum and tensor-to-scalar ratio on CMB scales. 
In standard Higgs inflation, the expressions of ${\cal P}_{\zeta}$, $n_s$, and 
$r$ are given, respectively, by Eqs.~(\ref{power1}), (\ref{ns}), and (\ref{r}), where 
$N$ is the number of e-foldings counted backward from the end of inflation. 
In the presence of the scalar-GB coupling, these observables are subject to 
the modification $N \to N_{\rm CMB}-\Delta N_c$, where $N_{\rm CMB}$ 
corresponds to the number of e-foldings on CMB scales. 
This means that, for smaller $\Delta N_c$, the models can exhibit better 
compatibility with the observational bounds on $n_s$ and $r$.
As we see in Fig.~\ref{fignsr}, the bump-type with $\Delta N_c=2.3$ (Set 2) is 
well inside the $1\sigma$ observational contour constrained from CMB 
and other data, while the 
plateau-type with 
$\Delta N_c=18.6$ (Set 1) is outside the $2\sigma$ contour. 
The intermediate-type model with $\Delta N_c=9.7$ (Set 3) is between 
$1\sigma$ and $2\sigma$ contours. These results 
show that the bump-type is 
generally favored over the plateau-type from the CMB constraints.

In Fig.~\ref{figpbh}, we plot the PBH fraction function $f(M)$ versus 
the mass $M$ for three sets of model parameters 
in Table~\ref{table1}.
In all these cases we have $f(M) \simeq 1$ in the mass range 
$10^{-16}M_{\odot} \lesssim M \lesssim 10^{-13}M_{\odot}$. 
Thus, the GB corrected Higgs inflation allows the possibility for 
generating the amount of PBHs serving as almost all DM, 
while being compatible with the observed temperature anisotropies 
on CMB scales especially for the bump-type effective potential.
We note that, in our model, it is also possible to generate PBHs 
in the larger mass range $M>10^{-13}M_{\odot}$ 
with the fraction $f(M) \lesssim 0.1$.

The enhanced primordial curvature perturbations on particular scales may induce gravitational waves at nonlinear order, which affects the spectrum of gravitational wave background \cite{Kohri:2018awv} 
(see also Ref.~\cite{Papanikolaou:2020qtd}). 
In addition, while a Gaussian distribution was used for
the coarse-grained density fluctuation in this paper, 
it was reported that the distribution function of curvature fluctuations may have a peculiar shape deviating from the Gaussian \cite{Namjoo:2012aa,Franciolini:2018vbk,Cai:2018dkf,Atal:2018neu,Atal:2019cdz,Atal:2019erb,Passaglia:2018ixg,Taoso:2021uvl,Biagetti:2021eep,Davies:2021loj,Cai:2021zsp,Cai:2022erk}.
Recently, it was shown that the PBH scenario in single-field inflation with an USR regime can induce large one-loop corrections to the scalar power spectrum on CMB scales \cite{Kristiano:2022maq,
Inomata:2022yte}. Since the analysis is limited to a canonical 
scalar field with a plateau-type potential, 
the results of Ref.~\cite{Kristiano:2022maq} are not  
applied to our model in which the enhancement of 
curvature perturbations occurs by the presence of 
the inflaton-GB coupling. However, it is worth 
computing one-loop corrections to ${\cal P}_{\zeta}$ 
in our model, especially for the bump-type 
effective potential.
We leave these interesting issues for future works.

\section*{Acknowledgments} 
We thank Tomohiro Fujita for useful discussions and comments. 
We also thank Bradley Kavanagh for giving us permission to use 
observational constraints 
given in Fig.~\ref{figpbh}.
ST is supported by the Grant-in-Aid for Scientific Research 
Fund of the JSPS Nos.~19K03854 and 22K03642.


\appendix

\section{Appendix: Correspondence with Horndeski theory}
\label{Horndeskitheory}

The inflationary model studied in this paper belongs to a subclass of Horndeski theory given by the action \cite{Horndeski,Def11,KYY,Charmousis:2011bf}
\be
{\cal S}=\int {\rm d}^4 x \sqrt{-g}\,{\cal L}_H\,,
\label{action}
\ee
where 
\ba
\hspace{-0.5cm}
{\cal L}_H
&=&
G_2(\phi,X)-G_{3}(\phi,X)\square\phi 
+G_{4}(\phi,X)\, R +G_{4,X}(\phi,X)\left[ (\square \phi)^{2}
-(\nabla_{\mu}\nabla_{\nu} \phi)
(\nabla^{\mu}\nabla^{\nu} \phi) \right]
+G_{5}(\phi,X)G_{\mu \nu} \nabla^{\mu}\nabla^{\nu} \phi
\notag\\
&&
-\frac{1}{6}G_{5,X}(\phi,X)
\left[ (\square \phi )^{3}-3(\square \phi)\,
(\nabla_{\mu}\nabla_{\nu} \phi)
(\nabla^{\mu}\nabla^{\nu} \phi)
+2(\nabla^{\mu}\nabla_{\alpha} \phi)
(\nabla^{\alpha}\nabla_{\beta} \phi)
(\nabla^{\beta}\nabla_{\mu} \phi) \right]\,,
\label{LH}
\ea
with $G_{\mu \nu}$ being the Einstein tensor.
Four functions $G_{j}$'s ($j=2,3,4,5$) depend on the scalar field $\phi$ and 
its kinetic term 
$X=-g^{\mu\nu}\nabla_{\mu}
\phi\nabla_{\nu}\phi/2$, 
where we use the notation 
$G_{j,X} \equiv {\rm d} G_j/{\rm d} X$.
The action (\ref{action1}) can be accommodated by the following coupling 
functions \cite{KYY}
\ba
& &
G_{2}(\phi,X)=X-\frac{1}{4}\lambda\phi^{4}
+8\mu^{(4)}(\phi) X^{2}
(3-\ln |X|)\,,
\label{G2}\\
& &
G_{3}(\phi,X)=4\mu^{(3)}(\phi) 
X(7-3\ln |X|)\,,
\label{G3}\\
& &
G_{4}(\phi,X)=\frac{\Mpl^{2}}{2}+\frac{1}{2}\xi\phi^{2}+4\mu^{(2)}(\phi) 
X(2-\ln |X|)\,,
\label{G4}\\
& &
G_{5}(\phi,X)=-4\mu^{(1)}(\phi) 
\ln |X|\,,
\label{G5}
\ea
where $\mu^{(i)}(\phi) \equiv
{{\rm d}^{i}\mu/{\rm d}\phi^{i}}$. 
In full Horndeski theories, the background and perturbation equations of motion 
on the flat FLRW background were already derived in Refs.~\cite{KYY,DeFelice:2011uc,DeFelice:2011hq}.
In this paper, we applied those results 
to the theory given by the coupling 
functions (\ref{G2})-(\ref{G5}).

\bibliographystyle{mybibstyle}
\bibliography{bib}

\begin{thebibliography}{145}%
\makeatletter
\providecommand \@ifxundefined [1]{%
 \@ifx{#1\undefined}
}%
\providecommand \@ifnum [1]{%
 \ifnum #1\expandafter \@firstoftwo
 \else \expandafter \@secondoftwo
 \fi
}%
\providecommand \@ifx [1]{%
 \ifx #1\expandafter \@firstoftwo
 \else \expandafter \@secondoftwo
 \fi
}%
\providecommand \natexlab [1]{#1}%
\providecommand \enquote  [1]{``#1''}%
\providecommand \bibnamefont  [1]{#1}%
\providecommand \bibfnamefont [1]{#1}%
\providecommand \citenamefont [1]{#1}%
\providecommand \href@noop [0]{\@secondoftwo}%
\providecommand \href [0]{\begingroup \@sanitize@url \@href}%
\providecommand \@href[1]{\@@startlink{#1}\@@href}%
\providecommand \@@href[1]{\endgroup#1\@@endlink}%
\providecommand \@sanitize@url [0]{\catcode `\\12\catcode `\$12\catcode
  `\&12\catcode `\#12\catcode `\^12\catcode `\_12\catcode `\%12\relax}%
\providecommand \@@startlink[1]{}%
\providecommand \@@endlink[0]{}%
\providecommand \url  [0]{\begingroup\@sanitize@url \@url }%
\providecommand \@url [1]{\endgroup\@href {#1}{\urlprefix }}%
\providecommand \urlprefix  [0]{URL }%
\providecommand \Eprint [0]{\href }%
\providecommand \doibase [0]{http://dx.doi.org/}%
\providecommand \selectlanguage [0]{\@gobble}%
\providecommand \bibinfo  [0]{\@secondoftwo}%
\providecommand \bibfield  [0]{\@secondoftwo}%
\providecommand \translation [1]{[#1]}%
\providecommand \BibitemOpen [0]{}%
\providecommand \bibitemStop [0]{}%
\providecommand \bibitemNoStop [0]{.\EOS\space}%
\providecommand \EOS [0]{\spacefactor3000\relax}%
\providecommand \BibitemShut  [1]{\csname bibitem#1\endcsname}%
\let\auto@bib@innerbib\@empty
\bibitem [{\citenamefont {Zel'dovich}\ and\ \citenamefont
  {Novikov}(1967)}]{zel1967hypothesis}%
  \BibitemOpen
  \bibfield  {author} {\bibinfo {author} {\bibfnamefont {Y.~B.}\ \bibnamefont
  {Zel'dovich}} and \bibinfo {author} {\bibfnamefont {I.~D.}\ \bibnamefont
  {Novikov}},\ }\href@noop {} {\bibfield  {journal} {\bibinfo  {journal} {\emph
  {Soviet Astronomy}}\ }\textbf {\bibinfo {volume} {10}},\ \bibinfo {pages}
  {602} (\bibinfo {year} {1967})}\BibitemShut {NoStop}%
\bibitem [{\citenamefont {Hawking}(1971)}]{Hawking:1971ei}%
  \BibitemOpen
  \bibfield  {author} {\bibinfo {author} {\bibfnamefont {S.}~\bibnamefont
  {Hawking}},\ }\href@noop {} {\bibfield  {journal} {\bibinfo  {journal} {\emph
  {Mon. Not. Roy. Astron. Soc.}}\ }\textbf {\bibinfo {volume} {152}},\ \bibinfo
  {pages} {75} (\bibinfo {year} {1971})}\BibitemShut {NoStop}%
\bibitem [{\citenamefont {Carr}\ and\ \citenamefont
  {Hawking}(1974)}]{Carr:1974nx}%
  \BibitemOpen
  \bibfield  {author} {\bibinfo {author} {\bibfnamefont {B.~J.}\ \bibnamefont
  {Carr}} and \bibinfo {author} {\bibfnamefont {S.~W.}\ \bibnamefont
  {Hawking}},\ }\href {\doibase 10.1093/mnras/168.2.399} {\bibfield  {journal}
  {\bibinfo  {journal} {\emph {Mon. Not. Roy. Astron. Soc.}}\ }\textbf
  {\bibinfo {volume} {168}},\ \bibinfo {pages} {399} (\bibinfo {year}
  {1974})}\BibitemShut {NoStop}%
\bibitem [{\citenamefont {Chapline}(1975)}]{Chapline:1975ojl}%
  \BibitemOpen
  \bibfield  {author} {\bibinfo {author} {\bibfnamefont {G.~F.}\ \bibnamefont
  {Chapline}},\ }\href {\doibase 10.1038/253251a0} {\bibfield  {journal}
  {\bibinfo  {journal} {\emph {Nature}}\ }\textbf {\bibinfo {volume} {253}},\
  \bibinfo {pages} {251} (\bibinfo {year} {1975})}\BibitemShut {NoStop}%
\bibitem [{\citenamefont {Meszaros}(1975)}]{Meszaros:1975ef}%
  \BibitemOpen
  \bibfield  {author} {\bibinfo {author} {\bibfnamefont {P.}~\bibnamefont
  {Meszaros}},\ }\href@noop {} {\bibfield  {journal} {\bibinfo  {journal}
  {\emph {Astron. Astrophys.}}\ }\textbf {\bibinfo {volume} {38}},\ \bibinfo
  {pages} {5} (\bibinfo {year} {1975})}\BibitemShut {NoStop}%
\bibitem [{\citenamefont {Khlopov}(2010)}]{Khlopov:2008qy}%
  \BibitemOpen
  \bibfield  {author} {\bibinfo {author} {\bibfnamefont {M.~Y.}\ \bibnamefont
  {Khlopov}},\ }\href {\doibase 10.1088/1674-4527/10/6/001} {\bibfield
  {journal} {\bibinfo  {journal} {\emph {Res. Astron. Astrophys.}}\ }\textbf
  {\bibinfo {volume} {10}},\ \bibinfo {pages} {495} (\bibinfo {year} {2010})},\
  \Eprint {http://arxiv.org/abs/0801.0116} {arXiv:0801.0116 [astro-ph]}
  \BibitemShut {NoStop}%
\bibitem [{\citenamefont {Sasaki}\ \emph {et~al.}(2018)\citenamefont {Sasaki},
  \citenamefont {Suyama}, \citenamefont {Tanaka},\ and\ \citenamefont
  {Yokoyama}}]{Sasaki:2018dmp}%
  \BibitemOpen
  \bibfield  {author} {\bibinfo {author} {\bibfnamefont {M.}~\bibnamefont
  {Sasaki}}, \bibinfo {author} {\bibfnamefont {T.}~\bibnamefont {Suyama}},
  \bibinfo {author} {\bibfnamefont {T.}~\bibnamefont {Tanaka}},  and \bibinfo
  {author} {\bibfnamefont {S.}~\bibnamefont {Yokoyama}},\ }\href {\doibase
  10.1088/1361-6382/aaa7b4} {\bibfield  {journal} {\bibinfo  {journal} {\emph
  {Class. Quant. Grav.}}\ }\textbf {\bibinfo {volume} {35}},\ \bibinfo {pages}
  {063001} (\bibinfo {year} {2018})},\ \Eprint
  {http://arxiv.org/abs/1801.05235} {arXiv:1801.05235 [astro-ph.CO]}
  \BibitemShut {NoStop}%
\bibitem [{\citenamefont {Carr}\ and\ \citenamefont
  {Kuhnel}(2020)}]{Carr:2020xqk}%
  \BibitemOpen
  \bibfield  {author} {\bibinfo {author} {\bibfnamefont {B.}~\bibnamefont
  {Carr}} and \bibinfo {author} {\bibfnamefont {F.}~\bibnamefont {Kuhnel}},\
  }\href {\doibase 10.1146/annurev-nucl-050520-125911} {\bibfield  {journal}
  {\bibinfo  {journal} {\emph {Ann. Rev. Nucl. Part. Sci.}}\ }\textbf {\bibinfo
  {volume} {70}},\ \bibinfo {pages} {355} (\bibinfo {year} {2020})},\ \Eprint
  {http://arxiv.org/abs/2006.02838} {arXiv:2006.02838 [astro-ph.CO]}
  \BibitemShut {NoStop}%
\bibitem [{\citenamefont {Green}\ and\ \citenamefont
  {Kavanagh}(2021)}]{Green:2020jor}%
  \BibitemOpen
  \bibfield  {author} {\bibinfo {author} {\bibfnamefont {A.~M.}\ \bibnamefont
  {Green}} and \bibinfo {author} {\bibfnamefont {B.~J.}\ \bibnamefont
  {Kavanagh}},\ }\href {\doibase 10.1088/1361-6471/abc534} {\bibfield
  {journal} {\bibinfo  {journal} {\emph {J. Phys. G}}\ }\textbf {\bibinfo
  {volume} {48}},\ \bibinfo {pages} {043001} (\bibinfo {year} {2021})},\
  \Eprint {http://arxiv.org/abs/2007.10722} {arXiv:2007.10722 [astro-ph.CO]}
  \BibitemShut {NoStop}%
\bibitem [{\citenamefont {Villanueva-Domingo}\ \emph
  {et~al.}(2021)\citenamefont {Villanueva-Domingo}, \citenamefont {Mena},\ and\
  \citenamefont {Palomares-Ruiz}}]{Villanueva-Domingo:2021spv}%
  \BibitemOpen
  \bibfield  {author} {\bibinfo {author} {\bibfnamefont {P.}~\bibnamefont
  {Villanueva-Domingo}}, \bibinfo {author} {\bibfnamefont {O.}~\bibnamefont
  {Mena}},  and \bibinfo {author} {\bibfnamefont {S.}~\bibnamefont
  {Palomares-Ruiz}},\ }\href {\doibase 10.3389/fspas.2021.681084} {\bibfield
  {journal} {\bibinfo  {journal} {\emph {Front. Astron. Space Sci.}}\ }\textbf
  {\bibinfo {volume} {8}},\ \bibinfo {pages} {87} (\bibinfo {year} {2021})},\
  \Eprint {http://arxiv.org/abs/2103.12087} {arXiv:2103.12087 [astro-ph.CO]}
  \BibitemShut {NoStop}%
\bibitem [{\citenamefont {Carr}\ and\ \citenamefont
  {Kuhnel}(2022)}]{Carr:2021bzv}%
  \BibitemOpen
  \bibfield  {author} {\bibinfo {author} {\bibfnamefont {B.}~\bibnamefont
  {Carr}} and \bibinfo {author} {\bibfnamefont {F.}~\bibnamefont {Kuhnel}},\
  }\href {\doibase 10.21468/SciPostPhysLectNotes.48} {\bibfield  {journal}
  {\bibinfo  {journal} {\emph {SciPost Phys. Lect. Notes}}\ }\textbf {\bibinfo
  {volume} {48}},\ \bibinfo {pages} {1} (\bibinfo {year} {2022})},\ \Eprint
  {http://arxiv.org/abs/2110.02821} {arXiv:2110.02821 [astro-ph.CO]}
  \BibitemShut {NoStop}%
\bibitem [{\citenamefont {Escriv\`a}\ \emph {et~al.}(2022)\citenamefont
  {Escriv\`a}, \citenamefont {Kuhnel},\ and\ \citenamefont
  {Tada}}]{Escriva:2022duf}%
  \BibitemOpen
  \bibfield  {author} {\bibinfo {author} {\bibfnamefont {A.}~\bibnamefont
  {Escriv\`a}}, \bibinfo {author} {\bibfnamefont {F.}~\bibnamefont {Kuhnel}},
  and \bibinfo {author} {\bibfnamefont {Y.}~\bibnamefont {Tada}},\ }\Eprint
  {http://arxiv.org/abs/2211.05767} {arXiv:2211.05767 [astro-ph.CO]}
  \BibitemShut {NoStop}%
\bibitem [{\citenamefont {Karam}\ \emph {et~al.}(2022)\citenamefont {Karam},
  \citenamefont {Koivunen}, \citenamefont {Tomberg}, \citenamefont {Vaskonen},\
  and\ \citenamefont {Veerm\"ae}}]{Karam:2022nym}%
  \BibitemOpen
  \bibfield  {author} {\bibinfo {author} {\bibfnamefont {A.}~\bibnamefont
  {Karam}}, \bibinfo {author} {\bibfnamefont {N.}~\bibnamefont {Koivunen}},
  \bibinfo {author} {\bibfnamefont {E.}~\bibnamefont {Tomberg}}, \bibinfo
  {author} {\bibfnamefont {V.}~\bibnamefont {Vaskonen}},  and \bibinfo {author}
  {\bibfnamefont {H.}~\bibnamefont {Veerm\"ae}},\ }\Eprint
  {http://arxiv.org/abs/2205.13540} {arXiv:2205.13540 [astro-ph.CO]}
  \BibitemShut {NoStop}%
\bibitem [{\citenamefont {Abbott}\ \emph {et~al.}(2016)\citenamefont {Abbott}
  \emph {et~al.}}]{LIGOScientific:2016aoc}%
  \BibitemOpen
  \bibfield  {author} {\bibinfo {author} {\bibfnamefont {B.~P.}\ \bibnamefont
  {Abbott}} \emph {et~al.} (\bibinfo {collaboration} {LIGO Scientific,
  Virgo}),\ }\href {\doibase 10.1103/PhysRevLett.116.061102} {\bibfield
  {journal} {\bibinfo  {journal} {\emph {Phys. Rev. Lett.}}\ }\textbf {\bibinfo
  {volume} {116}},\ \bibinfo {pages} {061102} (\bibinfo {year} {2016})},\
  \Eprint {http://arxiv.org/abs/1602.03837} {arXiv:1602.03837 [gr-qc]}
  \BibitemShut {NoStop}%
\bibitem [{\citenamefont {Abbott}\ \emph {et~al.}(2019)\citenamefont {Abbott}
  \emph {et~al.}}]{LIGOScientific:2018mvr}%
  \BibitemOpen
  \bibfield  {author} {\bibinfo {author} {\bibfnamefont {B.~P.}\ \bibnamefont
  {Abbott}} \emph {et~al.} (\bibinfo {collaboration} {LIGO Scientific,
  Virgo}),\ }\href {\doibase 10.1103/PhysRevX.9.031040} {\bibfield  {journal}
  {\bibinfo  {journal} {\emph {Phys. Rev. X}}\ }\textbf {\bibinfo {volume}
  {9}},\ \bibinfo {pages} {031040} (\bibinfo {year} {2019})},\ \Eprint
  {http://arxiv.org/abs/1811.12907} {arXiv:1811.12907 [astro-ph.HE]}
  \BibitemShut {NoStop}%
\bibitem [{\citenamefont {Abbott}\ \emph
  {et~al.}(2021{\natexlab{a}})\citenamefont {Abbott} \emph
  {et~al.}}]{LIGOScientific:2020ibl}%
  \BibitemOpen
  \bibfield  {author} {\bibinfo {author} {\bibfnamefont {R.}~\bibnamefont
  {Abbott}} \emph {et~al.} (\bibinfo {collaboration} {LIGO Scientific,
  Virgo}),\ }\href {\doibase 10.1103/PhysRevX.11.021053} {\bibfield  {journal}
  {\bibinfo  {journal} {\emph {Phys. Rev. X}}\ }\textbf {\bibinfo {volume}
  {11}},\ \bibinfo {pages} {021053} (\bibinfo {year} {2021}{\natexlab{a}})},\
  \Eprint {http://arxiv.org/abs/2010.14527} {arXiv:2010.14527 [gr-qc]}
  \BibitemShut {NoStop}%
\bibitem [{\citenamefont {Abbott}\ \emph
  {et~al.}(2021{\natexlab{b}})\citenamefont {Abbott} \emph
  {et~al.}}]{LIGOScientific:2021djp}%
  \BibitemOpen
  \bibfield  {author} {\bibinfo {author} {\bibfnamefont {R.}~\bibnamefont
  {Abbott}} \emph {et~al.} (\bibinfo {collaboration} {LIGO Scientific, VIRGO,
  KAGRA}),\ }\Eprint {http://arxiv.org/abs/2111.03606} {arXiv:2111.03606
  [gr-qc]} \BibitemShut {NoStop}%
\bibitem [{\citenamefont {Bird}\ \emph {et~al.}(2016)\citenamefont {Bird},
  \citenamefont {Cholis}, \citenamefont {Mu\~noz}, \citenamefont
  {Ali-Ha\"\i{}moud}, \citenamefont {Kamionkowski}, \citenamefont {Kovetz},
  \citenamefont {Raccanelli},\ and\ \citenamefont {Riess}}]{Bird:2016dcv}%
  \BibitemOpen
  \bibfield  {author} {\bibinfo {author} {\bibfnamefont {S.}~\bibnamefont
  {Bird}}, \bibinfo {author} {\bibfnamefont {I.}~\bibnamefont {Cholis}},
  \bibinfo {author} {\bibfnamefont {J.~B.}\ \bibnamefont {Mu\~noz}}, \bibinfo
  {author} {\bibfnamefont {Y.}~\bibnamefont {Ali-Ha\"\i{}moud}}, \bibinfo
  {author} {\bibfnamefont {M.}~\bibnamefont {Kamionkowski}}, \bibinfo {author}
  {\bibfnamefont {E.~D.}\ \bibnamefont {Kovetz}}, \bibinfo {author}
  {\bibfnamefont {A.}~\bibnamefont {Raccanelli}},  and \bibinfo {author}
  {\bibfnamefont {A.~G.}\ \bibnamefont {Riess}},\ }\href {\doibase
  10.1103/PhysRevLett.116.201301} {\bibfield  {journal} {\bibinfo  {journal}
  {\emph {Phys. Rev. Lett.}}\ }\textbf {\bibinfo {volume} {116}},\ \bibinfo
  {pages} {201301} (\bibinfo {year} {2016})},\ \Eprint
  {http://arxiv.org/abs/1603.00464} {arXiv:1603.00464 [astro-ph.CO]}
  \BibitemShut {NoStop}%
\bibitem [{\citenamefont {Sasaki}\ \emph {et~al.}(2016)\citenamefont {Sasaki},
  \citenamefont {Suyama}, \citenamefont {Tanaka},\ and\ \citenamefont
  {Yokoyama}}]{Sasaki:2016jop}%
  \BibitemOpen
  \bibfield  {author} {\bibinfo {author} {\bibfnamefont {M.}~\bibnamefont
  {Sasaki}}, \bibinfo {author} {\bibfnamefont {T.}~\bibnamefont {Suyama}},
  \bibinfo {author} {\bibfnamefont {T.}~\bibnamefont {Tanaka}},  and \bibinfo
  {author} {\bibfnamefont {S.}~\bibnamefont {Yokoyama}},\ }\href {\doibase
  10.1103/PhysRevLett.117.061101} {\bibfield  {journal} {\bibinfo  {journal}
  {\emph {Phys. Rev. Lett.}}\ }\textbf {\bibinfo {volume} {117}},\ \bibinfo
  {pages} {061101} (\bibinfo {year} {2016})},\ \bibinfo {note} {[Erratum:
  Phys.Rev.Lett. 121, 059901 (2018)]},\ \Eprint
  {http://arxiv.org/abs/1603.08338} {arXiv:1603.08338 [astro-ph.CO]}
  \BibitemShut {NoStop}%
\bibitem [{\citenamefont {Clesse}\ and\ \citenamefont
  {Garc\'\i{}a-Bellido}(2017)}]{Clesse:2016vqa}%
  \BibitemOpen
  \bibfield  {author} {\bibinfo {author} {\bibfnamefont {S.}~\bibnamefont
  {Clesse}} and \bibinfo {author} {\bibfnamefont {J.}~\bibnamefont
  {Garc\'\i{}a-Bellido}},\ }\href {\doibase 10.1016/j.dark.2016.10.002}
  {\bibfield  {journal} {\bibinfo  {journal} {\emph {Phys. Dark Univ.}}\
  }\textbf {\bibinfo {volume} {15}},\ \bibinfo {pages} {142} (\bibinfo {year}
  {2017})},\ \Eprint {http://arxiv.org/abs/1603.05234} {arXiv:1603.05234
  [astro-ph.CO]} \BibitemShut {NoStop}%
\bibitem [{\citenamefont {Wang}\ \emph {et~al.}(2018)\citenamefont {Wang},
  \citenamefont {Wang}, \citenamefont {Huang},\ and\ \citenamefont
  {Li}}]{Wang:2016ana}%
  \BibitemOpen
  \bibfield  {author} {\bibinfo {author} {\bibfnamefont {S.}~\bibnamefont
  {Wang}}, \bibinfo {author} {\bibfnamefont {Y.-F.}\ \bibnamefont {Wang}},
  \bibinfo {author} {\bibfnamefont {Q.-G.}\ \bibnamefont {Huang}},  and
  \bibinfo {author} {\bibfnamefont {T.~G.~F.}\ \bibnamefont {Li}},\ }\href
  {\doibase 10.1103/PhysRevLett.120.191102} {\bibfield  {journal} {\bibinfo
  {journal} {\emph {Phys. Rev. Lett.}}\ }\textbf {\bibinfo {volume} {120}},\
  \bibinfo {pages} {191102} (\bibinfo {year} {2018})},\ \Eprint
  {http://arxiv.org/abs/1610.08725} {arXiv:1610.08725 [astro-ph.CO]}
  \BibitemShut {NoStop}%
\bibitem [{\citenamefont {Bean}\ and\ \citenamefont
  {Magueijo}(2002)}]{Bean:2002kx}%
  \BibitemOpen
  \bibfield  {author} {\bibinfo {author} {\bibfnamefont {R.}~\bibnamefont
  {Bean}} and \bibinfo {author} {\bibfnamefont {J.}~\bibnamefont {Magueijo}},\
  }\href {\doibase 10.1103/PhysRevD.66.063505} {\bibfield  {journal} {\bibinfo
  {journal} {\emph {Phys. Rev. D}}\ }\textbf {\bibinfo {volume} {66}},\
  \bibinfo {pages} {063505} (\bibinfo {year} {2002})},\ \Eprint
  {http://arxiv.org/abs/astro-ph/0204486} {arXiv:astro-ph/0204486} \BibitemShut
  {NoStop}%
\bibitem [{\citenamefont {Carr}\ \emph {et~al.}(2021)\citenamefont {Carr},
  \citenamefont {Kohri}, \citenamefont {Sendouda},\ and\ \citenamefont
  {Yokoyama}}]{Carr:2020gox}%
  \BibitemOpen
  \bibfield  {author} {\bibinfo {author} {\bibfnamefont {B.}~\bibnamefont
  {Carr}}, \bibinfo {author} {\bibfnamefont {K.}~\bibnamefont {Kohri}},
  \bibinfo {author} {\bibfnamefont {Y.}~\bibnamefont {Sendouda}},  and \bibinfo
  {author} {\bibfnamefont {J.}~\bibnamefont {Yokoyama}},\ }\href {\doibase
  10.1088/1361-6633/ac1e31} {\bibfield  {journal} {\bibinfo  {journal} {\emph
  {Rept. Prog. Phys.}}\ }\textbf {\bibinfo {volume} {84}},\ \bibinfo {pages}
  {116902} (\bibinfo {year} {2021})},\ \Eprint
  {http://arxiv.org/abs/2002.12778} {arXiv:2002.12778 [astro-ph.CO]}
  \BibitemShut {NoStop}%
\bibitem [{\citenamefont {Ivanov}\ \emph {et~al.}(1994)\citenamefont {Ivanov},
  \citenamefont {Naselsky},\ and\ \citenamefont {Novikov}}]{Ivanov:1994pa}%
  \BibitemOpen
  \bibfield  {author} {\bibinfo {author} {\bibfnamefont {P.}~\bibnamefont
  {Ivanov}}, \bibinfo {author} {\bibfnamefont {P.}~\bibnamefont {Naselsky}},
  and \bibinfo {author} {\bibfnamefont {I.}~\bibnamefont {Novikov}},\ }\href
  {\doibase 10.1103/PhysRevD.50.7173} {\bibfield  {journal} {\bibinfo
  {journal} {\emph {Phys. Rev. D}}\ }\textbf {\bibinfo {volume} {50}},\
  \bibinfo {pages} {7173} (\bibinfo {year} {1994})}\BibitemShut {NoStop}%
\bibitem [{\citenamefont {Garcia-Bellido}\ \emph {et~al.}(1996)\citenamefont
  {Garcia-Bellido}, \citenamefont {Linde},\ and\ \citenamefont
  {Wands}}]{Garcia-Bellido:1996mdl}%
  \BibitemOpen
  \bibfield  {author} {\bibinfo {author} {\bibfnamefont {J.}~\bibnamefont
  {Garcia-Bellido}}, \bibinfo {author} {\bibfnamefont {A.~D.}\ \bibnamefont
  {Linde}},  and \bibinfo {author} {\bibfnamefont {D.}~\bibnamefont {Wands}},\
  }\href {\doibase 10.1103/PhysRevD.54.6040} {\bibfield  {journal} {\bibinfo
  {journal} {\emph {Phys. Rev. D}}\ }\textbf {\bibinfo {volume} {54}},\
  \bibinfo {pages} {6040} (\bibinfo {year} {1996})},\ \Eprint
  {http://arxiv.org/abs/astro-ph/9605094} {arXiv:astro-ph/9605094} \BibitemShut
  {NoStop}%
\bibitem [{\citenamefont {Bullock}\ and\ \citenamefont
  {Primack}(1997)}]{Bullock:1996at}%
  \BibitemOpen
  \bibfield  {author} {\bibinfo {author} {\bibfnamefont {J.~S.}\ \bibnamefont
  {Bullock}} and \bibinfo {author} {\bibfnamefont {J.~R.}\ \bibnamefont
  {Primack}},\ }\href {\doibase 10.1103/PhysRevD.55.7423} {\bibfield  {journal}
  {\bibinfo  {journal} {\emph {Phys. Rev. D}}\ }\textbf {\bibinfo {volume}
  {55}},\ \bibinfo {pages} {7423} (\bibinfo {year} {1997})},\ \Eprint
  {http://arxiv.org/abs/astro-ph/9611106} {arXiv:astro-ph/9611106} \BibitemShut
  {NoStop}%
\bibitem [{\citenamefont {Yokoyama}(1997)}]{Yokoyama:1995ex}%
  \BibitemOpen
  \bibfield  {author} {\bibinfo {author} {\bibfnamefont {J.}~\bibnamefont
  {Yokoyama}},\ }\href@noop {} {\bibfield  {journal} {\bibinfo  {journal}
  {\emph {Astron. Astrophys.}}\ }\textbf {\bibinfo {volume} {318}},\ \bibinfo
  {pages} {673} (\bibinfo {year} {1997})},\ \Eprint
  {http://arxiv.org/abs/astro-ph/9509027} {arXiv:astro-ph/9509027} \BibitemShut
  {NoStop}%
\bibitem [{\citenamefont {Yokoyama}(1998)}]{Yokoyama:1998pt}%
  \BibitemOpen
  \bibfield  {author} {\bibinfo {author} {\bibfnamefont {J.}~\bibnamefont
  {Yokoyama}},\ }\href {\doibase 10.1103/PhysRevD.58.083510} {\bibfield
  {journal} {\bibinfo  {journal} {\emph {Phys. Rev. D}}\ }\textbf {\bibinfo
  {volume} {58}},\ \bibinfo {pages} {083510} (\bibinfo {year} {1998})},\
  \Eprint {http://arxiv.org/abs/astro-ph/9802357} {arXiv:astro-ph/9802357}
  \BibitemShut {NoStop}%
\bibitem [{\citenamefont {Kawasaki}\ \emph {et~al.}(1998)\citenamefont
  {Kawasaki}, \citenamefont {Sugiyama},\ and\ \citenamefont
  {Yanagida}}]{Kawasaki:1997ju}%
  \BibitemOpen
  \bibfield  {author} {\bibinfo {author} {\bibfnamefont {M.}~\bibnamefont
  {Kawasaki}}, \bibinfo {author} {\bibfnamefont {N.}~\bibnamefont {Sugiyama}},
  and \bibinfo {author} {\bibfnamefont {T.}~\bibnamefont {Yanagida}},\ }\href
  {\doibase 10.1103/PhysRevD.57.6050} {\bibfield  {journal} {\bibinfo
  {journal} {\emph {Phys. Rev. D}}\ }\textbf {\bibinfo {volume} {57}},\
  \bibinfo {pages} {6050} (\bibinfo {year} {1998})},\ \Eprint
  {http://arxiv.org/abs/hep-ph/9710259} {arXiv:hep-ph/9710259} \BibitemShut
  {NoStop}%
\bibitem [{\citenamefont {Kawasaki}\ \emph {et~al.}(2006)\citenamefont
  {Kawasaki}, \citenamefont {Takayama}, \citenamefont {Yamaguchi},\ and\
  \citenamefont {Yokoyama}}]{Kawasaki:2006zv}%
  \BibitemOpen
  \bibfield  {author} {\bibinfo {author} {\bibfnamefont {M.}~\bibnamefont
  {Kawasaki}}, \bibinfo {author} {\bibfnamefont {T.}~\bibnamefont {Takayama}},
  \bibinfo {author} {\bibfnamefont {M.}~\bibnamefont {Yamaguchi}},  and
  \bibinfo {author} {\bibfnamefont {J.}~\bibnamefont {Yokoyama}},\ }\href
  {\doibase 10.1103/PhysRevD.74.043525} {\bibfield  {journal} {\bibinfo
  {journal} {\emph {Phys. Rev. D}}\ }\textbf {\bibinfo {volume} {74}},\
  \bibinfo {pages} {043525} (\bibinfo {year} {2006})},\ \Eprint
  {http://arxiv.org/abs/hep-ph/0605271} {arXiv:hep-ph/0605271} \BibitemShut
  {NoStop}%
\bibitem [{\citenamefont {Kohri}\ \emph {et~al.}(2008)\citenamefont {Kohri},
  \citenamefont {Lyth},\ and\ \citenamefont {Melchiorri}}]{Kohri:2007qn}%
  \BibitemOpen
  \bibfield  {author} {\bibinfo {author} {\bibfnamefont {K.}~\bibnamefont
  {Kohri}}, \bibinfo {author} {\bibfnamefont {D.~H.}\ \bibnamefont {Lyth}},
  and \bibinfo {author} {\bibfnamefont {A.}~\bibnamefont {Melchiorri}},\ }\href
  {\doibase 10.1088/1475-7516/2008/04/038} {\bibfield  {journal} {\bibinfo
  {journal} {\emph {JCAP}}\ }\textbf {\bibinfo {volume} {04}},\ \bibinfo
  {pages} {038} (\bibinfo {year} {2008})},\ \Eprint
  {http://arxiv.org/abs/0711.5006} {arXiv:0711.5006 [hep-ph]} \BibitemShut
  {NoStop}%
\bibitem [{\citenamefont {Saito}\ \emph {et~al.}(2008)\citenamefont {Saito},
  \citenamefont {Yokoyama},\ and\ \citenamefont {Nagata}}]{Saito:2008em}%
  \BibitemOpen
  \bibfield  {author} {\bibinfo {author} {\bibfnamefont {R.}~\bibnamefont
  {Saito}}, \bibinfo {author} {\bibfnamefont {J.}~\bibnamefont {Yokoyama}},
  and \bibinfo {author} {\bibfnamefont {R.}~\bibnamefont {Nagata}},\ }\href
  {\doibase 10.1088/1475-7516/2008/06/024} {\bibfield  {journal} {\bibinfo
  {journal} {\emph {JCAP}}\ }\textbf {\bibinfo {volume} {06}},\ \bibinfo
  {pages} {024} (\bibinfo {year} {2008})},\ \Eprint
  {http://arxiv.org/abs/0804.3470} {arXiv:0804.3470 [astro-ph]} \BibitemShut
  {NoStop}%
\bibitem [{\citenamefont {Bugaev}\ and\ \citenamefont
  {Klimai}(2008)}]{Bugaev:2008bi}%
  \BibitemOpen
  \bibfield  {author} {\bibinfo {author} {\bibfnamefont {E.}~\bibnamefont
  {Bugaev}} and \bibinfo {author} {\bibfnamefont {P.}~\bibnamefont {Klimai}},\
  }\href {\doibase 10.1103/PhysRevD.78.063515} {\bibfield  {journal} {\bibinfo
  {journal} {\emph {Phys. Rev. D}}\ }\textbf {\bibinfo {volume} {78}},\
  \bibinfo {pages} {063515} (\bibinfo {year} {2008})},\ \Eprint
  {http://arxiv.org/abs/0806.4541} {arXiv:0806.4541 [astro-ph]} \BibitemShut
  {NoStop}%
\bibitem [{\citenamefont {Alabidi}\ and\ \citenamefont
  {Kohri}(2009)}]{Alabidi:2009bk}%
  \BibitemOpen
  \bibfield  {author} {\bibinfo {author} {\bibfnamefont {L.}~\bibnamefont
  {Alabidi}} and \bibinfo {author} {\bibfnamefont {K.}~\bibnamefont {Kohri}},\
  }\href {\doibase 10.1103/PhysRevD.80.063511} {\bibfield  {journal} {\bibinfo
  {journal} {\emph {Phys. Rev. D}}\ }\textbf {\bibinfo {volume} {80}},\
  \bibinfo {pages} {063511} (\bibinfo {year} {2009})},\ \Eprint
  {http://arxiv.org/abs/0906.1398} {arXiv:0906.1398 [astro-ph.CO]} \BibitemShut
  {NoStop}%
\bibitem [{\citenamefont {Drees}\ and\ \citenamefont
  {Erfani}(2011)}]{Drees:2011hb}%
  \BibitemOpen
  \bibfield  {author} {\bibinfo {author} {\bibfnamefont {M.}~\bibnamefont
  {Drees}} and \bibinfo {author} {\bibfnamefont {E.}~\bibnamefont {Erfani}},\
  }\href {\doibase 10.1088/1475-7516/2011/04/005} {\bibfield  {journal}
  {\bibinfo  {journal} {\emph {JCAP}}\ }\textbf {\bibinfo {volume} {04}},\
  \bibinfo {pages} {005} (\bibinfo {year} {2011})},\ \Eprint
  {http://arxiv.org/abs/1102.2340} {arXiv:1102.2340 [hep-ph]} \BibitemShut
  {NoStop}%
\bibitem [{\citenamefont {Drees}\ and\ \citenamefont
  {Erfani}(2012)}]{Drees:2011yz}%
  \BibitemOpen
  \bibfield  {author} {\bibinfo {author} {\bibfnamefont {M.}~\bibnamefont
  {Drees}} and \bibinfo {author} {\bibfnamefont {E.}~\bibnamefont {Erfani}},\
  }\href {\doibase 10.1088/1475-7516/2012/01/035} {\bibfield  {journal}
  {\bibinfo  {journal} {\emph {JCAP}}\ }\textbf {\bibinfo {volume} {01}},\
  \bibinfo {pages} {035} (\bibinfo {year} {2012})},\ \Eprint
  {http://arxiv.org/abs/1110.6052} {arXiv:1110.6052 [astro-ph.CO]} \BibitemShut
  {NoStop}%
\bibitem [{\citenamefont {Martin}\ \emph {et~al.}(2013)\citenamefont {Martin},
  \citenamefont {Motohashi},\ and\ \citenamefont {Suyama}}]{Martin:2012pe}%
  \BibitemOpen
  \bibfield  {author} {\bibinfo {author} {\bibfnamefont {J.}~\bibnamefont
  {Martin}}, \bibinfo {author} {\bibfnamefont {H.}~\bibnamefont {Motohashi}},
  and \bibinfo {author} {\bibfnamefont {T.}~\bibnamefont {Suyama}},\ }\href
  {\doibase 10.1103/PhysRevD.87.023514} {\bibfield  {journal} {\bibinfo
  {journal} {\emph {Phys. Rev. D}}\ }\textbf {\bibinfo {volume} {87}},\
  \bibinfo {pages} {023514} (\bibinfo {year} {2013})},\ \Eprint
  {http://arxiv.org/abs/1211.0083} {arXiv:1211.0083 [astro-ph.CO]} \BibitemShut
  {NoStop}%
\bibitem [{\citenamefont {Kohri}\ \emph {et~al.}(2013)\citenamefont {Kohri},
  \citenamefont {Lin},\ and\ \citenamefont {Matsuda}}]{Kohri:2012yw}%
  \BibitemOpen
  \bibfield  {author} {\bibinfo {author} {\bibfnamefont {K.}~\bibnamefont
  {Kohri}}, \bibinfo {author} {\bibfnamefont {C.-M.}\ \bibnamefont {Lin}},  and
  \bibinfo {author} {\bibfnamefont {T.}~\bibnamefont {Matsuda}},\ }\href
  {\doibase 10.1103/PhysRevD.87.103527} {\bibfield  {journal} {\bibinfo
  {journal} {\emph {Phys. Rev. D}}\ }\textbf {\bibinfo {volume} {87}},\
  \bibinfo {pages} {103527} (\bibinfo {year} {2013})},\ \Eprint
  {http://arxiv.org/abs/1211.2371} {arXiv:1211.2371 [hep-ph]} \BibitemShut
  {NoStop}%
\bibitem [{\citenamefont {Kawasaki}\ \emph {et~al.}(2013)\citenamefont
  {Kawasaki}, \citenamefont {Kitajima},\ and\ \citenamefont
  {Yanagida}}]{Kawasaki:2012wr}%
  \BibitemOpen
  \bibfield  {author} {\bibinfo {author} {\bibfnamefont {M.}~\bibnamefont
  {Kawasaki}}, \bibinfo {author} {\bibfnamefont {N.}~\bibnamefont {Kitajima}},
  and \bibinfo {author} {\bibfnamefont {T.~T.}\ \bibnamefont {Yanagida}},\
  }\href {\doibase 10.1103/PhysRevD.87.063519} {\bibfield  {journal} {\bibinfo
  {journal} {\emph {Phys. Rev. D}}\ }\textbf {\bibinfo {volume} {87}},\
  \bibinfo {pages} {063519} (\bibinfo {year} {2013})},\ \Eprint
  {http://arxiv.org/abs/1207.2550} {arXiv:1207.2550 [hep-ph]} \BibitemShut
  {NoStop}%
\bibitem [{\citenamefont {Clesse}\ and\ \citenamefont
  {Garc\'\i{}a-Bellido}(2015)}]{Clesse:2015wea}%
  \BibitemOpen
  \bibfield  {author} {\bibinfo {author} {\bibfnamefont {S.}~\bibnamefont
  {Clesse}} and \bibinfo {author} {\bibfnamefont {J.}~\bibnamefont
  {Garc\'\i{}a-Bellido}},\ }\href {\doibase 10.1103/PhysRevD.92.023524}
  {\bibfield  {journal} {\bibinfo  {journal} {\emph {Phys. Rev. D}}\ }\textbf
  {\bibinfo {volume} {92}},\ \bibinfo {pages} {023524} (\bibinfo {year}
  {2015})},\ \Eprint {http://arxiv.org/abs/1501.07565} {arXiv:1501.07565
  [astro-ph.CO]} \BibitemShut {NoStop}%
\bibitem [{\citenamefont {Kawasaki}\ and\ \citenamefont
  {Tada}(2016)}]{Kawasaki:2015ppx}%
  \BibitemOpen
  \bibfield  {author} {\bibinfo {author} {\bibfnamefont {M.}~\bibnamefont
  {Kawasaki}} and \bibinfo {author} {\bibfnamefont {Y.}~\bibnamefont {Tada}},\
  }\href {\doibase 10.1088/1475-7516/2016/08/041} {\bibfield  {journal}
  {\bibinfo  {journal} {\emph {JCAP}}\ }\textbf {\bibinfo {volume} {08}},\
  \bibinfo {pages} {041} (\bibinfo {year} {2016})},\ \Eprint
  {http://arxiv.org/abs/1512.03515} {arXiv:1512.03515 [astro-ph.CO]}
  \BibitemShut {NoStop}%
\bibitem [{\citenamefont {Kawasaki}\ \emph {et~al.}(2016)\citenamefont
  {Kawasaki}, \citenamefont {Kusenko}, \citenamefont {Tada},\ and\
  \citenamefont {Yanagida}}]{Kawasaki:2016pql}%
  \BibitemOpen
  \bibfield  {author} {\bibinfo {author} {\bibfnamefont {M.}~\bibnamefont
  {Kawasaki}}, \bibinfo {author} {\bibfnamefont {A.}~\bibnamefont {Kusenko}},
  \bibinfo {author} {\bibfnamefont {Y.}~\bibnamefont {Tada}},  and \bibinfo
  {author} {\bibfnamefont {T.~T.}\ \bibnamefont {Yanagida}},\ }\href {\doibase
  10.1103/PhysRevD.94.083523} {\bibfield  {journal} {\bibinfo  {journal} {\emph
  {Phys. Rev. D}}\ }\textbf {\bibinfo {volume} {94}},\ \bibinfo {pages}
  {083523} (\bibinfo {year} {2016})},\ \Eprint
  {http://arxiv.org/abs/1606.07631} {arXiv:1606.07631 [astro-ph.CO]}
  \BibitemShut {NoStop}%
\bibitem [{\citenamefont {Pi}\ \emph {et~al.}(2018)\citenamefont {Pi},
  \citenamefont {Zhang}, \citenamefont {Huang},\ and\ \citenamefont
  {Sasaki}}]{Pi:2017gih}%
  \BibitemOpen
  \bibfield  {author} {\bibinfo {author} {\bibfnamefont {S.}~\bibnamefont
  {Pi}}, \bibinfo {author} {\bibfnamefont {Y.-l.}\ \bibnamefont {Zhang}},
  \bibinfo {author} {\bibfnamefont {Q.-G.}\ \bibnamefont {Huang}},  and
  \bibinfo {author} {\bibfnamefont {M.}~\bibnamefont {Sasaki}},\ }\href
  {\doibase 10.1088/1475-7516/2018/05/042} {\bibfield  {journal} {\bibinfo
  {journal} {\emph {JCAP}}\ }\textbf {\bibinfo {volume} {05}},\ \bibinfo
  {pages} {042} (\bibinfo {year} {2018})},\ \Eprint
  {http://arxiv.org/abs/1712.09896} {arXiv:1712.09896 [astro-ph.CO]}
  \BibitemShut {NoStop}%
\bibitem [{\citenamefont {Garcia-Bellido}\ and\ \citenamefont
  {Ruiz~Morales}(2017)}]{Garcia-Bellido:2017mdw}%
  \BibitemOpen
  \bibfield  {author} {\bibinfo {author} {\bibfnamefont {J.}~\bibnamefont
  {Garcia-Bellido}} and \bibinfo {author} {\bibfnamefont {E.}~\bibnamefont
  {Ruiz~Morales}},\ }\href {\doibase 10.1016/j.dark.2017.09.007} {\bibfield
  {journal} {\bibinfo  {journal} {\emph {Phys. Dark Univ.}}\ }\textbf {\bibinfo
  {volume} {18}},\ \bibinfo {pages} {47} (\bibinfo {year} {2017})},\ \Eprint
  {http://arxiv.org/abs/1702.03901} {arXiv:1702.03901 [astro-ph.CO]}
  \BibitemShut {NoStop}%
\bibitem [{\citenamefont {Kannike}\ \emph {et~al.}(2017)\citenamefont
  {Kannike}, \citenamefont {Marzola}, \citenamefont {Raidal},\ and\
  \citenamefont {Veerm\"ae}}]{Kannike:2017bxn}%
  \BibitemOpen
  \bibfield  {author} {\bibinfo {author} {\bibfnamefont {K.}~\bibnamefont
  {Kannike}}, \bibinfo {author} {\bibfnamefont {L.}~\bibnamefont {Marzola}},
  \bibinfo {author} {\bibfnamefont {M.}~\bibnamefont {Raidal}},  and \bibinfo
  {author} {\bibfnamefont {H.}~\bibnamefont {Veerm\"ae}},\ }\href {\doibase
  10.1088/1475-7516/2017/09/020} {\bibfield  {journal} {\bibinfo  {journal}
  {\emph {JCAP}}\ }\textbf {\bibinfo {volume} {09}},\ \bibinfo {pages} {020}
  (\bibinfo {year} {2017})},\ \Eprint {http://arxiv.org/abs/1705.06225}
  {arXiv:1705.06225 [astro-ph.CO]} \BibitemShut {NoStop}%
\bibitem [{\citenamefont {Germani}\ and\ \citenamefont
  {Prokopec}(2017)}]{Germani:2017bcs}%
  \BibitemOpen
  \bibfield  {author} {\bibinfo {author} {\bibfnamefont {C.}~\bibnamefont
  {Germani}} and \bibinfo {author} {\bibfnamefont {T.}~\bibnamefont
  {Prokopec}},\ }\href {\doibase 10.1016/j.dark.2017.09.001} {\bibfield
  {journal} {\bibinfo  {journal} {\emph {Phys. Dark Univ.}}\ }\textbf {\bibinfo
  {volume} {18}},\ \bibinfo {pages} {6} (\bibinfo {year} {2017})},\ \Eprint
  {http://arxiv.org/abs/1706.04226} {arXiv:1706.04226 [astro-ph.CO]}
  \BibitemShut {NoStop}%
\bibitem [{\citenamefont {Ando}\ \emph {et~al.}(2018)\citenamefont {Ando},
  \citenamefont {Inomata}, \citenamefont {Kawasaki}, \citenamefont {Mukaida},\
  and\ \citenamefont {Yanagida}}]{Ando:2017veq}%
  \BibitemOpen
  \bibfield  {author} {\bibinfo {author} {\bibfnamefont {K.}~\bibnamefont
  {Ando}}, \bibinfo {author} {\bibfnamefont {K.}~\bibnamefont {Inomata}},
  \bibinfo {author} {\bibfnamefont {M.}~\bibnamefont {Kawasaki}}, \bibinfo
  {author} {\bibfnamefont {K.}~\bibnamefont {Mukaida}},  and \bibinfo {author}
  {\bibfnamefont {T.~T.}\ \bibnamefont {Yanagida}},\ }\href {\doibase
  10.1103/PhysRevD.97.123512} {\bibfield  {journal} {\bibinfo  {journal} {\emph
  {Phys. Rev. D}}\ }\textbf {\bibinfo {volume} {97}},\ \bibinfo {pages}
  {123512} (\bibinfo {year} {2018})},\ \Eprint
  {http://arxiv.org/abs/1711.08956} {arXiv:1711.08956 [astro-ph.CO]}
  \BibitemShut {NoStop}%
\bibitem [{\citenamefont {Ezquiaga}\ \emph {et~al.}(2018)\citenamefont
  {Ezquiaga}, \citenamefont {Garcia-Bellido},\ and\ \citenamefont
  {Ruiz~Morales}}]{Ezquiaga:2017fvi}%
  \BibitemOpen
  \bibfield  {author} {\bibinfo {author} {\bibfnamefont {J.~M.}\ \bibnamefont
  {Ezquiaga}}, \bibinfo {author} {\bibfnamefont {J.}~\bibnamefont
  {Garcia-Bellido}},  and \bibinfo {author} {\bibfnamefont {E.}~\bibnamefont
  {Ruiz~Morales}},\ }\href {\doibase 10.1016/j.physletb.2017.11.039} {\bibfield
   {journal} {\bibinfo  {journal} {\emph {Phys. Lett. B}}\ }\textbf {\bibinfo
  {volume} {776}},\ \bibinfo {pages} {345} (\bibinfo {year} {2018})},\ \Eprint
  {http://arxiv.org/abs/1705.04861} {arXiv:1705.04861 [astro-ph.CO]}
  \BibitemShut {NoStop}%
\bibitem [{\citenamefont {Motohashi}\ and\ \citenamefont
  {Hu}(2017)}]{Motohashi:2017kbs}%
  \BibitemOpen
  \bibfield  {author} {\bibinfo {author} {\bibfnamefont {H.}~\bibnamefont
  {Motohashi}} and \bibinfo {author} {\bibfnamefont {W.}~\bibnamefont {Hu}},\
  }\href {\doibase 10.1103/PhysRevD.96.063503} {\bibfield  {journal} {\bibinfo
  {journal} {\emph {Phys. Rev. D}}\ }\textbf {\bibinfo {volume} {96}},\
  \bibinfo {pages} {063503} (\bibinfo {year} {2017})},\ \Eprint
  {http://arxiv.org/abs/1706.06784} {arXiv:1706.06784 [astro-ph.CO]}
  \BibitemShut {NoStop}%
\bibitem [{\citenamefont {Di}\ and\ \citenamefont {Gong}(2018)}]{Di:2017ndc}%
  \BibitemOpen
  \bibfield  {author} {\bibinfo {author} {\bibfnamefont {H.}~\bibnamefont {Di}}
  and \bibinfo {author} {\bibfnamefont {Y.}~\bibnamefont {Gong}},\ }\href
  {\doibase 10.1088/1475-7516/2018/07/007} {\bibfield  {journal} {\bibinfo
  {journal} {\emph {JCAP}}\ }\textbf {\bibinfo {volume} {07}},\ \bibinfo
  {pages} {007} (\bibinfo {year} {2018})},\ \Eprint
  {http://arxiv.org/abs/1707.09578} {arXiv:1707.09578 [astro-ph.CO]}
  \BibitemShut {NoStop}%
\bibitem [{\citenamefont {Ballesteros}\ and\ \citenamefont
  {Taoso}(2018)}]{Ballesteros:2017fsr}%
  \BibitemOpen
  \bibfield  {author} {\bibinfo {author} {\bibfnamefont {G.}~\bibnamefont
  {Ballesteros}} and \bibinfo {author} {\bibfnamefont {M.}~\bibnamefont
  {Taoso}},\ }\href {\doibase 10.1103/PhysRevD.97.023501} {\bibfield  {journal}
  {\bibinfo  {journal} {\emph {Phys. Rev. D}}\ }\textbf {\bibinfo {volume}
  {97}},\ \bibinfo {pages} {023501} (\bibinfo {year} {2018})},\ \Eprint
  {http://arxiv.org/abs/1709.05565} {arXiv:1709.05565 [hep-ph]} \BibitemShut
  {NoStop}%
\bibitem [{\citenamefont {Garcia-Bellido}\ \emph {et~al.}(2017)\citenamefont
  {Garcia-Bellido}, \citenamefont {Peloso},\ and\ \citenamefont
  {Unal}}]{Garcia-Bellido:2017aan}%
  \BibitemOpen
  \bibfield  {author} {\bibinfo {author} {\bibfnamefont {J.}~\bibnamefont
  {Garcia-Bellido}}, \bibinfo {author} {\bibfnamefont {M.}~\bibnamefont
  {Peloso}},  and \bibinfo {author} {\bibfnamefont {C.}~\bibnamefont {Unal}},\
  }\href {\doibase 10.1088/1475-7516/2017/09/013} {\bibfield  {journal}
  {\bibinfo  {journal} {\emph {JCAP}}\ }\textbf {\bibinfo {volume} {09}},\
  \bibinfo {pages} {013} (\bibinfo {year} {2017})},\ \Eprint
  {http://arxiv.org/abs/1707.02441} {arXiv:1707.02441 [astro-ph.CO]}
  \BibitemShut {NoStop}%
\bibitem [{\citenamefont {Hertzberg}\ and\ \citenamefont
  {Yamada}(2018)}]{Hertzberg:2017dkh}%
  \BibitemOpen
  \bibfield  {author} {\bibinfo {author} {\bibfnamefont {M.~P.}\ \bibnamefont
  {Hertzberg}} and \bibinfo {author} {\bibfnamefont {M.}~\bibnamefont
  {Yamada}},\ }\href {\doibase 10.1103/PhysRevD.97.083509} {\bibfield
  {journal} {\bibinfo  {journal} {\emph {Phys. Rev. D}}\ }\textbf {\bibinfo
  {volume} {97}},\ \bibinfo {pages} {083509} (\bibinfo {year} {2018})},\
  \Eprint {http://arxiv.org/abs/1712.09750} {arXiv:1712.09750 [astro-ph.CO]}
  \BibitemShut {NoStop}%
\bibitem [{\citenamefont {Inomata}\ \emph {et~al.}(2018)\citenamefont
  {Inomata}, \citenamefont {Kawasaki}, \citenamefont {Mukaida},\ and\
  \citenamefont {Yanagida}}]{Inomata:2018cht}%
  \BibitemOpen
  \bibfield  {author} {\bibinfo {author} {\bibfnamefont {K.}~\bibnamefont
  {Inomata}}, \bibinfo {author} {\bibfnamefont {M.}~\bibnamefont {Kawasaki}},
  \bibinfo {author} {\bibfnamefont {K.}~\bibnamefont {Mukaida}},  and \bibinfo
  {author} {\bibfnamefont {T.~T.}\ \bibnamefont {Yanagida}},\ }\href {\doibase
  10.1103/PhysRevD.97.043514} {\bibfield  {journal} {\bibinfo  {journal} {\emph
  {Phys. Rev. D}}\ }\textbf {\bibinfo {volume} {97}},\ \bibinfo {pages}
  {043514} (\bibinfo {year} {2018})},\ \Eprint
  {http://arxiv.org/abs/1711.06129} {arXiv:1711.06129 [astro-ph.CO]}
  \BibitemShut {NoStop}%
\bibitem [{\citenamefont {Cai}\ \emph {et~al.}(2018{\natexlab{a}})\citenamefont
  {Cai}, \citenamefont {Tong}, \citenamefont {Wang},\ and\ \citenamefont
  {Yan}}]{Cai:2018tuh}%
  \BibitemOpen
  \bibfield  {author} {\bibinfo {author} {\bibfnamefont {Y.-F.}\ \bibnamefont
  {Cai}}, \bibinfo {author} {\bibfnamefont {X.}~\bibnamefont {Tong}}, \bibinfo
  {author} {\bibfnamefont {D.-G.}\ \bibnamefont {Wang}},  and \bibinfo {author}
  {\bibfnamefont {S.-F.}\ \bibnamefont {Yan}},\ }\href {\doibase
  10.1103/PhysRevLett.121.081306} {\bibfield  {journal} {\bibinfo  {journal}
  {\emph {Phys. Rev. Lett.}}\ }\textbf {\bibinfo {volume} {121}},\ \bibinfo
  {pages} {081306} (\bibinfo {year} {2018}{\natexlab{a}})},\ \Eprint
  {http://arxiv.org/abs/1805.03639} {arXiv:1805.03639 [astro-ph.CO]}
  \BibitemShut {NoStop}%
\bibitem [{\citenamefont {Drees}\ and\ \citenamefont
  {Xu}(2021)}]{Drees:2019xpp}%
  \BibitemOpen
  \bibfield  {author} {\bibinfo {author} {\bibfnamefont {M.}~\bibnamefont
  {Drees}} and \bibinfo {author} {\bibfnamefont {Y.}~\bibnamefont {Xu}},\
  }\href {\doibase 10.1140/epjc/s10052-021-08976-2} {\bibfield  {journal}
  {\bibinfo  {journal} {\emph {Eur. Phys. J. C}}\ }\textbf {\bibinfo {volume}
  {81}},\ \bibinfo {pages} {182} (\bibinfo {year} {2021})},\ \Eprint
  {http://arxiv.org/abs/1905.13581} {arXiv:1905.13581 [hep-ph]} \BibitemShut
  {NoStop}%
\bibitem [{\citenamefont {Atal}\ \emph {et~al.}(2019)\citenamefont {Atal},
  \citenamefont {Garriga},\ and\ \citenamefont
  {Marcos-Caballero}}]{Atal:2019cdz}%
  \BibitemOpen
  \bibfield  {author} {\bibinfo {author} {\bibfnamefont {V.}~\bibnamefont
  {Atal}}, \bibinfo {author} {\bibfnamefont {J.}~\bibnamefont {Garriga}},  and
  \bibinfo {author} {\bibfnamefont {A.}~\bibnamefont {Marcos-Caballero}},\
  }\href {\doibase 10.1088/1475-7516/2019/09/073} {\bibfield  {journal}
  {\bibinfo  {journal} {\emph {JCAP}}\ }\textbf {\bibinfo {volume} {09}},\
  \bibinfo {pages} {073} (\bibinfo {year} {2019})},\ \Eprint
  {http://arxiv.org/abs/1905.13202} {arXiv:1905.13202 [astro-ph.CO]}
  \BibitemShut {NoStop}%
\bibitem [{\citenamefont {Atal}\ \emph {et~al.}(2020)\citenamefont {Atal},
  \citenamefont {Cid}, \citenamefont {Escriv\`a},\ and\ \citenamefont
  {Garriga}}]{Atal:2019erb}%
  \BibitemOpen
  \bibfield  {author} {\bibinfo {author} {\bibfnamefont {V.}~\bibnamefont
  {Atal}}, \bibinfo {author} {\bibfnamefont {J.}~\bibnamefont {Cid}}, \bibinfo
  {author} {\bibfnamefont {A.}~\bibnamefont {Escriv\`a}},  and \bibinfo
  {author} {\bibfnamefont {J.}~\bibnamefont {Garriga}},\ }\href {\doibase
  10.1088/1475-7516/2020/05/022} {\bibfield  {journal} {\bibinfo  {journal}
  {\emph {JCAP}}\ }\textbf {\bibinfo {volume} {05}},\ \bibinfo {pages} {022}
  (\bibinfo {year} {2020})},\ \Eprint {http://arxiv.org/abs/1908.11357}
  {arXiv:1908.11357 [astro-ph.CO]} \BibitemShut {NoStop}%
\bibitem [{\citenamefont {Mishra}\ and\ \citenamefont
  {Sahni}(2020)}]{Mishra:2019pzq}%
  \BibitemOpen
  \bibfield  {author} {\bibinfo {author} {\bibfnamefont {S.~S.}\ \bibnamefont
  {Mishra}} and \bibinfo {author} {\bibfnamefont {V.}~\bibnamefont {Sahni}},\
  }\href {\doibase 10.1088/1475-7516/2020/04/007} {\bibfield  {journal}
  {\bibinfo  {journal} {\emph {JCAP}}\ }\textbf {\bibinfo {volume} {04}},\
  \bibinfo {pages} {007} (\bibinfo {year} {2020})},\ \Eprint
  {http://arxiv.org/abs/1911.00057} {arXiv:1911.00057 [gr-qc]} \BibitemShut
  {NoStop}%
\bibitem [{\citenamefont {Cheong}\ \emph {et~al.}(2021)\citenamefont {Cheong},
  \citenamefont {Lee},\ and\ \citenamefont {Park}}]{Cheong:2019vzl}%
  \BibitemOpen
  \bibfield  {author} {\bibinfo {author} {\bibfnamefont {D.~Y.}\ \bibnamefont
  {Cheong}}, \bibinfo {author} {\bibfnamefont {S.~M.}\ \bibnamefont {Lee}},
  and \bibinfo {author} {\bibfnamefont {S.~C.}\ \bibnamefont {Park}},\ }\href
  {\doibase 10.1088/1475-7516/2021/01/032} {\bibfield  {journal} {\bibinfo
  {journal} {\emph {JCAP}}\ }\textbf {\bibinfo {volume} {01}},\ \bibinfo
  {pages} {032} (\bibinfo {year} {2021})},\ \Eprint
  {http://arxiv.org/abs/1912.12032} {arXiv:1912.12032 [hep-ph]} \BibitemShut
  {NoStop}%
\bibitem [{\citenamefont {Fu}\ \emph {et~al.}(2019)\citenamefont {Fu},
  \citenamefont {Wu},\ and\ \citenamefont {Yu}}]{Fu:2019ttf}%
  \BibitemOpen
  \bibfield  {author} {\bibinfo {author} {\bibfnamefont {C.}~\bibnamefont
  {Fu}}, \bibinfo {author} {\bibfnamefont {P.}~\bibnamefont {Wu}},  and
  \bibinfo {author} {\bibfnamefont {H.}~\bibnamefont {Yu}},\ }\href {\doibase
  10.1103/PhysRevD.100.063532} {\bibfield  {journal} {\bibinfo  {journal}
  {\emph {Phys. Rev. D}}\ }\textbf {\bibinfo {volume} {100}},\ \bibinfo {pages}
  {063532} (\bibinfo {year} {2019})},\ \Eprint
  {http://arxiv.org/abs/1907.05042} {arXiv:1907.05042 [astro-ph.CO]}
  \BibitemShut {NoStop}%
\bibitem [{\citenamefont {Dalianis}\ \emph {et~al.}(2020)\citenamefont
  {Dalianis}, \citenamefont {Karydas},\ and\ \citenamefont
  {Papantonopoulos}}]{Dalianis:2019vit}%
  \BibitemOpen
  \bibfield  {author} {\bibinfo {author} {\bibfnamefont {I.}~\bibnamefont
  {Dalianis}}, \bibinfo {author} {\bibfnamefont {S.}~\bibnamefont {Karydas}},
  and \bibinfo {author} {\bibfnamefont {E.}~\bibnamefont {Papantonopoulos}},\
  }\href {\doibase 10.1088/1475-7516/2020/06/040} {\bibfield  {journal}
  {\bibinfo  {journal} {\emph {JCAP}}\ }\textbf {\bibinfo {volume} {06}},\
  \bibinfo {pages} {040} (\bibinfo {year} {2020})},\ \Eprint
  {http://arxiv.org/abs/1910.00622} {arXiv:1910.00622 [astro-ph.CO]}
  \BibitemShut {NoStop}%
\bibitem [{\citenamefont {Ashoorioon}\ \emph {et~al.}(2021)\citenamefont
  {Ashoorioon}, \citenamefont {Rostami},\ and\ \citenamefont
  {Firouzjaee}}]{Ashoorioon:2019xqc}%
  \BibitemOpen
  \bibfield  {author} {\bibinfo {author} {\bibfnamefont {A.}~\bibnamefont
  {Ashoorioon}}, \bibinfo {author} {\bibfnamefont {A.}~\bibnamefont {Rostami}},
   and \bibinfo {author} {\bibfnamefont {J.~T.}\ \bibnamefont {Firouzjaee}},\
  }\href {\doibase 10.1007/JHEP07(2021)087} {\bibfield  {journal} {\bibinfo
  {journal} {\emph {JHEP}}\ }\textbf {\bibinfo {volume} {07}},\ \bibinfo
  {pages} {087} (\bibinfo {year} {2021})},\ \Eprint
  {http://arxiv.org/abs/1912.13326} {arXiv:1912.13326 [astro-ph.CO]}
  \BibitemShut {NoStop}%
\bibitem [{\citenamefont {Lin}\ \emph {et~al.}(2020)\citenamefont {Lin},
  \citenamefont {Gao}, \citenamefont {Gong}, \citenamefont {Lu}, \citenamefont
  {Zhang},\ and\ \citenamefont {Zhang}}]{Lin:2020goi}%
  \BibitemOpen
  \bibfield  {author} {\bibinfo {author} {\bibfnamefont {J.}~\bibnamefont
  {Lin}}, \bibinfo {author} {\bibfnamefont {Q.}~\bibnamefont {Gao}}, \bibinfo
  {author} {\bibfnamefont {Y.}~\bibnamefont {Gong}}, \bibinfo {author}
  {\bibfnamefont {Y.}~\bibnamefont {Lu}}, \bibinfo {author} {\bibfnamefont
  {C.}~\bibnamefont {Zhang}},  and \bibinfo {author} {\bibfnamefont
  {F.}~\bibnamefont {Zhang}},\ }\href {\doibase 10.1103/PhysRevD.101.103515}
  {\bibfield  {journal} {\bibinfo  {journal} {\emph {Phys. Rev. D}}\ }\textbf
  {\bibinfo {volume} {101}},\ \bibinfo {pages} {103515} (\bibinfo {year}
  {2020})},\ \Eprint {http://arxiv.org/abs/2001.05909} {arXiv:2001.05909
  [gr-qc]} \BibitemShut {NoStop}%
\bibitem [{\citenamefont {Yi}\ \emph {et~al.}(2021)\citenamefont {Yi},
  \citenamefont {Gong}, \citenamefont {Wang},\ and\ \citenamefont
  {Zhu}}]{Yi:2020kmq}%
  \BibitemOpen
  \bibfield  {author} {\bibinfo {author} {\bibfnamefont {Z.}~\bibnamefont
  {Yi}}, \bibinfo {author} {\bibfnamefont {Y.}~\bibnamefont {Gong}}, \bibinfo
  {author} {\bibfnamefont {B.}~\bibnamefont {Wang}},  and \bibinfo {author}
  {\bibfnamefont {Z.-h.}\ \bibnamefont {Zhu}},\ }\href {\doibase
  10.1103/PhysRevD.103.063535} {\bibfield  {journal} {\bibinfo  {journal}
  {\emph {Phys. Rev. D}}\ }\textbf {\bibinfo {volume} {103}},\ \bibinfo {pages}
  {063535} (\bibinfo {year} {2021})},\ \Eprint
  {http://arxiv.org/abs/2007.09957} {arXiv:2007.09957 [gr-qc]} \BibitemShut
  {NoStop}%
\bibitem [{\citenamefont {Palma}\ \emph {et~al.}(2020)\citenamefont {Palma},
  \citenamefont {Sypsas},\ and\ \citenamefont {Zenteno}}]{Palma:2020ejf}%
  \BibitemOpen
  \bibfield  {author} {\bibinfo {author} {\bibfnamefont {G.~A.}\ \bibnamefont
  {Palma}}, \bibinfo {author} {\bibfnamefont {S.}~\bibnamefont {Sypsas}},  and
  \bibinfo {author} {\bibfnamefont {C.}~\bibnamefont {Zenteno}},\ }\href
  {\doibase 10.1103/PhysRevLett.125.121301} {\bibfield  {journal} {\bibinfo
  {journal} {\emph {Phys. Rev. Lett.}}\ }\textbf {\bibinfo {volume} {125}},\
  \bibinfo {pages} {121301} (\bibinfo {year} {2020})},\ \Eprint
  {http://arxiv.org/abs/2004.06106} {arXiv:2004.06106 [astro-ph.CO]}
  \BibitemShut {NoStop}%
\bibitem [{\citenamefont {Braglia}\ \emph {et~al.}(2020)\citenamefont
  {Braglia}, \citenamefont {Hazra}, \citenamefont {Finelli}, \citenamefont
  {Smoot}, \citenamefont {Sriramkumar},\ and\ \citenamefont
  {Starobinsky}}]{Braglia:2020eai}%
  \BibitemOpen
  \bibfield  {author} {\bibinfo {author} {\bibfnamefont {M.}~\bibnamefont
  {Braglia}}, \bibinfo {author} {\bibfnamefont {D.~K.}\ \bibnamefont {Hazra}},
  \bibinfo {author} {\bibfnamefont {F.}~\bibnamefont {Finelli}}, \bibinfo
  {author} {\bibfnamefont {G.~F.}\ \bibnamefont {Smoot}}, \bibinfo {author}
  {\bibfnamefont {L.}~\bibnamefont {Sriramkumar}},  and \bibinfo {author}
  {\bibfnamefont {A.~A.}\ \bibnamefont {Starobinsky}},\ }\href {\doibase
  10.1088/1475-7516/2020/08/001} {\bibfield  {journal} {\bibinfo  {journal}
  {\emph {JCAP}}\ }\textbf {\bibinfo {volume} {08}},\ \bibinfo {pages} {001}
  (\bibinfo {year} {2020})},\ \Eprint {http://arxiv.org/abs/2005.02895}
  {arXiv:2005.02895 [astro-ph.CO]} \BibitemShut {NoStop}%
\bibitem [{\citenamefont {Kefala}\ \emph {et~al.}(2021)\citenamefont {Kefala},
  \citenamefont {Kodaxis}, \citenamefont {Stamou},\ and\ \citenamefont
  {Tetradis}}]{Kefala:2020xsx}%
  \BibitemOpen
  \bibfield  {author} {\bibinfo {author} {\bibfnamefont {K.}~\bibnamefont
  {Kefala}}, \bibinfo {author} {\bibfnamefont {G.~P.}\ \bibnamefont {Kodaxis}},
  \bibinfo {author} {\bibfnamefont {I.~D.}\ \bibnamefont {Stamou}},  and
  \bibinfo {author} {\bibfnamefont {N.}~\bibnamefont {Tetradis}},\ }\href
  {\doibase 10.1103/PhysRevD.104.023506} {\bibfield  {journal} {\bibinfo
  {journal} {\emph {Phys. Rev. D}}\ }\textbf {\bibinfo {volume} {104}},\
  \bibinfo {pages} {023506} (\bibinfo {year} {2021})},\ \Eprint
  {http://arxiv.org/abs/2010.12483} {arXiv:2010.12483 [astro-ph.CO]}
  \BibitemShut {NoStop}%
\bibitem [{\citenamefont {Ballesteros}\ \emph {et~al.}(2020)\citenamefont
  {Ballesteros}, \citenamefont {Rey}, \citenamefont {Taoso},\ and\
  \citenamefont {Urbano}}]{Ballesteros:2020qam}%
  \BibitemOpen
  \bibfield  {author} {\bibinfo {author} {\bibfnamefont {G.}~\bibnamefont
  {Ballesteros}}, \bibinfo {author} {\bibfnamefont {J.}~\bibnamefont {Rey}},
  \bibinfo {author} {\bibfnamefont {M.}~\bibnamefont {Taoso}},  and \bibinfo
  {author} {\bibfnamefont {A.}~\bibnamefont {Urbano}},\ }\href {\doibase
  10.1088/1475-7516/2020/07/025} {\bibfield  {journal} {\bibinfo  {journal}
  {\emph {JCAP}}\ }\textbf {\bibinfo {volume} {07}},\ \bibinfo {pages} {025}
  (\bibinfo {year} {2020})},\ \Eprint {http://arxiv.org/abs/2001.08220}
  {arXiv:2001.08220 [astro-ph.CO]} \BibitemShut {NoStop}%
\bibitem [{\citenamefont {Aldabergenov}\ \emph {et~al.}(2020)\citenamefont
  {Aldabergenov}, \citenamefont {Addazi},\ and\ \citenamefont
  {Ketov}}]{Aldabergenov:2020bpt}%
  \BibitemOpen
  \bibfield  {author} {\bibinfo {author} {\bibfnamefont {Y.}~\bibnamefont
  {Aldabergenov}}, \bibinfo {author} {\bibfnamefont {A.}~\bibnamefont
  {Addazi}},  and \bibinfo {author} {\bibfnamefont {S.~V.}\ \bibnamefont
  {Ketov}},\ }\href {\doibase 10.1140/epjc/s10052-020-08506-6} {\bibfield
  {journal} {\bibinfo  {journal} {\emph {Eur. Phys. J. C}}\ }\textbf {\bibinfo
  {volume} {80}},\ \bibinfo {pages} {917} (\bibinfo {year} {2020})},\ \Eprint
  {http://arxiv.org/abs/2006.16641} {arXiv:2006.16641 [hep-th]} \BibitemShut
  {NoStop}%
\bibitem [{\citenamefont {Aldabergenov}\ \emph {et~al.}(2021)\citenamefont
  {Aldabergenov}, \citenamefont {Addazi},\ and\ \citenamefont
  {Ketov}}]{Aldabergenov:2020yok}%
  \BibitemOpen
  \bibfield  {author} {\bibinfo {author} {\bibfnamefont {Y.}~\bibnamefont
  {Aldabergenov}}, \bibinfo {author} {\bibfnamefont {A.}~\bibnamefont
  {Addazi}},  and \bibinfo {author} {\bibfnamefont {S.~V.}\ \bibnamefont
  {Ketov}},\ }\href {\doibase 10.1016/j.physletb.2021.136069} {\bibfield
  {journal} {\bibinfo  {journal} {\emph {Phys. Lett. B}}\ }\textbf {\bibinfo
  {volume} {814}},\ \bibinfo {pages} {136069} (\bibinfo {year} {2021})},\
  \Eprint {http://arxiv.org/abs/2008.10476} {arXiv:2008.10476 [hep-th]}
  \BibitemShut {NoStop}%
\bibitem [{\citenamefont {Inomata}\ \emph {et~al.}(2021)\citenamefont
  {Inomata}, \citenamefont {McDonough},\ and\ \citenamefont
  {Hu}}]{Inomata:2021uqj}%
  \BibitemOpen
  \bibfield  {author} {\bibinfo {author} {\bibfnamefont {K.}~\bibnamefont
  {Inomata}}, \bibinfo {author} {\bibfnamefont {E.}~\bibnamefont {McDonough}},
  and \bibinfo {author} {\bibfnamefont {W.}~\bibnamefont {Hu}},\ }\href
  {\doibase 10.1103/PhysRevD.104.123553} {\bibfield  {journal} {\bibinfo
  {journal} {\emph {Phys. Rev. D}}\ }\textbf {\bibinfo {volume} {104}},\
  \bibinfo {pages} {123553} (\bibinfo {year} {2021})},\ \Eprint
  {http://arxiv.org/abs/2104.03972} {arXiv:2104.03972 [astro-ph.CO]}
  \BibitemShut {NoStop}%
\bibitem [{\citenamefont {Inomata}\ \emph
  {et~al.}(2022{\natexlab{a}})\citenamefont {Inomata}, \citenamefont
  {McDonough},\ and\ \citenamefont {Hu}}]{Inomata:2021tpx}%
  \BibitemOpen
  \bibfield  {author} {\bibinfo {author} {\bibfnamefont {K.}~\bibnamefont
  {Inomata}}, \bibinfo {author} {\bibfnamefont {E.}~\bibnamefont {McDonough}},
  and \bibinfo {author} {\bibfnamefont {W.}~\bibnamefont {Hu}},\ }\href
  {\doibase 10.1088/1475-7516/2022/02/031} {\bibfield  {journal} {\bibinfo
  {journal} {\emph {JCAP}}\ }\textbf {\bibinfo {volume} {02}},\ \bibinfo
  {pages} {031} (\bibinfo {year} {2022}{\natexlab{a}})},\ \Eprint
  {http://arxiv.org/abs/2110.14641} {arXiv:2110.14641 [astro-ph.CO]}
  \BibitemShut {NoStop}%
\bibitem [{\citenamefont {Dalianis}\ \emph {et~al.}(2021)\citenamefont
  {Dalianis}, \citenamefont {Kodaxis}, \citenamefont {Stamou}, \citenamefont
  {Tetradis},\ and\ \citenamefont {Tsigkas-Kouvelis}}]{Dalianis:2021iig}%
  \BibitemOpen
  \bibfield  {author} {\bibinfo {author} {\bibfnamefont {I.}~\bibnamefont
  {Dalianis}}, \bibinfo {author} {\bibfnamefont {G.~P.}\ \bibnamefont
  {Kodaxis}}, \bibinfo {author} {\bibfnamefont {I.~D.}\ \bibnamefont {Stamou}},
  \bibinfo {author} {\bibfnamefont {N.}~\bibnamefont {Tetradis}},  and \bibinfo
  {author} {\bibfnamefont {A.}~\bibnamefont {Tsigkas-Kouvelis}},\ }\href
  {\doibase 10.1103/PhysRevD.104.103510} {\bibfield  {journal} {\bibinfo
  {journal} {\emph {Phys. Rev. D}}\ }\textbf {\bibinfo {volume} {104}},\
  \bibinfo {pages} {103510} (\bibinfo {year} {2021})},\ \Eprint
  {http://arxiv.org/abs/2106.02467} {arXiv:2106.02467 [astro-ph.CO]}
  \BibitemShut {NoStop}%
\bibitem [{\citenamefont {Cai}\ \emph {et~al.}(2022{\natexlab{a}})\citenamefont
  {Cai}, \citenamefont {Ma}, \citenamefont {Sasaki}, \citenamefont {Wang},\
  and\ \citenamefont {Zhou}}]{Cai:2021zsp}%
  \BibitemOpen
  \bibfield  {author} {\bibinfo {author} {\bibfnamefont {Y.-F.}\ \bibnamefont
  {Cai}}, \bibinfo {author} {\bibfnamefont {X.-H.}\ \bibnamefont {Ma}},
  \bibinfo {author} {\bibfnamefont {M.}~\bibnamefont {Sasaki}}, \bibinfo
  {author} {\bibfnamefont {D.-G.}\ \bibnamefont {Wang}},  and \bibinfo {author}
  {\bibfnamefont {Z.}~\bibnamefont {Zhou}},\ }\href {\doibase
  10.1016/j.physletb.2022.137461} {\bibfield  {journal} {\bibinfo  {journal}
  {\emph {Phys. Lett. B}}\ }\textbf {\bibinfo {volume} {834}},\ \bibinfo
  {pages} {137461} (\bibinfo {year} {2022}{\natexlab{a}})},\ \Eprint
  {http://arxiv.org/abs/2112.13836} {arXiv:2112.13836 [astro-ph.CO]}
  \BibitemShut {NoStop}%
\bibitem [{\citenamefont {Lin}\ \emph {et~al.}(2021)\citenamefont {Lin},
  \citenamefont {Gao}, \citenamefont {Gong}, \citenamefont {Lu}, \citenamefont
  {Wang},\ and\ \citenamefont {Zhang}}]{Lin:2021vwc}%
  \BibitemOpen
  \bibfield  {author} {\bibinfo {author} {\bibfnamefont {J.}~\bibnamefont
  {Lin}}, \bibinfo {author} {\bibfnamefont {S.}~\bibnamefont {Gao}}, \bibinfo
  {author} {\bibfnamefont {Y.}~\bibnamefont {Gong}}, \bibinfo {author}
  {\bibfnamefont {Y.}~\bibnamefont {Lu}}, \bibinfo {author} {\bibfnamefont
  {Z.}~\bibnamefont {Wang}},  and \bibinfo {author} {\bibfnamefont
  {F.}~\bibnamefont {Zhang}},\ }\Eprint {http://arxiv.org/abs/2111.01362}
  {arXiv:2111.01362 [gr-qc]} \BibitemShut {NoStop}%
\bibitem [{\citenamefont {Kawai}\ and\ \citenamefont
  {Kim}(2021{\natexlab{a}})}]{Kawai:2021edk}%
  \BibitemOpen
  \bibfield  {author} {\bibinfo {author} {\bibfnamefont {S.}~\bibnamefont
  {Kawai}} and \bibinfo {author} {\bibfnamefont {J.}~\bibnamefont {Kim}},\
  }\href {\doibase 10.1103/PhysRevD.104.083545} {\bibfield  {journal} {\bibinfo
   {journal} {\emph {Phys. Rev. D}}\ }\textbf {\bibinfo {volume} {104}},\
  \bibinfo {pages} {083545} (\bibinfo {year} {2021}{\natexlab{a}})},\ \Eprint
  {http://arxiv.org/abs/2108.01340} {arXiv:2108.01340 [astro-ph.CO]}
  \BibitemShut {NoStop}%
\bibitem [{\citenamefont {Zhang}(2022)}]{Zhang:2021rqs}%
  \BibitemOpen
  \bibfield  {author} {\bibinfo {author} {\bibfnamefont {F.}~\bibnamefont
  {Zhang}},\ }\href {\doibase 10.1103/PhysRevD.105.063539} {\bibfield
  {journal} {\bibinfo  {journal} {\emph {Phys. Rev. D}}\ }\textbf {\bibinfo
  {volume} {105}},\ \bibinfo {pages} {063539} (\bibinfo {year} {2022})},\
  \Eprint {http://arxiv.org/abs/2112.10516} {arXiv:2112.10516 [gr-qc]}
  \BibitemShut {NoStop}%
\bibitem [{\citenamefont {Ahmed}\ \emph {et~al.}(2022)\citenamefont {Ahmed},
  \citenamefont {Junaid},\ and\ \citenamefont {Zubair}}]{Ahmed:2021ucx}%
  \BibitemOpen
  \bibfield  {author} {\bibinfo {author} {\bibfnamefont {W.}~\bibnamefont
  {Ahmed}}, \bibinfo {author} {\bibfnamefont {M.}~\bibnamefont {Junaid}},  and
  \bibinfo {author} {\bibfnamefont {U.}~\bibnamefont {Zubair}},\ }\href
  {\doibase 10.1016/j.nuclphysb.2022.115968} {\bibfield  {journal} {\bibinfo
  {journal} {\emph {Nucl. Phys. B}}\ }\textbf {\bibinfo {volume} {984}},\
  \bibinfo {pages} {115968} (\bibinfo {year} {2022})},\ \Eprint
  {http://arxiv.org/abs/2109.14838} {arXiv:2109.14838 [astro-ph.CO]}
  \BibitemShut {NoStop}%
\bibitem [{\citenamefont {Cai}\ \emph {et~al.}(2022{\natexlab{b}})\citenamefont
  {Cai}, \citenamefont {Ma}, \citenamefont {Sasaki}, \citenamefont {Wang},\
  and\ \citenamefont {Zhou}}]{Cai:2022erk}%
  \BibitemOpen
  \bibfield  {author} {\bibinfo {author} {\bibfnamefont {Y.-F.}\ \bibnamefont
  {Cai}}, \bibinfo {author} {\bibfnamefont {X.-H.}\ \bibnamefont {Ma}},
  \bibinfo {author} {\bibfnamefont {M.}~\bibnamefont {Sasaki}}, \bibinfo
  {author} {\bibfnamefont {D.-G.}\ \bibnamefont {Wang}},  and \bibinfo {author}
  {\bibfnamefont {Z.}~\bibnamefont {Zhou}},\ }\Eprint
  {http://arxiv.org/abs/2207.11910} {arXiv:2207.11910 [astro-ph.CO]}
  \BibitemShut {NoStop}%
\bibitem [{\citenamefont {Pi}\ and\ \citenamefont {Wang}(2022)}]{Pi:2022zxs}%
  \BibitemOpen
  \bibfield  {author} {\bibinfo {author} {\bibfnamefont {S.}~\bibnamefont {Pi}}
  and \bibinfo {author} {\bibfnamefont {J.}~\bibnamefont {Wang}},\ }\Eprint
  {http://arxiv.org/abs/2209.14183} {arXiv:2209.14183 [astro-ph.CO]}
  \BibitemShut {NoStop}%
\bibitem [{\citenamefont {Cheong}\ \emph {et~al.}(2022)\citenamefont {Cheong},
  \citenamefont {Kohri},\ and\ \citenamefont {Park}}]{Cheong:2022gfc}%
  \BibitemOpen
  \bibfield  {author} {\bibinfo {author} {\bibfnamefont {D.~Y.}\ \bibnamefont
  {Cheong}}, \bibinfo {author} {\bibfnamefont {K.}~\bibnamefont {Kohri}},  and
  \bibinfo {author} {\bibfnamefont {S.~C.}\ \bibnamefont {Park}},\ }\href
  {\doibase 10.1088/1475-7516/2022/10/015} {\bibfield  {journal} {\bibinfo
  {journal} {\emph {JCAP}}\ }\textbf {\bibinfo {volume} {10}},\ \bibinfo
  {pages} {015} (\bibinfo {year} {2022})},\ \Eprint
  {http://arxiv.org/abs/2205.14813} {arXiv:2205.14813 [hep-ph]} \BibitemShut
  {NoStop}%
\bibitem [{\citenamefont {Kawai}\ and\ \citenamefont
  {Kim}(2022)}]{Kawai:2022emp}%
  \BibitemOpen
  \bibfield  {author} {\bibinfo {author} {\bibfnamefont {S.}~\bibnamefont
  {Kawai}} and \bibinfo {author} {\bibfnamefont {J.}~\bibnamefont {Kim}},\
  }\Eprint {http://arxiv.org/abs/2209.15343} {arXiv:2209.15343 [astro-ph.CO]}
  \BibitemShut {NoStop}%
\bibitem [{\citenamefont {Akrami}\ \emph {et~al.}(2020)\citenamefont {Akrami}
  \emph {et~al.}}]{Planck:2018jri}%
  \BibitemOpen
  \bibfield  {author} {\bibinfo {author} {\bibfnamefont {Y.}~\bibnamefont
  {Akrami}} \emph {et~al.} (\bibinfo {collaboration} {Planck}),\ }\href
  {\doibase 10.1051/0004-6361/201833887} {\bibfield  {journal} {\bibinfo
  {journal} {\emph {Astron. Astrophys.}}\ }\textbf {\bibinfo {volume} {641}},\
  \bibinfo {pages} {A10} (\bibinfo {year} {2020})},\ \Eprint
  {http://arxiv.org/abs/1807.06211} {arXiv:1807.06211 [astro-ph.CO]}
  \BibitemShut {NoStop}%
\bibitem [{\citenamefont {Green}\ and\ \citenamefont
  {Malik}(2001)}]{Green:2000he}%
  \BibitemOpen
  \bibfield  {author} {\bibinfo {author} {\bibfnamefont {A.~M.}\ \bibnamefont
  {Green}} and \bibinfo {author} {\bibfnamefont {K.~A.}\ \bibnamefont
  {Malik}},\ }\href {\doibase 10.1103/PhysRevD.64.021301} {\bibfield  {journal}
  {\bibinfo  {journal} {\emph {Phys. Rev. D}}\ }\textbf {\bibinfo {volume}
  {64}},\ \bibinfo {pages} {021301} (\bibinfo {year} {2001})},\ \Eprint
  {http://arxiv.org/abs/hep-ph/0008113} {arXiv:hep-ph/0008113} \BibitemShut
  {NoStop}%
\bibitem [{\citenamefont {Bassett}\ and\ \citenamefont
  {Tsujikawa}(2001)}]{Bassett:2000ha}%
  \BibitemOpen
  \bibfield  {author} {\bibinfo {author} {\bibfnamefont {B.~A.}\ \bibnamefont
  {Bassett}} and \bibinfo {author} {\bibfnamefont {S.}~\bibnamefont
  {Tsujikawa}},\ }\href {\doibase 10.1103/PhysRevD.63.123503} {\bibfield
  {journal} {\bibinfo  {journal} {\emph {Phys. Rev. D}}\ }\textbf {\bibinfo
  {volume} {63}},\ \bibinfo {pages} {123503} (\bibinfo {year} {2001})},\
  \Eprint {http://arxiv.org/abs/hep-ph/0008328} {arXiv:hep-ph/0008328}
  \BibitemShut {NoStop}%
\bibitem [{\citenamefont {Suyama}\ \emph {et~al.}(2005)\citenamefont {Suyama},
  \citenamefont {Tanaka}, \citenamefont {Bassett},\ and\ \citenamefont
  {Kudoh}}]{Suyama:2004mz}%
  \BibitemOpen
  \bibfield  {author} {\bibinfo {author} {\bibfnamefont {T.}~\bibnamefont
  {Suyama}}, \bibinfo {author} {\bibfnamefont {T.}~\bibnamefont {Tanaka}},
  \bibinfo {author} {\bibfnamefont {B.}~\bibnamefont {Bassett}},  and \bibinfo
  {author} {\bibfnamefont {H.}~\bibnamefont {Kudoh}},\ }\href {\doibase
  10.1103/PhysRevD.71.063507} {\bibfield  {journal} {\bibinfo  {journal} {\emph
  {Phys. Rev. D}}\ }\textbf {\bibinfo {volume} {71}},\ \bibinfo {pages}
  {063507} (\bibinfo {year} {2005})},\ \Eprint
  {http://arxiv.org/abs/hep-ph/0410247} {arXiv:hep-ph/0410247} \BibitemShut
  {NoStop}%
\bibitem [{\citenamefont {Martin}\ \emph {et~al.}(2020)\citenamefont {Martin},
  \citenamefont {Papanikolaou},\ and\ \citenamefont {Vennin}}]{Martin:2019nuw}%
  \BibitemOpen
  \bibfield  {author} {\bibinfo {author} {\bibfnamefont {J.}~\bibnamefont
  {Martin}}, \bibinfo {author} {\bibfnamefont {T.}~\bibnamefont
  {Papanikolaou}},  and \bibinfo {author} {\bibfnamefont {V.}~\bibnamefont
  {Vennin}},\ }\href {\doibase 10.1088/1475-7516/2020/01/024} {\bibfield
  {journal} {\bibinfo  {journal} {\emph {JCAP}}\ }\textbf {\bibinfo {volume}
  {01}},\ \bibinfo {pages} {024} (\bibinfo {year} {2020})},\ \Eprint
  {http://arxiv.org/abs/1907.04236} {arXiv:1907.04236 [astro-ph.CO]}
  \BibitemShut {NoStop}%
\bibitem [{\citenamefont {Bezrukov}\ and\ \citenamefont
  {Shaposhnikov}(2008)}]{Bezrukov:2007ep}%
  \BibitemOpen
  \bibfield  {author} {\bibinfo {author} {\bibfnamefont {F.~L.}\ \bibnamefont
  {Bezrukov}} and \bibinfo {author} {\bibfnamefont {M.}~\bibnamefont
  {Shaposhnikov}},\ }\href {\doibase 10.1016/j.physletb.2007.11.072} {\bibfield
   {journal} {\bibinfo  {journal} {\emph {Phys. Lett. B}}\ }\textbf {\bibinfo
  {volume} {659}},\ \bibinfo {pages} {703} (\bibinfo {year} {2008})},\ \Eprint
  {http://arxiv.org/abs/0710.3755} {arXiv:0710.3755 [hep-th]} \BibitemShut
  {NoStop}%
\bibitem [{\citenamefont {Bezrukov}\ \emph {et~al.}(2009)\citenamefont
  {Bezrukov}, \citenamefont {Magnin},\ and\ \citenamefont
  {Shaposhnikov}}]{Bezrukov:2008ej}%
  \BibitemOpen
  \bibfield  {author} {\bibinfo {author} {\bibfnamefont {F.~L.}\ \bibnamefont
  {Bezrukov}}, \bibinfo {author} {\bibfnamefont {A.}~\bibnamefont {Magnin}},
  and \bibinfo {author} {\bibfnamefont {M.}~\bibnamefont {Shaposhnikov}},\
  }\href {\doibase 10.1016/j.physletb.2009.03.035} {\bibfield  {journal}
  {\bibinfo  {journal} {\emph {Phys. Lett. B}}\ }\textbf {\bibinfo {volume}
  {675}},\ \bibinfo {pages} {88} (\bibinfo {year} {2009})},\ \Eprint
  {http://arxiv.org/abs/0812.4950} {arXiv:0812.4950 [hep-ph]} \BibitemShut
  {NoStop}%
\bibitem [{\citenamefont {Bezrukov}\ \emph {et~al.}(2011)\citenamefont
  {Bezrukov}, \citenamefont {Magnin}, \citenamefont {Shaposhnikov},\ and\
  \citenamefont {Sibiryakov}}]{Bezrukov:2010jz}%
  \BibitemOpen
  \bibfield  {author} {\bibinfo {author} {\bibfnamefont {F.}~\bibnamefont
  {Bezrukov}}, \bibinfo {author} {\bibfnamefont {A.}~\bibnamefont {Magnin}},
  \bibinfo {author} {\bibfnamefont {M.}~\bibnamefont {Shaposhnikov}},  and
  \bibinfo {author} {\bibfnamefont {S.}~\bibnamefont {Sibiryakov}},\ }\href
  {\doibase 10.1007/JHEP01(2011)016} {\bibfield  {journal} {\bibinfo  {journal}
  {\emph {JHEP}}\ }\textbf {\bibinfo {volume} {01}},\ \bibinfo {pages} {016}
  (\bibinfo {year} {2011})},\ \Eprint {http://arxiv.org/abs/1008.5157}
  {arXiv:1008.5157 [hep-ph]} \BibitemShut {NoStop}%
\bibitem [{\citenamefont {Futamase}\ and\ \citenamefont
  {Maeda}(1989)}]{Futamase:1987ua}%
  \BibitemOpen
  \bibfield  {author} {\bibinfo {author} {\bibfnamefont {T.}~\bibnamefont
  {Futamase}} and \bibinfo {author} {\bibfnamefont {K.-i.}\ \bibnamefont
  {Maeda}},\ }\href {\doibase 10.1103/PhysRevD.39.399} {\bibfield  {journal}
  {\bibinfo  {journal} {\emph {Phys. Rev. D}}\ }\textbf {\bibinfo {volume}
  {39}},\ \bibinfo {pages} {399} (\bibinfo {year} {1989})}\BibitemShut
  {NoStop}%
\bibitem [{\citenamefont {Fakir}\ and\ \citenamefont
  {Unruh}(1990)}]{Fakir:1990eg}%
  \BibitemOpen
  \bibfield  {author} {\bibinfo {author} {\bibfnamefont {R.}~\bibnamefont
  {Fakir}} and \bibinfo {author} {\bibfnamefont {W.~G.}\ \bibnamefont
  {Unruh}},\ }\href {\doibase 10.1103/PhysRevD.41.1783} {\bibfield  {journal}
  {\bibinfo  {journal} {\emph {Phys. Rev. D}}\ }\textbf {\bibinfo {volume}
  {41}},\ \bibinfo {pages} {1783} (\bibinfo {year} {1990})}\BibitemShut
  {NoStop}%
\bibitem [{\citenamefont {Komatsu}\ and\ \citenamefont
  {Futamase}(1999)}]{Komatsu:1999mt}%
  \BibitemOpen
  \bibfield  {author} {\bibinfo {author} {\bibfnamefont {E.}~\bibnamefont
  {Komatsu}} and \bibinfo {author} {\bibfnamefont {T.}~\bibnamefont
  {Futamase}},\ }\href {\doibase 10.1103/PhysRevD.59.064029} {\bibfield
  {journal} {\bibinfo  {journal} {\emph {Phys. Rev. D}}\ }\textbf {\bibinfo
  {volume} {59}},\ \bibinfo {pages} {064029} (\bibinfo {year} {1999})},\
  \Eprint {http://arxiv.org/abs/astro-ph/9901127} {arXiv:astro-ph/9901127}
  \BibitemShut {NoStop}%
\bibitem [{\citenamefont {Tsujikawa}\ and\ \citenamefont
  {Gumjudpai}(2004)}]{Tsujikawa:2004my}%
  \BibitemOpen
  \bibfield  {author} {\bibinfo {author} {\bibfnamefont {S.}~\bibnamefont
  {Tsujikawa}} and \bibinfo {author} {\bibfnamefont {B.}~\bibnamefont
  {Gumjudpai}},\ }\href {\doibase 10.1103/PhysRevD.69.123523} {\bibfield
  {journal} {\bibinfo  {journal} {\emph {Phys. Rev. D}}\ }\textbf {\bibinfo
  {volume} {69}},\ \bibinfo {pages} {123523} (\bibinfo {year} {2004})},\
  \Eprint {http://arxiv.org/abs/astro-ph/0402185} {arXiv:astro-ph/0402185}
  \BibitemShut {NoStop}%
\bibitem [{\citenamefont {Linde}\ \emph {et~al.}(2011)\citenamefont {Linde},
  \citenamefont {Noorbala},\ and\ \citenamefont {Westphal}}]{Linde:2011nh}%
  \BibitemOpen
  \bibfield  {author} {\bibinfo {author} {\bibfnamefont {A.}~\bibnamefont
  {Linde}}, \bibinfo {author} {\bibfnamefont {M.}~\bibnamefont {Noorbala}},
  and \bibinfo {author} {\bibfnamefont {A.}~\bibnamefont {Westphal}},\ }\href
  {\doibase 10.1088/1475-7516/2011/03/013} {\bibfield  {journal} {\bibinfo
  {journal} {\emph {JCAP}}\ }\textbf {\bibinfo {volume} {03}},\ \bibinfo
  {pages} {013} (\bibinfo {year} {2011})},\ \Eprint
  {http://arxiv.org/abs/1101.2652} {arXiv:1101.2652 [hep-th]} \BibitemShut
  {NoStop}%
\bibitem [{\citenamefont {Ade}\ \emph {et~al.}(2014)\citenamefont {Ade} \emph
  {et~al.}}]{Planck:2013jfk}%
  \BibitemOpen
  \bibfield  {author} {\bibinfo {author} {\bibfnamefont {P.~A.~R.}\
  \bibnamefont {Ade}} \emph {et~al.} (\bibinfo {collaboration} {Planck}),\
  }\href {\doibase 10.1051/0004-6361/201321569} {\bibfield  {journal} {\bibinfo
   {journal} {\emph {Astron. Astrophys.}}\ }\textbf {\bibinfo {volume} {571}},\
  \bibinfo {pages} {A22} (\bibinfo {year} {2014})},\ \Eprint
  {http://arxiv.org/abs/1303.5082} {arXiv:1303.5082 [astro-ph.CO]} \BibitemShut
  {NoStop}%
\bibitem [{\citenamefont {Tsujikawa}\ \emph {et~al.}(2013)\citenamefont
  {Tsujikawa}, \citenamefont {Ohashi}, \citenamefont {Kuroyanagi},\ and\
  \citenamefont {De~Felice}}]{Tsujikawa:2013ila}%
  \BibitemOpen
  \bibfield  {author} {\bibinfo {author} {\bibfnamefont {S.}~\bibnamefont
  {Tsujikawa}}, \bibinfo {author} {\bibfnamefont {J.}~\bibnamefont {Ohashi}},
  \bibinfo {author} {\bibfnamefont {S.}~\bibnamefont {Kuroyanagi}},  and
  \bibinfo {author} {\bibfnamefont {A.}~\bibnamefont {De~Felice}},\ }\href
  {\doibase 10.1103/PhysRevD.88.023529} {\bibfield  {journal} {\bibinfo
  {journal} {\emph {Phys. Rev. D}}\ }\textbf {\bibinfo {volume} {88}},\
  \bibinfo {pages} {023529} (\bibinfo {year} {2013})},\ \Eprint
  {http://arxiv.org/abs/1305.3044} {arXiv:1305.3044 [astro-ph.CO]} \BibitemShut
  {NoStop}%
\bibitem [{\citenamefont {Hwang}\ and\ \citenamefont
  {Noh}(2005)}]{Hwang:2005hb}%
  \BibitemOpen
  \bibfield  {author} {\bibinfo {author} {\bibfnamefont {J.-c.}\ \bibnamefont
  {Hwang}} and \bibinfo {author} {\bibfnamefont {H.}~\bibnamefont {Noh}},\
  }\href {\doibase 10.1103/PhysRevD.71.063536} {\bibfield  {journal} {\bibinfo
  {journal} {\emph {Phys. Rev. D}}\ }\textbf {\bibinfo {volume} {71}},\
  \bibinfo {pages} {063536} (\bibinfo {year} {2005})},\ \Eprint
  {http://arxiv.org/abs/gr-qc/0412126} {arXiv:gr-qc/0412126} \BibitemShut
  {NoStop}%
\bibitem [{\citenamefont {Guo}\ \emph {et~al.}(2007)\citenamefont {Guo},
  \citenamefont {Ohta},\ and\ \citenamefont {Tsujikawa}}]{Guo:2006ct}%
  \BibitemOpen
  \bibfield  {author} {\bibinfo {author} {\bibfnamefont {Z.-K.}\ \bibnamefont
  {Guo}}, \bibinfo {author} {\bibfnamefont {N.}~\bibnamefont {Ohta}},  and
  \bibinfo {author} {\bibfnamefont {S.}~\bibnamefont {Tsujikawa}},\ }\href
  {\doibase 10.1103/PhysRevD.75.023520} {\bibfield  {journal} {\bibinfo
  {journal} {\emph {Phys. Rev. D}}\ }\textbf {\bibinfo {volume} {75}},\
  \bibinfo {pages} {023520} (\bibinfo {year} {2007})},\ \Eprint
  {http://arxiv.org/abs/hep-th/0610336} {arXiv:hep-th/0610336} \BibitemShut
  {NoStop}%
\bibitem [{\citenamefont {Satoh}\ and\ \citenamefont
  {Soda}(2008)}]{Satoh:2008ck}%
  \BibitemOpen
  \bibfield  {author} {\bibinfo {author} {\bibfnamefont {M.}~\bibnamefont
  {Satoh}} and \bibinfo {author} {\bibfnamefont {J.}~\bibnamefont {Soda}},\
  }\href {\doibase 10.1088/1475-7516/2008/09/019} {\bibfield  {journal}
  {\bibinfo  {journal} {\emph {JCAP}}\ }\textbf {\bibinfo {volume} {09}},\
  \bibinfo {pages} {019} (\bibinfo {year} {2008})},\ \Eprint
  {http://arxiv.org/abs/0806.4594} {arXiv:0806.4594 [astro-ph]} \BibitemShut
  {NoStop}%
\bibitem [{\citenamefont {Guo}\ and\ \citenamefont
  {Schwarz}(2009)}]{Guo:2009uk}%
  \BibitemOpen
  \bibfield  {author} {\bibinfo {author} {\bibfnamefont {Z.-K.}\ \bibnamefont
  {Guo}} and \bibinfo {author} {\bibfnamefont {D.~J.}\ \bibnamefont
  {Schwarz}},\ }\href {\doibase 10.1103/PhysRevD.80.063523} {\bibfield
  {journal} {\bibinfo  {journal} {\emph {Phys. Rev. D}}\ }\textbf {\bibinfo
  {volume} {80}},\ \bibinfo {pages} {063523} (\bibinfo {year} {2009})},\
  \Eprint {http://arxiv.org/abs/0907.0427} {arXiv:0907.0427 [hep-th]}
  \BibitemShut {NoStop}%
\bibitem [{\citenamefont {Guo}\ and\ \citenamefont
  {Schwarz}(2010)}]{Guo:2010jr}%
  \BibitemOpen
  \bibfield  {author} {\bibinfo {author} {\bibfnamefont {Z.-K.}\ \bibnamefont
  {Guo}} and \bibinfo {author} {\bibfnamefont {D.~J.}\ \bibnamefont
  {Schwarz}},\ }\href {\doibase 10.1103/PhysRevD.81.123520} {\bibfield
  {journal} {\bibinfo  {journal} {\emph {Phys. Rev. D}}\ }\textbf {\bibinfo
  {volume} {81}},\ \bibinfo {pages} {123520} (\bibinfo {year} {2010})},\
  \Eprint {http://arxiv.org/abs/1001.1897} {arXiv:1001.1897 [hep-th]}
  \BibitemShut {NoStop}%
\bibitem [{\citenamefont {Kawai}\ and\ \citenamefont
  {Kim}(2021{\natexlab{b}})}]{Kawai:2021bye}%
  \BibitemOpen
  \bibfield  {author} {\bibinfo {author} {\bibfnamefont {S.}~\bibnamefont
  {Kawai}} and \bibinfo {author} {\bibfnamefont {J.}~\bibnamefont {Kim}},\
  }\href {\doibase 10.1103/PhysRevD.104.043525} {\bibfield  {journal} {\bibinfo
   {journal} {\emph {Phys. Rev. D}}\ }\textbf {\bibinfo {volume} {104}},\
  \bibinfo {pages} {043525} (\bibinfo {year} {2021}{\natexlab{b}})},\ \Eprint
  {http://arxiv.org/abs/2105.04386} {arXiv:2105.04386 [hep-ph]} \BibitemShut
  {NoStop}%
\bibitem [{\citenamefont {Kobayashi}\ \emph {et~al.}(2011)\citenamefont
  {Kobayashi}, \citenamefont {Yamaguchi},\ and\ \citenamefont
  {Yokoyama}}]{KYY}%
  \BibitemOpen
  \bibfield  {author} {\bibinfo {author} {\bibfnamefont {T.}~\bibnamefont
  {Kobayashi}}, \bibinfo {author} {\bibfnamefont {M.}~\bibnamefont
  {Yamaguchi}},  and \bibinfo {author} {\bibfnamefont {J.}~\bibnamefont
  {Yokoyama}},\ }\href {\doibase 10.1143/PTP.126.511} {\bibfield  {journal}
  {\bibinfo  {journal} {\emph {Prog. Theor. Phys.}}\ }\textbf {\bibinfo
  {volume} {126}},\ \bibinfo {pages} {511} (\bibinfo {year} {2011})},\ \Eprint
  {http://arxiv.org/abs/1105.5723} {arXiv:1105.5723 [hep-th]} \BibitemShut
  {NoStop}%
\bibitem [{\citenamefont {Kase}\ and\ \citenamefont
  {Tsujikawa}(2019)}]{Kase:2018aps}%
  \BibitemOpen
  \bibfield  {author} {\bibinfo {author} {\bibfnamefont {R.}~\bibnamefont
  {Kase}} and \bibinfo {author} {\bibfnamefont {S.}~\bibnamefont {Tsujikawa}},\
  }\href {\doibase 10.1142/S0218271819420057} {\bibfield  {journal} {\bibinfo
  {journal} {\emph {Int. J. Mod. Phys. D}}\ }\textbf {\bibinfo {volume} {28}},\
  \bibinfo {pages} {1942005} (\bibinfo {year} {2019})},\ \Eprint
  {http://arxiv.org/abs/1809.08735} {arXiv:1809.08735 [gr-qc]} \BibitemShut
  {NoStop}%
\bibitem [{\citenamefont {Salopek}\ \emph {et~al.}(1989)\citenamefont
  {Salopek}, \citenamefont {Bond},\ and\ \citenamefont
  {Bardeen}}]{Salopek:1988qh}%
  \BibitemOpen
  \bibfield  {author} {\bibinfo {author} {\bibfnamefont {D.~S.}\ \bibnamefont
  {Salopek}}, \bibinfo {author} {\bibfnamefont {J.~R.}\ \bibnamefont {Bond}},
  and \bibinfo {author} {\bibfnamefont {J.~M.}\ \bibnamefont {Bardeen}},\
  }\href {\doibase 10.1103/PhysRevD.40.1753} {\bibfield  {journal} {\bibinfo
  {journal} {\emph {Phys. Rev. D}}\ }\textbf {\bibinfo {volume} {40}},\
  \bibinfo {pages} {1753} (\bibinfo {year} {1989})}\BibitemShut {NoStop}%
\bibitem [{\citenamefont {Makino}\ and\ \citenamefont
  {Sasaki}(1991)}]{Makino:1991sg}%
  \BibitemOpen
  \bibfield  {author} {\bibinfo {author} {\bibfnamefont {N.}~\bibnamefont
  {Makino}} and \bibinfo {author} {\bibfnamefont {M.}~\bibnamefont {Sasaki}},\
  }\href {\doibase 10.1143/PTP.86.103} {\bibfield  {journal} {\bibinfo
  {journal} {\emph {Prog. Theor. Phys.}}\ }\textbf {\bibinfo {volume} {86}},\
  \bibinfo {pages} {103} (\bibinfo {year} {1991})}\BibitemShut {NoStop}%
\bibitem [{\citenamefont {Kaiser}(1995)}]{Kaiser:1994vs}%
  \BibitemOpen
  \bibfield  {author} {\bibinfo {author} {\bibfnamefont {D.~I.}\ \bibnamefont
  {Kaiser}},\ }\href {\doibase 10.1103/PhysRevD.52.4295} {\bibfield  {journal}
  {\bibinfo  {journal} {\emph {Phys. Rev. D}}\ }\textbf {\bibinfo {volume}
  {52}},\ \bibinfo {pages} {4295} (\bibinfo {year} {1995})},\ \Eprint
  {http://arxiv.org/abs/astro-ph/9408044} {arXiv:astro-ph/9408044} \BibitemShut
  {NoStop}%
\bibitem [{\citenamefont {De~Simone}\ \emph {et~al.}(2009)\citenamefont
  {De~Simone}, \citenamefont {Hertzberg},\ and\ \citenamefont
  {Wilczek}}]{DeSimone:2008ei}%
  \BibitemOpen
  \bibfield  {author} {\bibinfo {author} {\bibfnamefont {A.}~\bibnamefont
  {De~Simone}}, \bibinfo {author} {\bibfnamefont {M.~P.}\ \bibnamefont
  {Hertzberg}},  and \bibinfo {author} {\bibfnamefont {F.}~\bibnamefont
  {Wilczek}},\ }\href {\doibase 10.1016/j.physletb.2009.05.054} {\bibfield
  {journal} {\bibinfo  {journal} {\emph {Phys. Lett. B}}\ }\textbf {\bibinfo
  {volume} {678}},\ \bibinfo {pages} {1} (\bibinfo {year} {2009})},\ \Eprint
  {http://arxiv.org/abs/0812.4946} {arXiv:0812.4946 [hep-ph]} \BibitemShut
  {NoStop}%
\bibitem [{\citenamefont {Hamada}\ \emph {et~al.}(2014)\citenamefont {Hamada},
  \citenamefont {Kawai}, \citenamefont {Oda},\ and\ \citenamefont
  {Park}}]{Hamada:2014iga}%
  \BibitemOpen
  \bibfield  {author} {\bibinfo {author} {\bibfnamefont {Y.}~\bibnamefont
  {Hamada}}, \bibinfo {author} {\bibfnamefont {H.}~\bibnamefont {Kawai}},
  \bibinfo {author} {\bibfnamefont {K.-y.}\ \bibnamefont {Oda}},  and \bibinfo
  {author} {\bibfnamefont {S.~C.}\ \bibnamefont {Park}},\ }\href {\doibase
  10.1103/PhysRevLett.112.241301} {\bibfield  {journal} {\bibinfo  {journal}
  {\emph {Phys. Rev. Lett.}}\ }\textbf {\bibinfo {volume} {112}},\ \bibinfo
  {pages} {241301} (\bibinfo {year} {2014})},\ \Eprint
  {http://arxiv.org/abs/1403.5043} {arXiv:1403.5043 [hep-ph]} \BibitemShut
  {NoStop}%
\bibitem [{\citenamefont {Hamada}\ \emph {et~al.}(2015)\citenamefont {Hamada},
  \citenamefont {Kawai}, \citenamefont {Oda},\ and\ \citenamefont
  {Park}}]{Hamada:2014wna}%
  \BibitemOpen
  \bibfield  {author} {\bibinfo {author} {\bibfnamefont {Y.}~\bibnamefont
  {Hamada}}, \bibinfo {author} {\bibfnamefont {H.}~\bibnamefont {Kawai}},
  \bibinfo {author} {\bibfnamefont {K.-y.}\ \bibnamefont {Oda}},  and \bibinfo
  {author} {\bibfnamefont {S.~C.}\ \bibnamefont {Park}},\ }\href {\doibase
  10.1103/PhysRevD.91.053008} {\bibfield  {journal} {\bibinfo  {journal} {\emph
  {Phys. Rev. D}}\ }\textbf {\bibinfo {volume} {91}},\ \bibinfo {pages}
  {053008} (\bibinfo {year} {2015})},\ \Eprint {http://arxiv.org/abs/1408.4864}
  {arXiv:1408.4864 [hep-ph]} \BibitemShut {NoStop}%
\bibitem [{\citenamefont {Bezrukov}\ \emph {et~al.}(2015)\citenamefont
  {Bezrukov}, \citenamefont {Rubio},\ and\ \citenamefont
  {Shaposhnikov}}]{Bezrukov:2014ipa}%
  \BibitemOpen
  \bibfield  {author} {\bibinfo {author} {\bibfnamefont {F.}~\bibnamefont
  {Bezrukov}}, \bibinfo {author} {\bibfnamefont {J.}~\bibnamefont {Rubio}},
  and \bibinfo {author} {\bibfnamefont {M.}~\bibnamefont {Shaposhnikov}},\
  }\href {\doibase 10.1103/PhysRevD.92.083512} {\bibfield  {journal} {\bibinfo
  {journal} {\emph {Phys. Rev. D}}\ }\textbf {\bibinfo {volume} {92}},\
  \bibinfo {pages} {083512} (\bibinfo {year} {2015})},\ \Eprint
  {http://arxiv.org/abs/1412.3811} {arXiv:1412.3811 [hep-ph]} \BibitemShut
  {NoStop}%
\bibitem [{\citenamefont {Bezrukov}\ \emph {et~al.}(2018)\citenamefont
  {Bezrukov}, \citenamefont {Pauly},\ and\ \citenamefont
  {Rubio}}]{Bezrukov:2017dyv}%
  \BibitemOpen
  \bibfield  {author} {\bibinfo {author} {\bibfnamefont {F.}~\bibnamefont
  {Bezrukov}}, \bibinfo {author} {\bibfnamefont {M.}~\bibnamefont {Pauly}},
  and \bibinfo {author} {\bibfnamefont {J.}~\bibnamefont {Rubio}},\ }\href
  {\doibase 10.1088/1475-7516/2018/02/040} {\bibfield  {journal} {\bibinfo
  {journal} {\emph {JCAP}}\ }\textbf {\bibinfo {volume} {02}},\ \bibinfo
  {pages} {040} (\bibinfo {year} {2018})},\ \Eprint
  {http://arxiv.org/abs/1706.05007} {arXiv:1706.05007 [hep-ph]} \BibitemShut
  {NoStop}%
\bibitem [{\citenamefont {Horndeski}(1974)}]{Horndeski}%
  \BibitemOpen
  \bibfield  {author} {\bibinfo {author} {\bibfnamefont {G.~W.}\ \bibnamefont
  {Horndeski}},\ }\href {\doibase 10.1007/BF01807638} {\bibfield  {journal}
  {\bibinfo  {journal} {\emph {Int. J. Theor. Phys.}}\ }\textbf {\bibinfo
  {volume} {10}},\ \bibinfo {pages} {363} (\bibinfo {year} {1974})}\BibitemShut
  {NoStop}%
\bibitem [{\citenamefont {Deffayet}\ \emph {et~al.}(2011)\citenamefont
  {Deffayet}, \citenamefont {Gao}, \citenamefont {Steer},\ and\ \citenamefont
  {Zahariade}}]{Def11}%
  \BibitemOpen
  \bibfield  {author} {\bibinfo {author} {\bibfnamefont {C.}~\bibnamefont
  {Deffayet}}, \bibinfo {author} {\bibfnamefont {X.}~\bibnamefont {Gao}},
  \bibinfo {author} {\bibfnamefont {D.~A.}\ \bibnamefont {Steer}},  and
  \bibinfo {author} {\bibfnamefont {G.}~\bibnamefont {Zahariade}},\ }\href
  {\doibase 10.1103/PhysRevD.84.064039} {\bibfield  {journal} {\bibinfo
  {journal} {\emph {Phys. Rev. D}}\ }\textbf {\bibinfo {volume} {84}},\
  \bibinfo {pages} {064039} (\bibinfo {year} {2011})},\ \Eprint
  {http://arxiv.org/abs/1103.3260} {arXiv:1103.3260 [hep-th]} \BibitemShut
  {NoStop}%
\bibitem [{\citenamefont {Charmousis}\ \emph {et~al.}(2012)\citenamefont
  {Charmousis}, \citenamefont {Copeland}, \citenamefont {Padilla},\ and\
  \citenamefont {Saffin}}]{Charmousis:2011bf}%
  \BibitemOpen
  \bibfield  {author} {\bibinfo {author} {\bibfnamefont {C.}~\bibnamefont
  {Charmousis}}, \bibinfo {author} {\bibfnamefont {E.~J.}\ \bibnamefont
  {Copeland}}, \bibinfo {author} {\bibfnamefont {A.}~\bibnamefont {Padilla}},
  and \bibinfo {author} {\bibfnamefont {P.~M.}\ \bibnamefont {Saffin}},\ }\href
  {\doibase 10.1103/PhysRevLett.108.051101} {\bibfield  {journal} {\bibinfo
  {journal} {\emph {Phys. Rev. Lett.}}\ }\textbf {\bibinfo {volume} {108}},\
  \bibinfo {pages} {051101} (\bibinfo {year} {2012})},\ \Eprint
  {http://arxiv.org/abs/1106.2000} {arXiv:1106.2000 [hep-th]} \BibitemShut
  {NoStop}%
\bibitem [{\citenamefont {De~Felice}\ and\ \citenamefont
  {Tsujikawa}(2011{\natexlab{a}})}]{DeFelice:2011zh}%
  \BibitemOpen
  \bibfield  {author} {\bibinfo {author} {\bibfnamefont {A.}~\bibnamefont
  {De~Felice}} and \bibinfo {author} {\bibfnamefont {S.}~\bibnamefont
  {Tsujikawa}},\ }\href {\doibase 10.1088/1475-7516/2011/04/029} {\bibfield
  {journal} {\bibinfo  {journal} {\emph {JCAP}}\ }\textbf {\bibinfo {volume}
  {04}},\ \bibinfo {pages} {029} (\bibinfo {year} {2011}{\natexlab{a}})},\
  \Eprint {http://arxiv.org/abs/1103.1172} {arXiv:1103.1172 [astro-ph.CO]}
  \BibitemShut {NoStop}%
\bibitem [{\citenamefont {Gross}\ and\ \citenamefont
  {Sloan}(1987)}]{Gross:1986mw}%
  \BibitemOpen
  \bibfield  {author} {\bibinfo {author} {\bibfnamefont {D.~J.}\ \bibnamefont
  {Gross}} and \bibinfo {author} {\bibfnamefont {J.~H.}\ \bibnamefont
  {Sloan}},\ }\href {\doibase 10.1016/0550-3213(87)90465-2} {\bibfield
  {journal} {\bibinfo  {journal} {\emph {Nucl. Phys. B}}\ }\textbf {\bibinfo
  {volume} {291}},\ \bibinfo {pages} {41} (\bibinfo {year} {1987})}\BibitemShut
  {NoStop}%
\bibitem [{\citenamefont {Metsaev}\ and\ \citenamefont
  {Tseytlin}(1987)}]{Metsaev:1987zx}%
  \BibitemOpen
  \bibfield  {author} {\bibinfo {author} {\bibfnamefont {R.~R.}\ \bibnamefont
  {Metsaev}} and \bibinfo {author} {\bibfnamefont {A.~A.}\ \bibnamefont
  {Tseytlin}},\ }\href {\doibase 10.1016/0550-3213(87)90077-0} {\bibfield
  {journal} {\bibinfo  {journal} {\emph {Nucl. Phys. B}}\ }\textbf {\bibinfo
  {volume} {293}},\ \bibinfo {pages} {385} (\bibinfo {year}
  {1987})}\BibitemShut {NoStop}%
\bibitem [{\citenamefont {Gasperini}\ \emph {et~al.}(1997)\citenamefont
  {Gasperini}, \citenamefont {Maggiore},\ and\ \citenamefont
  {Veneziano}}]{Gasperini:1996fu}%
  \BibitemOpen
  \bibfield  {author} {\bibinfo {author} {\bibfnamefont {M.}~\bibnamefont
  {Gasperini}}, \bibinfo {author} {\bibfnamefont {M.}~\bibnamefont {Maggiore}},
   and \bibinfo {author} {\bibfnamefont {G.}~\bibnamefont {Veneziano}},\ }\href
  {\doibase 10.1016/S0550-3213(97)00149-1} {\bibfield  {journal} {\bibinfo
  {journal} {\emph {Nucl. Phys. B}}\ }\textbf {\bibinfo {volume} {494}},\
  \bibinfo {pages} {315} (\bibinfo {year} {1997})},\ \Eprint
  {http://arxiv.org/abs/hep-th/9611039} {arXiv:hep-th/9611039} \BibitemShut
  {NoStop}%
\bibitem [{\citenamefont {Kawai}\ \emph {et~al.}(1998)\citenamefont {Kawai},
  \citenamefont {Sakagami},\ and\ \citenamefont {Soda}}]{Kawai:1998ab}%
  \BibitemOpen
  \bibfield  {author} {\bibinfo {author} {\bibfnamefont {S.}~\bibnamefont
  {Kawai}}, \bibinfo {author} {\bibfnamefont {M.-a.}\ \bibnamefont {Sakagami}},
   and \bibinfo {author} {\bibfnamefont {J.}~\bibnamefont {Soda}},\ }\href
  {\doibase 10.1016/S0370-2693(98)00925-3} {\bibfield  {journal} {\bibinfo
  {journal} {\emph {Phys. Lett. B}}\ }\textbf {\bibinfo {volume} {437}},\
  \bibinfo {pages} {284} (\bibinfo {year} {1998})},\ \Eprint
  {http://arxiv.org/abs/gr-qc/9802033} {arXiv:gr-qc/9802033} \BibitemShut
  {NoStop}%
\bibitem [{\citenamefont {Cartier}\ \emph {et~al.}(2001)\citenamefont
  {Cartier}, \citenamefont {Hwang},\ and\ \citenamefont
  {Copeland}}]{Cartier:2001is}%
  \BibitemOpen
  \bibfield  {author} {\bibinfo {author} {\bibfnamefont {C.}~\bibnamefont
  {Cartier}}, \bibinfo {author} {\bibfnamefont {J.-c.}\ \bibnamefont {Hwang}},
  and \bibinfo {author} {\bibfnamefont {E.~J.}\ \bibnamefont {Copeland}},\
  }\href {\doibase 10.1103/PhysRevD.64.103504} {\bibfield  {journal} {\bibinfo
  {journal} {\emph {Phys. Rev. D}}\ }\textbf {\bibinfo {volume} {64}},\
  \bibinfo {pages} {103504} (\bibinfo {year} {2001})},\ \Eprint
  {http://arxiv.org/abs/astro-ph/0106197} {arXiv:astro-ph/0106197} \BibitemShut
  {NoStop}%
\bibitem [{\citenamefont {Calcagni}\ \emph {et~al.}(2005)\citenamefont
  {Calcagni}, \citenamefont {Tsujikawa},\ and\ \citenamefont
  {Sami}}]{Calcagni:2005im}%
  \BibitemOpen
  \bibfield  {author} {\bibinfo {author} {\bibfnamefont {G.}~\bibnamefont
  {Calcagni}}, \bibinfo {author} {\bibfnamefont {S.}~\bibnamefont {Tsujikawa}},
   and \bibinfo {author} {\bibfnamefont {M.}~\bibnamefont {Sami}},\ }\href
  {\doibase 10.1088/0264-9381/22/19/011} {\bibfield  {journal} {\bibinfo
  {journal} {\emph {Class. Quant. Grav.}}\ }\textbf {\bibinfo {volume} {22}},\
  \bibinfo {pages} {3977} (\bibinfo {year} {2005})},\ \Eprint
  {http://arxiv.org/abs/hep-th/0505193} {arXiv:hep-th/0505193} \BibitemShut
  {NoStop}%
\bibitem [{\citenamefont {Khan}\ and\ \citenamefont
  {Yogesh}(2022)}]{Khan:2022odn}%
  \BibitemOpen
  \bibfield  {author} {\bibinfo {author} {\bibfnamefont {H.~A.}\ \bibnamefont
  {Khan}} and \bibinfo {author} {\bibnamefont {Yogesh}},\ }\href {\doibase
  10.1103/PhysRevD.105.063526} {\bibfield  {journal} {\bibinfo  {journal}
  {\emph {Phys. Rev. D}}\ }\textbf {\bibinfo {volume} {105}},\ \bibinfo {pages}
  {063526} (\bibinfo {year} {2022})},\ \Eprint
  {http://arxiv.org/abs/2201.06439} {arXiv:2201.06439 [astro-ph.CO]}
  \BibitemShut {NoStop}%
\bibitem [{\citenamefont {Green}\ \emph {et~al.}(2004)\citenamefont {Green},
  \citenamefont {Liddle}, \citenamefont {Malik},\ and\ \citenamefont
  {Sasaki}}]{Green:2004wb}%
  \BibitemOpen
  \bibfield  {author} {\bibinfo {author} {\bibfnamefont {A.~M.}\ \bibnamefont
  {Green}}, \bibinfo {author} {\bibfnamefont {A.~R.}\ \bibnamefont {Liddle}},
  \bibinfo {author} {\bibfnamefont {K.~A.}\ \bibnamefont {Malik}},  and
  \bibinfo {author} {\bibfnamefont {M.}~\bibnamefont {Sasaki}},\ }\href
  {\doibase 10.1103/PhysRevD.70.041502} {\bibfield  {journal} {\bibinfo
  {journal} {\emph {Phys. Rev. D}}\ }\textbf {\bibinfo {volume} {70}},\
  \bibinfo {pages} {041502} (\bibinfo {year} {2004})},\ \Eprint
  {http://arxiv.org/abs/astro-ph/0403181} {arXiv:astro-ph/0403181} \BibitemShut
  {NoStop}%
\bibitem [{\citenamefont {Young}\ \emph {et~al.}(2014)\citenamefont {Young},
  \citenamefont {Byrnes},\ and\ \citenamefont {Sasaki}}]{Young:2014ana}%
  \BibitemOpen
  \bibfield  {author} {\bibinfo {author} {\bibfnamefont {S.}~\bibnamefont
  {Young}}, \bibinfo {author} {\bibfnamefont {C.~T.}\ \bibnamefont {Byrnes}},
  and \bibinfo {author} {\bibfnamefont {M.}~\bibnamefont {Sasaki}},\ }\href
  {\doibase 10.1088/1475-7516/2014/07/045} {\bibfield  {journal} {\bibinfo
  {journal} {\emph {JCAP}}\ }\textbf {\bibinfo {volume} {07}},\ \bibinfo
  {pages} {045} (\bibinfo {year} {2014})},\ \Eprint
  {http://arxiv.org/abs/1405.7023} {arXiv:1405.7023 [gr-qc]} \BibitemShut
  {NoStop}%
\bibitem [{\citenamefont {Harada}\ \emph {et~al.}(2013)\citenamefont {Harada},
  \citenamefont {Yoo},\ and\ \citenamefont {Kohri}}]{Harada:2013epa}%
  \BibitemOpen
  \bibfield  {author} {\bibinfo {author} {\bibfnamefont {T.}~\bibnamefont
  {Harada}}, \bibinfo {author} {\bibfnamefont {C.-M.}\ \bibnamefont {Yoo}},
  and \bibinfo {author} {\bibfnamefont {K.}~\bibnamefont {Kohri}},\ }\href
  {\doibase 10.1103/PhysRevD.88.084051} {\bibfield  {journal} {\bibinfo
  {journal} {\emph {Phys. Rev. D}}\ }\textbf {\bibinfo {volume} {88}},\
  \bibinfo {pages} {084051} (\bibinfo {year} {2013})},\ \bibinfo {note}
  {[Erratum: Phys.Rev.D 89, 029903 (2014)]},\ \Eprint
  {http://arxiv.org/abs/1309.4201} {arXiv:1309.4201 [astro-ph.CO]} \BibitemShut
  {NoStop}%
\bibitem [{\citenamefont {Germani}\ and\ \citenamefont
  {Musco}(2019)}]{Germani:2018jgr}%
  \BibitemOpen
  \bibfield  {author} {\bibinfo {author} {\bibfnamefont {C.}~\bibnamefont
  {Germani}} and \bibinfo {author} {\bibfnamefont {I.}~\bibnamefont {Musco}},\
  }\href {\doibase 10.1103/PhysRevLett.122.141302} {\bibfield  {journal}
  {\bibinfo  {journal} {\emph {Phys. Rev. Lett.}}\ }\textbf {\bibinfo {volume}
  {122}},\ \bibinfo {pages} {141302} (\bibinfo {year} {2019})},\ \Eprint
  {http://arxiv.org/abs/1805.04087} {arXiv:1805.04087 [astro-ph.CO]}
  \BibitemShut {NoStop}%
\bibitem [{\citenamefont {Kavanagh}(2019)}]{bradley_j_kavanagh_2019_3538999}%
  \BibitemOpen
  \bibfield  {author} {\bibinfo {author} {\bibfnamefont {B.~J.}\ \bibnamefont
  {Kavanagh}},\ }\href {\doibase 10.5281/zenodo.3538999} {\enquote {\bibinfo
  {title} {bradkav/pbhbounds: Release version},}\ } (\bibinfo {year}
  {2019})\BibitemShut {NoStop}%
\bibitem [{\citenamefont {Carr}(1975)}]{Carr:1975qj}%
  \BibitemOpen
  \bibfield  {author} {\bibinfo {author} {\bibfnamefont {B.~J.}\ \bibnamefont
  {Carr}},\ }\href {\doibase 10.1086/153853} {\bibfield  {journal} {\bibinfo
  {journal} {\emph {Astrophys. J.}}\ }\textbf {\bibinfo {volume} {201}},\
  \bibinfo {pages} {1} (\bibinfo {year} {1975})}\BibitemShut {NoStop}%
\bibitem [{\citenamefont {Kohri}\ and\ \citenamefont
  {Terada}(2018)}]{Kohri:2018awv}%
  \BibitemOpen
  \bibfield  {author} {\bibinfo {author} {\bibfnamefont {K.}~\bibnamefont
  {Kohri}} and \bibinfo {author} {\bibfnamefont {T.}~\bibnamefont {Terada}},\
  }\href {\doibase 10.1103/PhysRevD.97.123532} {\bibfield  {journal} {\bibinfo
  {journal} {\emph {Phys. Rev. D}}\ }\textbf {\bibinfo {volume} {97}},\
  \bibinfo {pages} {123532} (\bibinfo {year} {2018})},\ \Eprint
  {http://arxiv.org/abs/1804.08577} {arXiv:1804.08577 [gr-qc]} \BibitemShut
  {NoStop}%
\bibitem [{\citenamefont {Papanikolaou}\ \emph {et~al.}(2021)\citenamefont
  {Papanikolaou}, \citenamefont {Vennin},\ and\ \citenamefont
  {Langlois}}]{Papanikolaou:2020qtd}%
  \BibitemOpen
  \bibfield  {author} {\bibinfo {author} {\bibfnamefont {T.}~\bibnamefont
  {Papanikolaou}}, \bibinfo {author} {\bibfnamefont {V.}~\bibnamefont
  {Vennin}},  and \bibinfo {author} {\bibfnamefont {D.}~\bibnamefont
  {Langlois}},\ }\href {\doibase 10.1088/1475-7516/2021/03/053} {\bibfield
  {journal} {\bibinfo  {journal} {\emph {JCAP}}\ }\textbf {\bibinfo {volume}
  {03}},\ \bibinfo {pages} {053} (\bibinfo {year} {2021})},\ \Eprint
  {http://arxiv.org/abs/2010.11573} {arXiv:2010.11573 [astro-ph.CO]}
  \BibitemShut {NoStop}%
\bibitem [{\citenamefont {Namjoo}\ \emph {et~al.}(2013)\citenamefont {Namjoo},
  \citenamefont {Firouzjahi},\ and\ \citenamefont {Sasaki}}]{Namjoo:2012aa}%
  \BibitemOpen
  \bibfield  {author} {\bibinfo {author} {\bibfnamefont {M.~H.}\ \bibnamefont
  {Namjoo}}, \bibinfo {author} {\bibfnamefont {H.}~\bibnamefont {Firouzjahi}},
  and \bibinfo {author} {\bibfnamefont {M.}~\bibnamefont {Sasaki}},\ }\href
  {\doibase 10.1209/0295-5075/101/39001} {\bibfield  {journal} {\bibinfo
  {journal} {\emph {EPL}}\ }\textbf {\bibinfo {volume} {101}},\ \bibinfo
  {pages} {39001} (\bibinfo {year} {2013})},\ \Eprint
  {http://arxiv.org/abs/1210.3692} {arXiv:1210.3692 [astro-ph.CO]} \BibitemShut
  {NoStop}%
\bibitem [{\citenamefont {Franciolini}\ \emph {et~al.}(2018)\citenamefont
  {Franciolini}, \citenamefont {Kehagias}, \citenamefont {Matarrese},\ and\
  \citenamefont {Riotto}}]{Franciolini:2018vbk}%
  \BibitemOpen
  \bibfield  {author} {\bibinfo {author} {\bibfnamefont {G.}~\bibnamefont
  {Franciolini}}, \bibinfo {author} {\bibfnamefont {A.}~\bibnamefont
  {Kehagias}}, \bibinfo {author} {\bibfnamefont {S.}~\bibnamefont {Matarrese}},
   and \bibinfo {author} {\bibfnamefont {A.}~\bibnamefont {Riotto}},\ }\href
  {\doibase 10.1088/1475-7516/2018/03/016} {\bibfield  {journal} {\bibinfo
  {journal} {\emph {JCAP}}\ }\textbf {\bibinfo {volume} {03}},\ \bibinfo
  {pages} {016} (\bibinfo {year} {2018})},\ \Eprint
  {http://arxiv.org/abs/1801.09415} {arXiv:1801.09415 [astro-ph.CO]}
  \BibitemShut {NoStop}%
\bibitem [{\citenamefont {Cai}\ \emph {et~al.}(2018{\natexlab{b}})\citenamefont
  {Cai}, \citenamefont {Chen}, \citenamefont {Namjoo}, \citenamefont {Sasaki},
  \citenamefont {Wang},\ and\ \citenamefont {Wang}}]{Cai:2018dkf}%
  \BibitemOpen
  \bibfield  {author} {\bibinfo {author} {\bibfnamefont {Y.-F.}\ \bibnamefont
  {Cai}}, \bibinfo {author} {\bibfnamefont {X.}~\bibnamefont {Chen}}, \bibinfo
  {author} {\bibfnamefont {M.~H.}\ \bibnamefont {Namjoo}}, \bibinfo {author}
  {\bibfnamefont {M.}~\bibnamefont {Sasaki}}, \bibinfo {author} {\bibfnamefont
  {D.-G.}\ \bibnamefont {Wang}},  and \bibinfo {author} {\bibfnamefont
  {Z.}~\bibnamefont {Wang}},\ }\href {\doibase 10.1088/1475-7516/2018/05/012}
  {\bibfield  {journal} {\bibinfo  {journal} {\emph {JCAP}}\ }\textbf {\bibinfo
  {volume} {05}},\ \bibinfo {pages} {012} (\bibinfo {year}
  {2018}{\natexlab{b}})},\ \Eprint {http://arxiv.org/abs/1712.09998}
  {arXiv:1712.09998 [astro-ph.CO]} \BibitemShut {NoStop}%
\bibitem [{\citenamefont {Atal}\ and\ \citenamefont
  {Germani}(2019)}]{Atal:2018neu}%
  \BibitemOpen
  \bibfield  {author} {\bibinfo {author} {\bibfnamefont {V.}~\bibnamefont
  {Atal}} and \bibinfo {author} {\bibfnamefont {C.}~\bibnamefont {Germani}},\
  }\href {\doibase 10.1016/j.dark.2019.100275} {\bibfield  {journal} {\bibinfo
  {journal} {\emph {Phys. Dark Univ.}}\ }\textbf {\bibinfo {volume} {24}},\
  \bibinfo {pages} {100275} (\bibinfo {year} {2019})},\ \Eprint
  {http://arxiv.org/abs/1811.07857} {arXiv:1811.07857 [astro-ph.CO]}
  \BibitemShut {NoStop}%
\bibitem [{\citenamefont {Passaglia}\ \emph {et~al.}(2019)\citenamefont
  {Passaglia}, \citenamefont {Hu},\ and\ \citenamefont
  {Motohashi}}]{Passaglia:2018ixg}%
  \BibitemOpen
  \bibfield  {author} {\bibinfo {author} {\bibfnamefont {S.}~\bibnamefont
  {Passaglia}}, \bibinfo {author} {\bibfnamefont {W.}~\bibnamefont {Hu}},  and
  \bibinfo {author} {\bibfnamefont {H.}~\bibnamefont {Motohashi}},\ }\href
  {\doibase 10.1103/PhysRevD.99.043536} {\bibfield  {journal} {\bibinfo
  {journal} {\emph {Phys. Rev. D}}\ }\textbf {\bibinfo {volume} {99}},\
  \bibinfo {pages} {043536} (\bibinfo {year} {2019})},\ \Eprint
  {http://arxiv.org/abs/1812.08243} {arXiv:1812.08243 [astro-ph.CO]}
  \BibitemShut {NoStop}%
\bibitem [{\citenamefont {Taoso}\ and\ \citenamefont
  {Urbano}(2021)}]{Taoso:2021uvl}%
  \BibitemOpen
  \bibfield  {author} {\bibinfo {author} {\bibfnamefont {M.}~\bibnamefont
  {Taoso}} and \bibinfo {author} {\bibfnamefont {A.}~\bibnamefont {Urbano}},\
  }\href {\doibase 10.1088/1475-7516/2021/08/016} {\bibfield  {journal}
  {\bibinfo  {journal} {\emph {JCAP}}\ }\textbf {\bibinfo {volume} {08}},\
  \bibinfo {pages} {016} (\bibinfo {year} {2021})},\ \Eprint
  {http://arxiv.org/abs/2102.03610} {arXiv:2102.03610 [astro-ph.CO]}
  \BibitemShut {NoStop}%
\bibitem [{\citenamefont {Biagetti}\ \emph {et~al.}(2021)\citenamefont
  {Biagetti}, \citenamefont {De~Luca}, \citenamefont {Franciolini},
  \citenamefont {Kehagias},\ and\ \citenamefont {Riotto}}]{Biagetti:2021eep}%
  \BibitemOpen
  \bibfield  {author} {\bibinfo {author} {\bibfnamefont {M.}~\bibnamefont
  {Biagetti}}, \bibinfo {author} {\bibfnamefont {V.}~\bibnamefont {De~Luca}},
  \bibinfo {author} {\bibfnamefont {G.}~\bibnamefont {Franciolini}}, \bibinfo
  {author} {\bibfnamefont {A.}~\bibnamefont {Kehagias}},  and \bibinfo {author}
  {\bibfnamefont {A.}~\bibnamefont {Riotto}},\ }\href {\doibase
  10.1016/j.physletb.2021.136602} {\bibfield  {journal} {\bibinfo  {journal}
  {\emph {Phys. Lett. B}}\ }\textbf {\bibinfo {volume} {820}},\ \bibinfo
  {pages} {136602} (\bibinfo {year} {2021})},\ \Eprint
  {http://arxiv.org/abs/2105.07810} {arXiv:2105.07810 [astro-ph.CO]}
  \BibitemShut {NoStop}%
\bibitem [{\citenamefont {Davies}\ \emph {et~al.}(2022)\citenamefont {Davies},
  \citenamefont {Carrilho},\ and\ \citenamefont {Mulryne}}]{Davies:2021loj}%
  \BibitemOpen
  \bibfield  {author} {\bibinfo {author} {\bibfnamefont {M.~W.}\ \bibnamefont
  {Davies}}, \bibinfo {author} {\bibfnamefont {P.}~\bibnamefont {Carrilho}},
  and \bibinfo {author} {\bibfnamefont {D.~J.}\ \bibnamefont {Mulryne}},\
  }\href {\doibase 10.1088/1475-7516/2022/06/019} {\bibfield  {journal}
  {\bibinfo  {journal} {\emph {JCAP}}\ }\textbf {\bibinfo {volume} {06}},\
  \bibinfo {pages} {019} (\bibinfo {year} {2022})},\ \Eprint
  {http://arxiv.org/abs/2110.08189} {arXiv:2110.08189 [astro-ph.CO]}
  \BibitemShut {NoStop}%
\bibitem [{\citenamefont {Kristiano}\ and\ \citenamefont
  {Yokoyama}(2022)}]{Kristiano:2022maq}%
  \BibitemOpen
  \bibfield  {author} {\bibinfo {author} {\bibfnamefont {J.}~\bibnamefont
  {Kristiano}} and \bibinfo {author} {\bibfnamefont {J.}~\bibnamefont
  {Yokoyama}},\ }\Eprint {http://arxiv.org/abs/2211.03395} {arXiv:2211.03395
  [hep-th]} \BibitemShut {NoStop}%
\bibitem [{\citenamefont {Inomata}\ \emph
  {et~al.}(2022{\natexlab{b}})\citenamefont {Inomata}, \citenamefont
  {Braglia},\ and\ \citenamefont {Chen}}]{Inomata:2022yte}%
  \BibitemOpen
  \bibfield  {author} {\bibinfo {author} {\bibfnamefont {K.}~\bibnamefont
  {Inomata}}, \bibinfo {author} {\bibfnamefont {M.}~\bibnamefont {Braglia}},
  and \bibinfo {author} {\bibfnamefont {X.}~\bibnamefont {Chen}},\ }\Eprint
  {http://arxiv.org/abs/2211.02586} {arXiv:2211.02586 [astro-ph.CO]}
  \BibitemShut {NoStop}%
\bibitem [{\citenamefont {De~Felice}\ and\ \citenamefont
  {Tsujikawa}(2011{\natexlab{b}})}]{DeFelice:2011uc}%
  \BibitemOpen
  \bibfield  {author} {\bibinfo {author} {\bibfnamefont {A.}~\bibnamefont
  {De~Felice}} and \bibinfo {author} {\bibfnamefont {S.}~\bibnamefont
  {Tsujikawa}},\ }\href {\doibase 10.1103/PhysRevD.84.083504} {\bibfield
  {journal} {\bibinfo  {journal} {\emph {Phys. Rev. D}}\ }\textbf {\bibinfo
  {volume} {84}},\ \bibinfo {pages} {083504} (\bibinfo {year}
  {2011}{\natexlab{b}})},\ \Eprint {http://arxiv.org/abs/1107.3917}
  {arXiv:1107.3917 [gr-qc]} \BibitemShut {NoStop}%
\bibitem [{\citenamefont {De~Felice}\ \emph {et~al.}(2011)\citenamefont
  {De~Felice}, \citenamefont {Kobayashi},\ and\ \citenamefont
  {Tsujikawa}}]{DeFelice:2011hq}%
  \BibitemOpen
  \bibfield  {author} {\bibinfo {author} {\bibfnamefont {A.}~\bibnamefont
  {De~Felice}}, \bibinfo {author} {\bibfnamefont {T.}~\bibnamefont
  {Kobayashi}},  and \bibinfo {author} {\bibfnamefont {S.}~\bibnamefont
  {Tsujikawa}},\ }\href {\doibase 10.1016/j.physletb.2011.11.028} {\bibfield
  {journal} {\bibinfo  {journal} {\emph {Phys. Lett. B}}\ }\textbf {\bibinfo
  {volume} {706}},\ \bibinfo {pages} {123} (\bibinfo {year} {2011})},\ \Eprint
  {http://arxiv.org/abs/1108.4242} {arXiv:1108.4242 [gr-qc]} \BibitemShut
  {NoStop}%
\end{thebibliography}%

\end{document}